\documentclass[review]{elsarticle}

\usepackage[colorlinks,citecolor=blue,linktoc=all,linkcolor=cyan]{hyperref}
\usepackage{graphicx}

\usepackage[T1]{fontenc}
\usepackage{dsfont}               % use mathds instead of mathbb for outline fonts
\usepackage{mathrsfs}             % provides mathscr without overwriting mathcal
\usepackage{slashed}              % For Dirac slash notation.
\usepackage{amsmath}
\usepackage{amssymb}
\usepackage{amsbsy}
\usepackage{amsfonts}

\usepackage{braket}
\usepackage{subcaption}
\usepackage{float}

\numberwithin{equation}{section}
\numberwithin{table}{section}
\numberwithin{figure}{section}

\journal{Progress in Particle and Nuclear Physics}

\usepackage{titlesec}
\usepackage{sectsty}
\titleformat{\section}{\normalfont\Large\bfseries}{\thesection}{1em}{}
\titleformat{\subsection}{\normalfont\large\bfseries}{\thesubsection}{1em}{}
\titleformat{\subsubsection}{\normalfont\normalsize\bfseries}{\thesubsubsection}{1em}{}
\begin{document}
	
	\begin{frontmatter}
		
		\title{Stochastic gravitational wave background: methods and Implications}

		%authors, affiliations, corresponding author mention 
		
		\author[UA]{Nick van Remortel\corref{Nick}}
		\cortext[Nick]{Corresponding author}
		\ead{nick.vanremortel@uantwerpen.be}
		
		\author[UA,Artemis]{Kamiel Janssens}

		\author[VUB,UA]{Kevin Turbang}
		
		\address[UA]{Universiteit Antwerpen, Prinsstraat 13, 2000 Antwerpen, Belgium}
		\address[Artemis]{Artemis, Universit\'e C\^ote d'Azur, Observatoire de la C\^ote d'Azur, CNRS, F-06304 Nice, France}
		\address[VUB]{Vrije Universiteit Brussel, Pleinlaan 2, 1050 Brussel, Belgium}

		\begin{abstract}
			Beyond individually resolvable gravitational wave events such as binary black hole and binary neutron star mergers, the superposition of many more weak signals coming from a multitude of sources is expected to contribute to an overall background, the so-called stochastic gravitational wave background. In this review, we give an overview of possible detection methods in the search for this background and provide a detailed review of the data-analysis techniques, focusing primarily on current Earth-based interferometric gravitational-wave detectors. In addition, various validation techniques aimed at reinforcing the claim of a detection of such a background are discussed as well. We conclude this review by listing some of the astrophysical and cosmological implications resulting from current upper limits on the stochastic background of gravitational waves.
		\end{abstract}
		
		
	\end{frontmatter}
	
	\newpage
	
	\thispagestyle{empty}
	\tableofcontents
	
	%to begin the line numbers: 
	%\linenumbers

	%beginning of the core of the manuscript
	\newpage
	
	\paragraph{List of abbreviations}~\\~\\
    ASAF: all-sky all-frequency\\
    ASD: amplitude spectral density \\
    BBH: binary black hole\\
    BBN: Big Bang nucleosynthesis\\
    BBR: broadband radiometer\\
    BNS: binary neutron star\\
    CBC: compact binary coalescence\\
    CMB: cosmic microwave background\\
    ET: Einstein Telescope\\
    FOPT: first order phase transition\\
    GW: gravitational wave\\
    LISA: Laser Interferometer Space Antenna\\
    LVK: LIGO-Virgo-KAGRA\\
    MSP: millisecond pulsar\\
    NBR: narrowband radiometer\\
    ORF: overlap reduction function\\
    PBH: primordial black holes\\
    PI: power-law integrated (sensitivity curve)\\
    PSD: power spectral density\\
    PTA: pulsar timing array\\
    SGWB: stochastic gravitational wave background\\
    SMBHB: supermassive binary black hole binaries\\
    SNR: signal-to-noise ratio\\
    TBS: the Bayesian search\\

	\newpage
%%%%%%%%%%%%%%%%%%%%
%%% Introduction %%%
%%%%%%%%%%%%%%%%%%%%

\section{Introduction to gravitational waves}
\label{sec:Introduction}

The discovery of the first gravitational wave (GW) signal due to the  merger of two black holes \cite{PhysRevLett.116.061102} made by the LIGO and Virgo collaborations in 2015 propelled the field of GW astronomy into a new era. Since then, several tens of binary coalescence signals have been detected, including the merger of two black holes with a wide and continuous range of composite stellar size masses \cite{GWTC1,PhysRevX.11.021053,GWTC2.1,theligoscientificcollaboration2022population}. In addition, evidences for the binary merger of a black hole and a neutron star \cite{Abbott_2021_NS} and two neutron stars \cite{PhysRevLett.119.161101,ApJL_892_L3} have been collected as well. The latter observation was followed up by several observations in the electromagnetic spectrum \cite{Abbott_2017, Abbott_2017_2}, which kick-started the field of gravitational-wave multi-messenger astronomy. The signals of all these events, which have characteristic transient features, are of relative short duration ($\mathcal{O}$(1-100 s)), corresponding to the time during which the signal remains within the sensitive frequency band of current earth bound interferometers, and have relatively large amplitudes that exceed the intrinsic noise levels of the detectors. Many other astrophysical sources are predicted to yield detectable GW signals that are either continuous in nature and which typically originate from asymmetrical rotating compact objects such as pulsars, or are burst-like, such as supernovae type objects. None of these signals 
have been detected so far, but could well be in reach of operational and planned Earth-based and satellite borne GW detectors~\cite{RILES2013}. 
Finally, there is also bound to be a stochastic gravitational wave background (SGWB) in our Universe that can contain several components. The first component is of astrophysical nature and consists of the random superposition of individually unresolved signals from the entire population of astrophysical sources listed above. In addition, signals from cosmological events or structures could be present as well~\cite{Caprini_2018}. The SGWB is persistent but can have an intermittent nature, has no phase coherence, and is in several experimental conditions, such as for unresolved binary coalescences and cosmological signals searched for by current earth based interferometers, buried under the intrinsic noise level of a single detector. Such a signal has a small but non-negligible contribution to the total energy content of our Universe. Its detection, and in particular a primordial or cosmological component, would be as significant as the discovery of the cosmic microwave background (CMB) \cite{1965ApJ_142_419P, Durrer_2015}. Its spectral structure will yield information on the dynamical properties of its contributors and on the cosmological evolution of our Universe, and up to its earliest time scales, way before the decoupling of the CMB.

In the past decades, many review papers on the SGWB have been published~\cite{PhysRevD.46.5250,PhysRevD.48.2389,PhysRevD.55.448,SGWBRevAllen,SGWBRevRomano,SGWBRevChristensen,galaxies10010034}, where the philosophy and explanation of the mathematical framework in this paper has the most overlap with the earlier work in \cite{SGWBRevRomano}. In this paper we discuss the properties of the possible components of the SGWB, their theoretical and experimental bounds, the state-of-the-art of detection techniques and an outlook for future observations. We will refrain from giving an extensive review of the astrophysical, cosmological and particle physics inspired models for the generation of a SGWB, as these have been presented in other recent reviews, see e.g. \cite{Regimbau_2008_rev,Regimbau_2011_rev,SGWBRevChristensen,Caprini_2018,Caprini_2009,PhysRevD.79.083519,Caprini_2016}. We try to complement previous work by giving an update on the latest results and upper limits achieved by Earth-based interferometric gravitational-wave detectors as well as pulsar timing arrays. The main focus of this paper will be on the analysis techniques that are used or being investigated for data analysis of Earth-based interferometers. Since we are nearing the first detection of a SGWB with the continuously increasing sensitivity of the detectors, we will discuss several techniques that can be used to prove the observed signal is due to GWs and not to a terrestrial or instrumental noise source.\\

	\newpage
%%%%%%%%%%%%%%%%%%%%%%%
%%% SGWB properties %%%
%%%%%%%%%%%%%%%%%%%%%%%

\section{Stochastic background: definitions hypotheses and properties}
\label{sec:SGWBProperties}
In addition to individually detectable GW sources with generally deterministic signal properties, the Universe is permeated by a SGWB.
Gradual understanding of the properties of this type of GW signal and of its detectability has been accumulated since the 1980's by the works of Michelson~\cite{10.1093/mnras/227.4.933}, Christensen~\cite{PhDThesisNelson,PhysRevD.46.5250,PhysRevD.55.448,SGWBRevChristensen} and Flanagan~\cite{PhysRevD.48.2389}, and has been extensively reviewed and expanded upon by Allen and Romano~\cite{SGWBRevAllen}, Romano and Cornish~\cite{SGWBRevRomano}, and more recently in \cite{galaxies10010034}. The discussion on the general properties of the SGWB is still ongoing, but in general terms it is the result of the incoherent sum of a large amount of weak, unresolvable sources. If you identify these sources with the way galaxies are distributed within the universe at its largest scales, they are distributed almost isotropically across the Universe. Depending on the intrinsic sensitivity of current and planned observatories, and the accessible GW frequency range, the isotropy of the detectable background is not completely obvious. For example, the LISA space mission \cite{amaroseoane2017laser} will observe a galactic 'foreground' of binary white dwarfs \cite{amaroseoane2017laser}, and Earth bound observatories could detect significant contributions of binary pulsars in nearby galaxy clusters~\cite{PhysRevD.89.084076}. In addition, the direct detection of binary mergers with deterministic signal characteristics should be subtracted from the weaker, unresolved background radiation. In that sense, 'weak' is a relative and evolving concept. By virtue of the central limit theorem, the amplitudes of the stochastic background originating from an incoherent superposition of independent sources should be Gaussian, but only in the limit of large numbers. Current estimates on the population and merger rates~\cite{theligoscientificcollaboration2022population} of binary systems of black holes, neutron stars and black hole - neutron star systems indicate that for current observatories, the binary black hole merger signals that are detectable given the limited frequency band at which the detectors have maximal sensitivity, have an intermittent nature whose time structure is more `popcorn' like, rather than stationary \cite{physrevd.79.062002,Coward_2006,Regimbau_2008_rev,Regimbau_2011_rev,PhysRevD.86.122001}. This translates in a so-called small duty cycle, which corresponds to the probability of occurrence of a GW signal from these sources in a data analysis time window.

The SGWB signal is in general approximated as a weak signal. Its power is small compared to the power spectral densities of individual detectors. One therefore relies in most cases on the cross-correlation of outputs from two interferometers, defining a baseline that is given by their locations, relative separation, and relative orientations of the interferometer arms~\cite{SGWBRevAllen,SGWBRevRomano}.  Using the cross-correlation technique, the property of independence and stationarity implies that any correlation depends on time differences, i.e. if $h_A(t)$ and $h_{A'}'(t')$ are the strain output of two detectors with respective polarizations $A$ and $A'$, then the statistical correlator $\braket{h_A(t)h_{A'}'(t')}$ will only depend on $t-t'$, or in the frequency domain
\begin{equation}
\braket{\tilde{h}_A(f)\tilde{h}_{A'}'(f')}\propto \delta (f-f')\delta_{AA'}
\end{equation}
If the amplitudes are Gaussian, then all N-point correlators should reduce to products of two-point correlators, or the expectation values of a single strain amplitude. These approximations are currently followed in the vast majority of SGWB data analyses.

In the absence of detection of a SGWB signal, one generally puts upper limits on the total energy density of the SGWB which, as originally outlined in~\cite{PhysRev.166.1263}, relates to the cross-correlation of first time derivative of the strain amplitudes in frequency space of two observatories that point at a location in the sky, $\hat{\bf{n}}$, at any given siderial time, $t$:
\begin{equation}
\rho_{GW}=\frac{c^2}{32\pi G}\braket{\dot{h}(\hat{\bf{n}},t) \dot{h}'(\hat{\bf{n}},t)}.
\end{equation}
%where $(\hat{\bf{n}},\hat{\bf{n}}')$ and $(t,t')$ represent %respectively the directional vectors from each of the two %observatories towards the source of the SGWB signal, and their %respective arrival times. 
The energy density, $\rho_{GW}$, is generally expressed as a dimensionless quantity by normalizing it to the critical density of the Friedmann Universe without net curvature
\begin{equation}
 \rho_c=\frac{3c^2H^2_0}{8\pi G},
\end{equation}
in order to obtain the dimensionless energy density of the SGWB per logaritmic interval of frequency, $f$,
\begin{equation}
\label{eq:omega_GW}
 \Omega_{GW}(f,\hat{\bf{n}})=\frac{f}{\rho_c}\frac{d^3\rho_{GW}}{df d^2\hat{\bf{n}}},
\end{equation}
The assumption of isotropy can be maintained or abandoned. The SGWB component attributed to the early Universe is generally assumed to be isotropically distributed in the same degree as the electromagnetic 2.7 K radiation attributed to the CMB, but if one considers the distribution of luminous matter within our local environment, and notably our own galaxy, large anisotropies can be expected, depending on the relative strength of the nearby sources and the dominant frequencies at which they emit GWs, as compared to the frequency band an observatory is sensitive at~\cite{PhysRevD.56.545}. 
Assuming that in the limit of weak signals one probes a limited set of GW source populations with known spectral shapes that will only exhibit angular dependency in their relative contributions to the overall observed frequency spectrum,  a factorization of the two-point correlators in a frequency and angular component is justified~\cite{PhysRevD.77.042002}. To our knowledge this  assumption is maintained in all exiting SGWB and CMB analyses. The angular component can be represented in a sky map that is either segmented in an array of pixels, or based on a spherical harmonic decomposition. The frequency dependence can be  integrated over a large frequency band or one can produce sky maps for specific narrow frequency bins.  Finally one can also target specific locations in the sky, such as the center of our galaxy, or known sources of possible GW activity in nearby galaxies. The final outcome of current data analyses is therefore either a set of sky maps that contain observations or upper bounds on the signal-to-noise ratio, or on correlated strain power. 
The different analysis techniques will be discussed in more detail in Sec. \ref{sec:AnalysisTechniques}.

	\newpage
%%%%%%%%%%%%%%%%%%%%%%%%%
%%% Detection Methods %%%
%%%%%%%%%%%%%%%%%%%%%%%%%

\section{Detection methods}
\label{sec:DetectionMethods}

A wide variety of data and experiments have been used to study GWs, and in particular SGWBs. We give a broad overview of three categories (resonant objects, interferometry, and cosmological measurements) and subsequently discuss planetary bodies, pulsar timing arrays, and laser interferometry in more detail.

A first method relies on resonant objects being excited by GWs, which was first conceived by Weber \cite{PhysRev.117.306}. 
If the GW has the same frequency as the natural frequency of the object, the object will be set into oscillation, similar to a tuning fork. 
Resonant bar detectors on Earth have been used to search for GWs in the kHz region \cite{Resonantbars2006}. Specifically searches for a SGWB in two bar detectors \cite{1999A&A...351..811A} as well as a bar detector and interferometric gravitational-wave detector, which will be shortly introduced, \cite{PhysRevD.76.022001} have been used.
However, also other objects such as the Earth or the Moon can be considered as a resonant object that can be excited by GWs, which will be discussed in more detail in Sec. \ref{sec:DetectionMethods_Others}. 
The Earth is sensitive to GWs with frequencies between 0.3 mHz and 5 mHz \cite{Coughlin_2014}, whereas the Moon would be sensitive to the frequency range between 1 mHz and 1 Hz \cite{Harms_2021}.

Thus far, the most successful method for GW detection relies on a beam of electromagnetic radiation traveling between two free falling test masses.
This beam can be used to measure the separation between the two test masses, which is altered when a GW passes.
Earth and space based interferometric gravitational-wave detectors, pulsar timing arrays as well as spacecraft Doppler tracking belong to this category. In Sec. \ref{sec:DetectionMethods_PTA} pulsar timing arrays are discussed, whereas in Sec. \ref{sec:DetectionMethods_LaserITF} we will discuss present day and future interferometric gravitational-wave detectors, which are often also referred to as `laser interferometers'. We will not discuss spacecraft Doppler tracking further, but refer the interested reader to \cite{10.12942/lrr-2006-1,SGWBRevRomano}. Neither will we discuss any of the proposed ground-based \cite{DIMOPOULOS200937,PhysRevD.93.021101} and space-based \cite{DIMOPOULOS200937,AGIS-LEO_2011,Gao_2018} atom interferometers.

A third method uses cosmological measurements to infer upper limits on a SGWB of cosmological origin. This particular type of SGWB could arise due to processes happening throughout the cosmological history of our Universe, from e.g. cosmic strings, phase transition or inflation (see \cite{Caprini_2018} for a complete review on the topic). Constraints can be placed on the SGWB using Cosmic Microwave Background (CMB) \cite{1965ApJ_142_419P, Durrer_2015} measurements, as well as primordial deuterium abundance from Big Bang Nucleosynthesis (BBN) \cite{Kolb:1990vq}. Indeed, the SGWB would contribute to the total radiation energy density of the Universe and thus, increase the expansion rate of the Universe. Therefore, one can put upper limits on this background to ensure consistency with CMB and BBN observations. Using data from the Planck experiment \cite{Planck2016}, the authors of \cite{PAGANO2016823} report 95\% upper limits on the dimensionless energy density of GWs $\Omega_{\rm GW}$: $h^2\Omega_{\rm GW}<1.2\times 10^6$ for frequencies $f>10^{-15}$ Hz., where $h^2$ parameterizes the value of the Hubble constant $H_0$, i.e. $H_0=h\times 100 \rm~km~s^{-1}~Mpc^{-1}$. 

We conclude this introduction to the detection methods by mentioning another possible detection method through astrometry. Here, the key concept of the method lies in the careful monitoring of the direction of light from distant sources, which would be affected by GWs. Therefore, this effect can be used by telescopes as a way to detect a SGWB. However, we do not provide further details and refer the interested reader to \cite{2011PhRvD.83b4024B,1996ApJ.465.566P}.

%-----------------------%
%-- Detection Methods --%
%-------- Others -------%
%-----------------------%

\subsection{Planetary bodies as resonant gravitational wave detectors}
\label{sec:DetectionMethods_Others}

In this section we will discuss how planetary bodies can be used as resonant objects to search for SGWBs, where we will use both the Earth and the Moon as examples.

GWs are expected to interact with elastic bodies and thus, with the Earth itself as well. Therefore, GWs can excite Earth's free oscillations, called normal modes. Using a network of gravimeters and seismometers, it is possible to monitor Earth's normal modes and excitations thereof by the passage of a GW \cite{Coughlin_2014}. By comparing this to the precise prediction for the Earth's response to the passage of a GW and the cross-correlation across various gravimeters and seismometers, one can put constraints on the amplitude of the SGWB. The authors of \cite{Coughlin_2014} use 10 years worth of seismometer and gravimeter data monitoring Earth's seismic activity to constrain the SGWB, resulting in the following constraints: $h^2\Omega_{\rm GW}<0.035-0.15$ for frequencies in the mHz band.

More recently, the idea to use the Moon as a GW experiment was put forward \cite{Harms_2021}. At the core of this idea lies the same principle as the one used for the Earth's normal modes measurements described above. The Lunar Gravitational-Wave Antenna experiment entails the deployment of seismometers on the Moon to measure its normal modes and measure or constrain GWs. This experiment would be sensitive to the normal modes of the Moon within the 1 mHz - 1 Hz frequency band excited by GWs. An advantage compared to using the Earth as a GW observatory is the lower seismic activity of the Moon. Indeed, the absence of of oceans and lower tidal activity increases the ability of seismometers on the Moon to detect GWs. The main seismic noise sources include Moonquakes and meteoroid impacts. The latter has been shown to be below the targeted sensitivity of the experiment \cite{Harms_2021}, while the annual rate of seismic energy release has been shown to be up to 8 orders of magnitude smaller than on Earth \cite{KHAN2013331}.

%-----------------------%
%-- Detection Methods --%
%---- Pulsar timing ----%
%-----------------------%

\subsection{Pulsar Timing}
\label{sec:DetectionMethods_PTA}

\begin{figure}[ht]
	\centering
	\includegraphics[scale=.65]{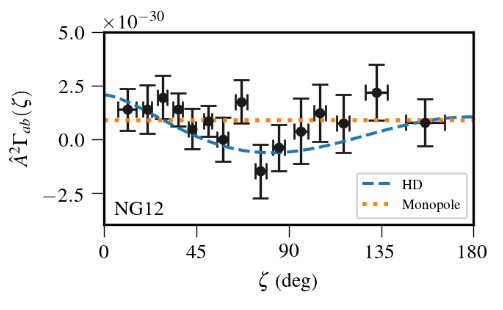}
	\caption{NANOGrav results for their 12.5 year dataset (results in black). The expected correlation between pulsars as a function of their angular separation $\zeta$ (as given by the Hellings and Downs curve) is shown in dashed red, whereas the monopole correlation is shown in dashed blue. The original version of this plot was presented in~\cite{2020}.}
	\label{Fig:HellingsAndDowns}
\end{figure}

Another method to detect a SGWB consists of using an array of pulsars. Pulsars are rotating neutron stars, whose rotation and magnetic axes are misaligned \cite{Lommen_2015}. Because of this misalignment, one observes regular radio pulses on Earth coinciding with the moment that the magnetic axis of the rotating pulsar is aligned with the Earth. Due to the regularity of these pulses, pulsars can be thought of as astrophysical clocks. In fact, deviations in the time of arrival of these pulses is what is sought for in pulsar timing arrays. Indeed, as a GW passes, it deforms spacetime, causing the time of arrival of pulses to deviate from their regular value. This effect is used to search for a SGWB.\\
A few examples of pulsar timing array collaborations are the \textit{North American Nanohertz Observatory of Gravitational Waves} (NANOGrav) \cite{McLaughlin:2013ira}, the \textit{European Pulsar Timing Array} (EPTA) \cite{EPTA2015} and the \textit{Parkes Pulsar Timing Array} (PPTA) \cite{PPTA:2013}. These experiments join efforts in the detection of GWs in the \textit{International Pulsar Timing Array} (IPTA) consortium \cite{IPTA2010}. \\
Pulsar timing arrays are most sensitive to frequency ranges around the 10 nHz band. Among the detectable sources by such experiments are super massive binary black hole mergers whose component masses are larger than $10^8~M_\odot$, which would emit a continuous signal over several thousands of years \cite{Lommen_2015}. In addition, a SGWB sourced by cosmic strings or primordial GWs could also be detectable at the frequencies accessible by pulsar timing arrays. \\
In order to look for a SGWB coming from the previously mentioned sources, one looks for deviations in times of arrival of pulsar pulses. These variations in times of arrival of the pulses are called timing residual. The perturbation in the residuals due to the passage of a GW can be parameterized as:
\begin{equation}
\label{Eq:PulsarRespons}
	R_{\rm GW}(t)=\frac{1}{2}\left(1+\cos\theta\right)\left(r_+(t)\cos(2\psi)+r_\times(t)\sin(2\psi)\right),
\end{equation}
where $\theta$ is the angle between the GW source and the pulsar and $\psi$ is the polarization angle of the GW source \cite{Lommen_2015}. Furthermore, one observes that in an ideal world, $R_{\rm GW}$ is simply the difference between the measured time of arrival of the pulse and the predicted time of arrival. However, in reality, this is not necessarily the case due to e.g. noise and deviations from timing model fits. The subscripts $\{+,\times\}$ stand for the two possible, 'plus' and 'cross', polarizations of the GWs as predicted by general relativity. In this expression, the functions $r_{+,\times}$ are related to the strain $h_{+,\times}$ as follows:
\begin{align}
	r_{+,\times}(t)&=r^e_{+,\times}(t)-r^p_{+,\times}(t)\\
	r^e_{+,\times}(t)&=\int_0^th^e_{+,\times}(\tau)d\tau\\
	r^p_{+,\times}(t)&=\int_0^th^p_{+,\times}\left(\tau-\frac{d}{c}(1-\cos\theta)\right)d\tau,
\end{align}
where $d$ is the distance between Earth and the pulsar, $t$ is the time variable, and $c$ is the speed of light. We refrain from going into details about the linear GW polarisations $h_{+,\times}$, but refer the interested reader to \cite{maggiore2008gravitational}. The superscripts $\{e,p\}$ denote the GW strain at Earth and at the pulsar, respectively. Note that the Earth term $r_{+,\times}^e$ and the pulsar term $r_{+,\times}^p$ are the same function, but evaluated at a different time which depends on the angle $\theta$ between the pulsar and the GW source as well as the distance to the pulsar.\\
Both the intrinsic pulsar noise as well as the passage of GWs would display a similar spectrum, such that the observation thereof is not sufficient to claim the detection of a SGWB. In order to claim a detection of the background, it is therefore essential to observe a spatial correlation across the various pulsars caused by the passage of GWs. Indeed, the Earth term $r_{+,\times}^e$ in Eq.~\eqref{Eq:PulsarRespons} will be correlated across all pulsars, whereas the pulsar term $r_{+,\times}^p$ is not. This spatial correlation is known as the Hellings and Downs correlation \cite{osti_6268184} and is given by:
\begin{equation}
	C(\gamma)=\frac{3}{2}\left(\frac{1-\cos\gamma}{2}\ln\frac{1-\cos\gamma}{2}-\frac{1}{6}\frac{1-\cos\gamma}{2}+\frac{1}{3}\right),
\end{equation}
where $\gamma$ denotes the angle in the sky between two pulsars. This correlation is illustrated by the dashed red line in Fig~\ref{Fig:HellingsAndDowns}. 
\\
To end this section, we report the results of the NANOGrav Collaboration which published their results of a 12.5 year pulsar-timing dataset \cite{2020}. The analysis is performed using the timing residual cross-correlation spectrum, which takes the form:
\begin{equation}
	S_{ab}(f)=\Gamma_{ab}\frac{A_{\rm GW}^2}{12\pi^2}\left(\frac{f}{f_{\rm yr}}\right)^{-\gamma}f_{\rm yr}^{-3},
\end{equation}
where $\Gamma_{ab}$ is the overlap reduction function (ORF), $f_{\rm yr}=1\text{yr}^{-1}$, and the spectral index $\gamma$ which is expected to be $13/3$ for a super massive binary black hole background \cite{phinney2001practical}. In general, an ORF parameterizes the decrease in sensitivity due to the relative orientation and separation of detectors in a cross-correlation search \cite{SGWBRevRomano}. For an isotropic SGWB, the ORF for pulsars is expected to follow the Hellings and Downs "quadrupolar" correlation as introduced previously \cite{osti_6268184}. Other spatially correlated effects are expected to display different ORFs, e.g. a monopole correlation for the errors in the timescale or a dipole correlation for the systematic errors in solar system ephemeris modeling.\\
A strong evidence for a power-law stochastic process with common amplitude and spectral index across all pulsars is found in the NANOGrav 12.5 year dataset \cite{2020}.
Indeed, using the cross-correlation spectrum above, a Bayesian posterior for a spectrum with $\gamma=13/3$ with median characteristic GW strain amplitude $1.92\times10^{-15}$ at a reference frequency $f_{\rm ref}=1\text{yr}^{-1}$ is found. As mentioned above, in order to claim a detection of a SGWB, it is crucial to observe the quadrupolar Hellings-Downs spatial correlation \cite{osti_6268184}. However, this correlation remains unobserved within the results of NANOGrav, as illustrated in Fig~\ref{Fig:HellingsAndDowns}, and therefore, a discovery of a SGWB cannot be claimed. Furthermore, we note that the non-observation of a SGWB by NANOGrav is consistent with the results reported by other pulsar timing array collaborations such as IPTA, EPTA and PPTA \cite{10.1093/mnras/stab3418, 10.1093/mnras/stab2833,Goncharov_2021_PPTA}. Nevertheless, for all pulsar timing arrays more data in the coming years will increase the evidence for or against the detection of a SGWB.

%------------------------%
%--- Detection Methods --%
%- Laser interferometry -%
%------------------------%

\subsection{Laser Interferometry}
\label{sec:DetectionMethods_LaserITF}

In this section we will discuss some of the basic concepts of laser interferometry \cite{RILES2013} and how it can be used for observing GWs.
The response of an interferometer to GWs can be expressed as a \textit{strain} measured in the two arms of an interferometer: $h=\frac{\Delta L(t)}{L}$, where $L$ is the arm length of the interferometer and $\Delta L$ is the change in length due to the passage of a GW.

The strain signal $s(t)$ introduced in a laser intereferometric detector by GWs -- in the long wavelength approximation, i.e. when the wavelength of the GWs is large compared to the detector's geometry -- considering multiple polarization modes $h_A(t)$ is given by \cite{2022PhRvD.105h4002W,anderson_2011}
\begin{equation}
\label{eq:strainresponse}
\begin{aligned}
s(t) &= \sum_m d_{ij} e_A^{ij} h_m(t),
\end{aligned}
\end{equation}
where we have assumed a sum over $i$,$j$ and $e_A^{ij}$ is the symmetric polarization tensor of mode $A$. The detector response is given by \cite{PhysRevD.86.122001,2022PhRvD.105h4002W}
\begin{equation}
\label{eq:detectorresponse_a}
\begin{aligned}
\bf{d} &= \frac{1}{2}(\bf{e^1}\otimes\bf{e^1} - \bf{e^2}\otimes\bf{e^2})\text{, with components}\\
d_{ij} &= \frac{1}{2}(e^1_i e^1_j - e^2_ie^2_j),
\end{aligned}
\end{equation}
where $\bf{e^1}$ and $\bf{e^2}$ are the unit vectors along the interferometer's arms and $\otimes$ represent a tensorial product. For interferometer I we can define
\begin{equation}
    \label{eq:antenaResponseF}
    F_I^A(\hat{\bf n}) = d_{I,ij} e_A^{ij}(\hat{\bf n}),
\end{equation}
where we have explicitly introduced the dependence on the location of the GW source, with $\hat{\textbf{n}}$ a vector pointing towards the source.

Earth-based interferometric gravitational-wave detectors have an arm length of several kilometers and are sensitive to GWs from a few Hz to a few kHz. To increase their sensitivity to passing GWs, they use a Fabry-Perot cavity to increase the effective path length of the laser light \cite{RILES2013}.

Currently there are five operational Earth-based interferometric gravitational-wave detectors: Advanced LIGO Hanford, Advanced LIGO Livingston~\cite{ADV_LIGO_2015}, Advanced Virgo~\cite{VIRGO:2014yos}, KAGRA~\cite{PhysRevD.88.043007}  and GEO600~\cite{Luck:2010rt}. However, both KAGRA and GEO600 did not reach the required sensitivity to observe the GW events observed by the LIGO and Virgo instruments during the most recent third observing run (O3) by the LIGO, Virgo and KAGRA (LVK) collaborations. KAGRA should reach a similar level of sensitivity over the coming years with the planned upgrades \cite{PhysRevD.88.043007,Akutsu_2020,10.1117/12.2560824}, whereas GEO600 will not. GEO600 focuses on researching technical challenges as well as taking data over long observation periods when other observatories (e.g. Advanced LIGO and Advanced Virgo) are being upgraded \cite{Dooley_2015}.

\begin{figure}[ht]
\centering
\includegraphics[width=0.6\textwidth]{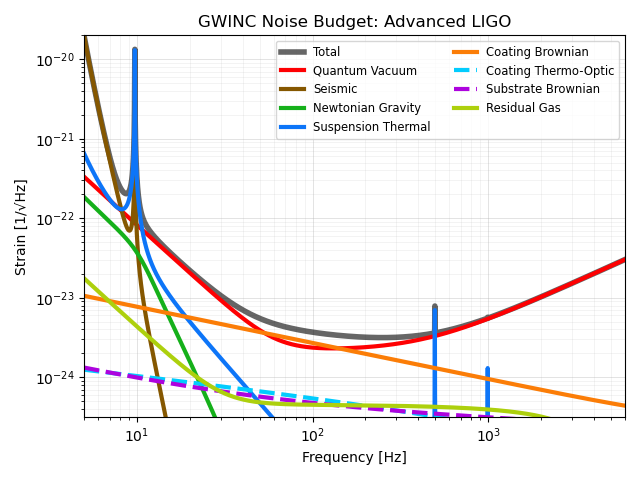}
\caption{Principal noise contributions for the nominal (high power, broadband) mode of
operation of Advanced LIGO. This figure was produced using the open source code pyGWINC \cite{2020ascl.soft07020R}. For more info on the principle noise sources, see \cite{ADV_LIGO_2015}.}
\label{fig:ALIGO_Sensitivity}
\end{figure}

As an example we show Advanced LIGO's design sensitivity curve~\cite{ADV_LIGO_2015} in Fig \ref{fig:ALIGO_Sensitivity}. This sensitivity curve is given by the detectors amplitude spectral density (ASD), which is the square root of the power spectral density (PSD) given by the auto-correlation of the detector's strain. Together with the design sensitivity, Fig \ref{fig:ALIGO_Sensitivity} also contains the budget of the main noise sources. At low frequencies ($\lesssim$ 11 Hz) seismic noise is dominant. Quantum noise is a combination of shot noise at high frequencies and radiation pressure at low frequencies. Together with thermal noise in the suspensions and mirror coatings ('Coating Brownian noise'), these are the most important noise sources building up Advanced LIGO's sensitivity curve.

To extend the Earth-based LVK network, there are plans to have a third LIGO detector in India, which could become operational later this decade \cite{LIGO-M1100296-v2,saleem2021science}. At the same time, there are also plans to build interferometers that will be ten times more sensitive in their measured $strain$ than LIGO and Virgo. Two such examples are the Einstein Telescope (ET) \cite{Punturo_2010,Hild:2010id,Maggiore_2020,doi:10.1063/5.0018414,ETdesignRep} and Cosmic Explorer (CE) \cite{Reitze2019Cosmic}, planned to become operational sometime in the next decade. Whereas Advanced LIGO and Advanced Virgo are referred to as second generation GW-detectors, ET and CE are referred to as third generation interferometric gravitational-wave detectors.

To reach larger distances between test masses in order to explore lower frequency ranges, one can build interferometers in space. Until today, there are no operational space-based laser interferometers. Nevertheless, the Laser Interferometer Space Antenna (LISA) is planned to launch somewhere in the 2030s \cite{Danzmann_1996,amaroseoane2017laser}. As of May 2022, the LISA mission design passed through Phase A, allowing to proceed for final design, opening up the path towards adoption of the mission by 2024. With its 2.5 million km long arms, LISA will be able to measure GWs in the 0.1 mHz-1 Hz frequency band.
The proposed TianQin \cite{Luo_2016,10.1093/ptep/ptaa114} interferometer planned to start operating around 2035, will be sensitive to GWs in the same frequency region as LISA.
Also in the 2030s, the launch of another space-based interferometer called B-DECIGO is planned \cite{10.1093/ptep/ptab019}. B-DECIGO is the scientific pathfinder for DECIGO, but at the same time will also collect scientific data, although with less sensitivity compared to DECIGO. B-DECIGO and DECIGO will be observing GWs in the frequency band from 0.1 Hz to 10 HZ, bridging the gap between the LISA and Earth-based detectors' frequency bands.

Some examples of sources that can be observed by space-based interferometers are: inspirals of galactic white dwarf binaries, binary coalescences of massive black holes and the pre-merger phase of the CBC events observed by Earth-based interferometric gravitational-wave detectors \cite{GWSources_2009Rev,RILES2013,amaroseoane2017laser}.

From all the current direct observations of binary coalescences by LIGO and Virgo~\cite{theligoscientificcollaboration2022population}, it is shown that the merger rates of binary neutron stars lie between 10 Gpc$^{-3}$ yr$^{-1}$ and 1700 Gpc$^{-3}$ yr$^{-1}$. The neutron star - black hole merger rate is estimated to be between 7.8 Gpc$^{-3}$ yr$^{-1}$ and 140 Gpc$^{-3}$ yr$^{-1}$, and the binary black hole (BBH) merger rate between 17.9 Gpc$^{-3}$ yr$^{-1}$ and 44 Gpc$^{-3}$ yr$^{-1}$ at a fiducial redshift of (z = 0.2).

\begin{figure}[ht]
	\centering
	\includegraphics[width=0.7\columnwidth]{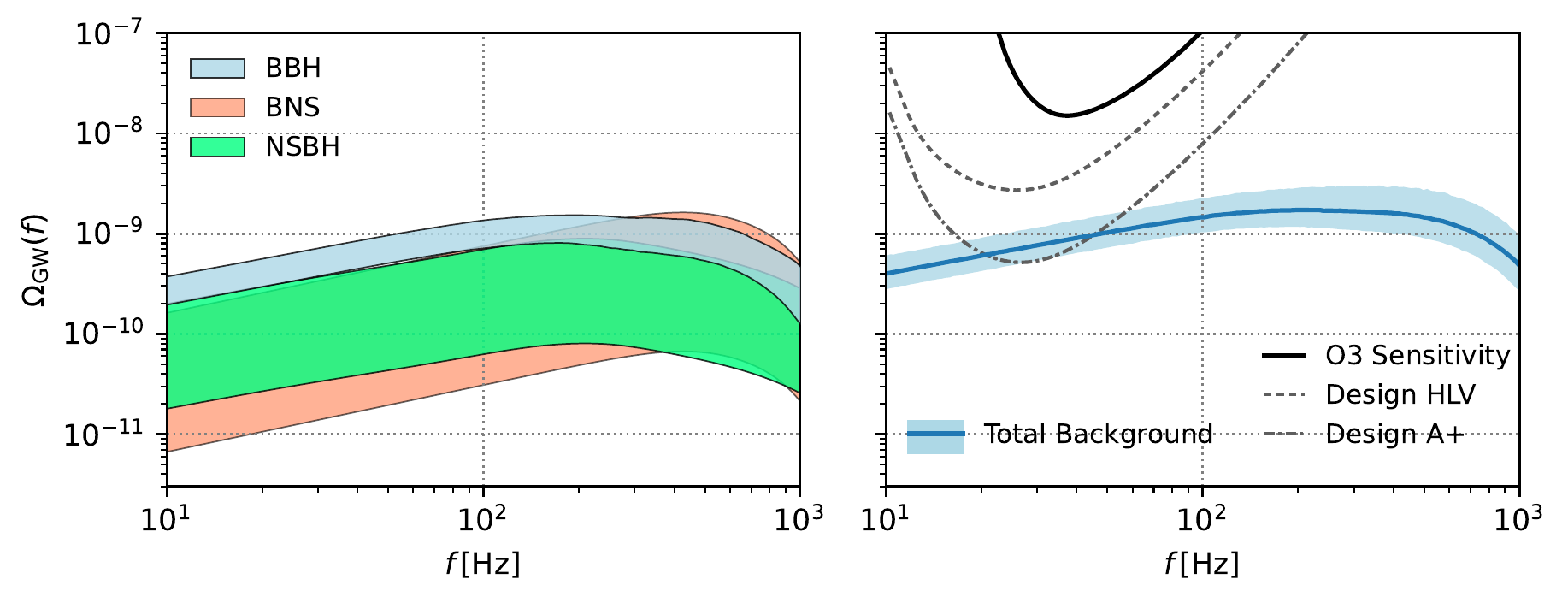}
	\caption{The predicted contribution to the astrophysical gravitational-wave background due to binary mergers following O3. \textbf{Left}: The individual
contributions expected from BNS, NSBH and BBH mergers. \textbf{Right}: Estimate of the total
gravitational-wave background (blue), as well as the  experimental current sensitivity (solid black) of the LVK detector network. For comparison,
we additionally show the expected sensitivities of the LIGO-Virgo network at design sensitivity, as well as that of LIGO’s
anticipated `A+' configuraton. The original version of this plot was presented in~\cite{theligoscientificcollaboration2022population}; the version shown here was obtained using open data published in \cite{ligo_scientific_collaboration_and_virgo_2021_5655785}}
	\label{Fig:pop-omega}
\end{figure}

However, the merger rate density of BBH is also observed to increase with redshift at a rate proportional to $(1+z)^\kappa$ with $\kappa = 2.7^{+1.8}_{-1.9}$.
While these densities are currently highly unconstrained for redshifts significantly greater than unity, they allow us to compute the expected contributions from these three classes of binary systems to $\Omega_{\rm GW}(f)$, integrated over the complete sky, as is shown in Fig~\ref{Fig:pop-omega}.
\\

The left panel of Fig~\ref{Fig:pop-omega} thus predicts the expected contribution, and corresponding uncertainties, of all CBC signals to the SGWB, while the right panel compares the total expected contribution to the power-integrated sensitivity curves, explained in sect.~\ref{sec:AnalysisTechniques_Isotropic}, of the Advanced LIGO-Virgo detector network during the most recent observation run (O3), to the design sensitivity of the Advanced detector configurations~\cite{ADV_LIGO_2015, VIRGO:2014yos}, and to the designed sensitivity of future upgrades of these detectors~\cite{PhysRevD.91.062005}. 
The A+ design sensitivity is expected to be reached during the fifth observing run O5 of the LVK collaborations. According to the LVK observing run plans (15 June 2022 update) \cite{LVKRunPlans} O5 should take place later this decade. The A+ power-integrated sensitivity curve in the right panel of Fig. \ref{Fig:pop-omega} represents a 2$\sigma$ detectability using 2 years of data with 50\% duty cycle. Under the assumption that the A+ design sensitivity will be reached in time and 2 years of data with 50\% duty cycle are collected this implies a preliminary 2$\sigma$ detection of the astrophysical component of the SGWB might be detectable before the end of this decade. Here we also have assumed that no other sources will contribute significantly to the detectable SGWB, and that no analysis improvements will be made that increase the detection sensitivity.

%%%%%%%%%%%%%%%%%%%%%%%%%%%
%%% Analysis techniques %%%
%%%%%%%%%%%%%%%%%%%%%%%%%%%

\section{Analysis techniques}
\label{sec:AnalysisTechniques}

In this section we will explain several analysis methods that can be used in the search for a SGWB, where we focus on techniques for Earth-based interferometric gravitational-wave detectors. For a review paper discussing a more general framework, we refer the interested reader to \cite{SGWBRevRomano}.\\
Starting from the general assumptions in Sec.~\ref{sec:SGWBProperties}, and taking into account the assumptions on stationarity and statistical independence, we will define proper statistical estimators for  $\Omega(f,\hat{\bf{n}})$:
\begin{equation}
\label{omega-spectral}
\Omega(f,\hat{\bf{n}})=\frac{2\pi^2}{3H_0^2}f^3 S(f,\hat{\bf{n}}),
\end{equation}
where $S(f,\hat{\bf{n}})$ is the spectral density obtained from the cross-correlation of Fourier transformed strain outputs of two detectors used in the analysis. An isotropic analysis
integrates over the full solid angle and $S(f,\hat{\bf{n}})$ reduces to $4\pi S(f)$. In anisotropic analyses $S(f,\hat{\bf{n}}) $ is decomposed in an eigenbasis as
\begin{equation}
  \label{PSD-spectral}
S(f,\hat{\bf{n}})=\sum_p S_p(f)\times e_p(\hat{\textbf{n}})
\end{equation}
where the  decomposition in eigenfunctions $e_p(\hat{\textbf{n}})$ is typically based on a pixel map or on spherical harmonic functions. In following subsections we will decribe in greater detail how a test statistic is constructed in each of these types of analyses and how one obtains a posterior likelihood or upper limits for $\Omega(f,\hat{\bf{n}})$.

In the following sections we will first focus on the cross-correlation analysis technique for an isotropic SGWB (Sec. \ref{sec:AnalysisTechniques_Isotropic}), followed by those for an anisotropic background (Sec. \ref{sec:AnalysisTechniques_Anisotropic}). We conclude this section on analysis techniques with a discussion on a search method under development to observe the SGWB coming from unresolved binary black hole coalesences (Sec. \ref{sec:AnalysisTechniques_Isotropic_TBS}). 
%-------------------------%
%-- Analysis techniques --%
%------- Isotropic -------%
%-------------------------%

\subsection{Cross-correlation analysis for the search for an isotropic SGWB}
\label{sec:AnalysisTechniques_Isotropic}

The cross-correlation analysis technique is currently used as the standard search for an isotropic SGWB. Typically one assumes the SGWB is Gaussian, unpolarized and stationary. 
To search for this SGWB one constructs a cross-correlation statistic $\hat{C}_{IJ}(f)$ \cite{SGWBRevAllen},

\begin{equation}
\label{eq:methods:bin_by_bin_estimator}
\hat{C}_{IJ}(f) =\frac{2}{T}  \frac{\Re {\rm e}[\tilde{s}_I^\star(f) \tilde{s}_J(f)]}{\gamma_{IJ}(f) S_0(f)},
\end{equation}
where $I$ and $J$ refer to the two detectors used in this cross-correlation search, $\tilde{s}_I(f)$ the Fourier transform of $s_I(t)$ -- the time domain strain data -- and $T$ the time segment duration of the Fourier transform, such that $\frac{2}{T}  \Re {\rm e}[\tilde{s}_I^\star(f) \tilde{s}_J(f)]$ is the cross power spectral density (CSD). $S_0(f)$ is a `spectral shape' for an $\Omega(f) =$ constant background, given by $S_0(f) = 3H_0^2(f)/(10\pi^2f^3)$, with $H_0$ being the Hubble constant.
Finally, $\gamma_{IJ}(f)$ -- the normalized ORF -- encodes the geometry of the detector baseline $IJ$ and is given by \cite{PhysRevD.46.5250}
\begin{equation}
\label{eq:ORF}
\gamma(f) = \frac{5}{8\pi}\sum_A\int_{\rm Sky} F^A_1(\hat{\textbf{n}})F^A_2(\hat{\textbf{n}})e^{2\pi if\Delta\textbf{x}\cdot\hat{\textbf{n}}/c}d\hat{\textbf{n}},
\end{equation}
where $A$ runs over the polarization of the GWs, $\hat{\textbf{n}}$ is a vector pointing towards the GW source, which is integrated over the entire sky for an isotropic signal, and $F^A_1(\hat{\textbf{n}})$ is the detector response function for a GW with polarization $A$ from a source in direction $\hat{\textbf{n}}$, as introduced in Eq.\ref{eq:antenaResponseF}.

Since the stationarity assumption breaks down for (very) long times, the data is divided into multiple segments for which the cross-correlation statistic $\hat{C}$ is calculated. The deviation from stationary is due to the time dependent levels of noise. For the purpose of stochastic searches performed by the LVK collaborations the data is expected to be stationary over segments $\sim \mathcal{O}(100s)$\cite{PhysRevD.104.022004}. This neglects the effect from short duration transient noise sources, often called 'glitches', which will be discussed more in Sec. \ref{sec:AnalysisTechniques_ValidationTechniques}.
Afterwards the multiple segments are combined, ideally using inverse noise weighting \cite{SGWBRevAllen,SGWBRevRomano}.
The variance of the cross-correlation statistic $\hat{C}$ in the small signal-to-noise ratio limit is given by 
\begin{equation}
\label{eq:def_sigma}
\sigma_{IJ}^2(f) \approx \frac{1}{2T \Delta f}  \frac{P_I(f) P_J(f)}{\gamma_{IJ}^2(f) S_0^2(f)},
\end{equation}
where $P_I(f)$ and $P_J(f)$ are the power spectral densities of detectors $I$ and $J$, respectively, and $\Delta f$ the used frequency resolution. 
The normalization is chosen such that $\hat{C}_{IJ}(f)$ is an estimator for the dimensionless fractional energy density in GWs $\Omega_{\rm GW}(f)$, integrated over all sky directions, as described in Eq.~\eqref{omega-spectral}.

The optimal estimator for a SGWB doesn't only depend on the cross-correlation, but also on the auto-correlation of both detectors $I$ and $J$. However, the noise power spectral densities of an instrument are often not known to the level of precision needed to accurately take the auto-correlation into account, as is the case for current generation ground based detectors Advanced LIGO, Advanced Virgo and KAGRA. Relying solely on the cross-correlation therefore yields an almost optimal estimator \cite{SGWBRevRomano}.

With this caveat the optimal way of combining the different frequency bins when performing a search for a SGWB -- with any given spectral shape -- is \cite{PhysRevD.104.022004}

\begin{eqnarray}
\label{eq:optimal-filter}
\hat C^{IJ} &=& \frac{\sum_k w(f_k) \hat C^{IJ}(f_k)\sigma^{-2}_{IJ}(f_k)}{\sum_k w(f_k)^{2}\sigma^{-2}_{IJ}(f_k)}, \nonumber \\
\sigma_{IJ}^{-2} &=& \sum_k w(f_k)^{2} \sigma^{-2}_{IJ}(f_k).
\end{eqnarray}
Here $f_k$ are a discrete set of frequencies and $w(f_k)$ are the optimal weights:
\begin{equation}
w(f) = \frac{\Omega_{\rm GW}(f)}{\Omega_{\rm GW}(f_{\rm ref})},
\label{eq:optimal-w}
\end{equation}
where $\Omega_{\rm GW}(f)$ is the spectral shape of the SGWB background for which the optimal filter is designed and $f_{\rm ref}$ is a fixed reference frequency, e.g. $f_{\rm ref} = 25$ Hz for LVK analyses \cite{PhysRevD.104.022004}. We want to note that the choice of $f_{\rm ref}$ is entirely arbitrary, the LVK collaborations have however chosen $f_{\rm ref} = 25$ Hz since it conveniently lies roughly at the start of their most sensitive region \cite{PhysRevD.104.022004}

When there are more than two detectors in the observing network this leads to more than one baseline. One can calculate a network cross-correlation statistic by taking the weighted average of the baseline cross-correlation statistics -- which is the preferred method over computing higher order correlations from more than two detectors -- \cite{SGWBRevAllen,Malaspinas_2005,Ballmer_2006,Mitra_2008,SGWBRevRomano}:
\begin{equation}
\label{eq:total_crossCorr}
\hat{C} = \frac{\sum_{IJ} \hat{C}^{IJ} \sigma_{IJ}^{-2}}{\sum_{IJ} \sigma_{IJ}^{-2}},\ \ \sigma^{-2} = \sum_{IJ} \sigma_{IJ}^{-2}.
\end{equation}

Often the expected signals behave (approximately) as a power-law and it is worth looking for signals using this assumption for $\Omega_{\rm GW}(f)$:
\begin{equation}
\label{eq:OmegaGW_PL}
\Omega_{\rm GW}(f) = \Omega_{\rm ref} \left(\frac{f}{f_{\rm ref}} \right)^{\alpha},
\end{equation}
where $\alpha$ is called the spectral index.
For the inspiral phase of compact binaries, it can be derived that this spectral index equals $\alpha=-2/3$~\cite{maggiore2008gravitational, PhysRevLett.116.131102, Regimbau_2011_rev}, while for core collapse supernovae and rotating neutron stars, its value is typically $\alpha=-3$~\cite{Regimbau2001, 10.1046/j.1365-8711.1999.02194.x}. For relics of inflationary cosmology, the spectral index is typically zero, but other cosmological models can result in broken power laws~\cite{PhysRevLett.98.111101, Caprini_2018}. More details on various astrophysical and cosmological model implications will be given in Section~\ref{sec:Implications}.
When searching for an isotropic SGWB, the likelihood used for parameter estimation adopts the form
\begin{equation} \label{eq:PE_likelihood}
p(\hat{C}_k^{IJ}|\mathbf{\Theta}) \propto \exp \left[ -\frac{1}{2} \sum_{IJ} \sum_k \left( \frac{\hat{C}_k^{IJ} - \Omega_{\rm GW}(f_k|\mathbf{\Theta})}{\sigma^2_{IJ}(f_k)}\right) \right],
\end{equation}
where $\hat{C}_k^{IJ}=\hat{C}^{IJ}(f_k)$ is the cross-correlation statistic as defined in Eq. \eqref{eq:methods:bin_by_bin_estimator}, $\sigma^2_{IJ}(f_k)$ its variance as defined in Eq. \eqref{eq:def_sigma} and $\Omega_{GW}(f_k|\bf{\Theta})$ refers to the desired model of the SGWB. Traditionally, one searches for signals with a power-law shape \cite{PhysRevD.104.022004}, as shown in Eq. \ref{eq:OmegaGW_PL}. However, more recently people are also considering broken power law models, often in the context of cosmological models such as first order phase transitions \cite{PhysRevLett.126.151301}, which will be discussed in more detail in Sec. \ref{sec:Implications_Cosmological}. By considering multiple signal models in Eq. \ref{eq:PE_likelihood} one can attempt to perform spectral separation \cite{PhysRevD.103.043023}, which will be shortly highlighted in Sec. \ref{sec:AnalysisTechniques_Isotropic_TBS}. Similarly it can also be used to account for correlated noise sources \cite{BayesianGWMag} as will be discussed in Sec. \ref{sec:AnalysisTechniques_ValidationTechniques_BayesianSchumann}.

At this time, it is worth defining a sensitivity curve which takes the broadband character of the signal into account. One can construct a curve such that at any frequency its tangent represents the sensitivity at which one could detect a power-law $\Omega_{\rm GW}(f)$ with an SNR of 1 for the given baseline~\cite{Thrane:2013oya}. This sensitivity curve is called the power-law integrated (PI) sensitivity curve and is often denoted by $\Omega^{\rm PI}_{IJ}(f)$ for the baseline $IJ$. The sensitivity curves in the right panel of Fig \ref{Fig:pop-omega} are PI sensitivity curves.

The results of this cross-correlation analysis technique applied to the latest data of the LIGO, Virgo and KAGRA collaborations allows one to constrain the SGWB \cite{PhysRevD.104.022004}. For the search for an isotropic SGWB, three different power-laws were considered for the dimensionless energy density of GWs, as introduced in Eq.\eqref{eq:OmegaGW_PL}, with $\alpha\in\{0, 2/3, 3\}$. The latest O3 analysis of the LVK collaboration constrains a power-law isotropic SGWB to have an amplitude $\Omega_{\rm ref}$ at a reference frequency $f_{\rm ref}=25$ Hz to be: $\Omega_{\rm ref}\le 5.8\times 10^{-9}$, $\Omega_{\rm ref}\le 3.4\times 10^{-9}$, $\Omega_{\rm ref}\le 3.9\times 10^{-10}$, for spectral indices $\alpha\in\{0, 2/3, 3\}$, respectively. \\~\\

%-------------------------%
%-- Analysis techniques --%
%------ Anisotropic ------%
%-------------------------%

\subsection{Anisotropic backgrounds}
\label{sec:AnalysisTechniques_Anisotropic}
At distance scales of the order of the diameter of superclusters of order $\thicksim 100$~Mpc, matter in our universe is isotropically distributed. On a local scale, in particular scales of the order of the diameter of our Local Group of galaxies of $\thicksim 3$~Mpc, and, corresponding to redshifts of $z\ll 1$, the distribution of astrophysical GW sources is not isotropic. This is generally known as the {\it cosmological principle}~\cite{ryden2017cosmology}.
Therefore, it is natural to assume that the SGWB will contain both an isotropic and anisotropic component. The relatively strong signals that are detectable from GW sources in our near environment are counterbalanced by the population growth of GW sources at larger distances. Their relative contribution is nontrivial and depends on the frequency range accessible by current and future experiments, on the redshift dependence of GW source populations and the expansion of our Universe.  

In addition to binary mergers, populations of millisecond pulsars could also contribute to the anisotropy in the SGWB in the frequency ranges accessible by the LVK detector network~\cite{PhysRevD.89.084076}, and binary white dwarfs in our own galaxy are estimated to be a sizable anisotropic astrophysical foreground in the frequency range accessible by the LISA mission \cite{amaroseoane2017laser}. GW signals emitted by a single asymmetrically rotating neutron star are continuous and typically peaked narrowly around a single frequency, but entire populations will give rise to a continuum of emissions over a broad range of frequencies.
Finally, just like in the case of the Cosmic Microwave Background~\cite{1992ApJ_396L_1S} the Earth's rotational and orbital motion can induce a  measurable modulation, and related dipole and higher order multi-pole asymmetry, in the detected signals from an isotropically distributed background. The total GW power associated with these multi-pole asymmetries decreases with increasing order and its detection depends naturally on the sensitivity of the detector network and the intensity of the total SGWB~\cite{PhysRevD.56.545}.

Anisotropic analyses relying on the radiometer principle adopt the formalism developed in~\cite{PhysRevD.56.545, Ballmer_2006, Cornish_2001, Mitra_2008, PhysRevD.80.122002, PhysRevD.76.082003}. For reasons explained at the bottom of Section~\ref{sec:SGWBProperties}, they all rely on the assumption that the spectral density $S(f,\hat{\bf{n}})$ from Eq.~\eqref{omega-spectral} factorizes as

\begin{equation}
\label{spectral_fact}
S(f,\hat{\bf{n}})=\mathcal{P}(\hat{\bf{n}})H(f),
\end{equation}
where $H(f)$ is a dimensionless function of frequency that describes the spectral shape, normalized so that $H(f_{\rm ref})=1$, where $f_{\rm ref}$ is a reference frequency, often taken around the most sensitive frequency of the detectors used in the cross-correlation analysis. In most analyses the spectral shape is modeled by a power law
\begin{equation}
\label{plaw}
H(f)=(f/f_{\rm ref})^\alpha,
\end{equation}
where $\alpha$ is referred to as the power law index or spectral index, as discussed in Section~\ref{sec:AnalysisTechniques_Isotropic}. 
In turn, $\mathcal{P}(\hat{\bf{n}})$ represents the  angular distribution of GW power. This distribution can be expanded in terms of a set of basis functions on the two-sphere

\begin{equation}
\label{eigenfunc_P_alpha}
\mathcal{P}(\hat{\bf{n}})=\sum_{a} \mathcal{P}_{a}\textbf{e}_{a}(\hat{\bf{n}}) ,
\end{equation}
where the choice of eigenbasis $\textbf{e}_{a}(\hat{\bf{n}})$ is determined by the angular features one is trying to detect.

In an anisotropic analysis, one uses a maximum-likelihood estimator for each of the $\mathcal{P}_{a}$ coefficients of the SGWB power, where the subscript $a$ can relate to the spherical harmonic function indices or the pixel index in a pixel map. The test statistic is, just like in the isotropic case, based on the cross-correlation spectrum obtained from two baselines, $I$ and $J$, evaluated at a set of discrete times, $t$, and discrete frequencies $f$

\begin{equation}
\label{teststat}
C_{f,t}=\frac{2}{T} \tilde{s}_I^\star(f,t) \tilde{s}_J(f,t),
\end{equation}
with an expectation value equal to

\begin{equation}
\label{expval}
\braket{C_{f,t}}=\sum_a H(f)\gamma_{a}(f,t)\mathcal{P}_{a},
\end{equation}
where we drop the subscripts $I$ and $J$ for clarity and where the index $a$ is the generic eigenbasis index from Eq.~\eqref{eigenfunc_P_alpha}.
The factor $\gamma_{a}(f,t)$ corresponds to the integrand of Eq.~\eqref{eq:ORF}, which now has a sidereal time dependence.
Eq.~\eqref{teststat} differs from the test statistic in Eq.~\eqref{eq:methods:bin_by_bin_estimator} by the fact that we consider the complete complex quantities $\tilde{s}_{I,J}(f)$, and by the denominator which is taken into account by the covariance matrix given by
\begin{equation}
\label{covariance}
N_{f,t}= P_I(f,t)P_J(f,t),
\end{equation}
where $P_{I,J}(f,t)$ are the one-sided power spectra of the detector outputs for time segment $t$.
It can be shown that the maximum likelihood estimators for $\mathcal{P}_{a}$ are given by

\begin{equation}
\label{estimators}
\hat{\mathcal{P}}_{a}=(\Gamma^{-1})_{ab}X_b,
\end{equation}
where
\begin{equation}
\label{xbeta}
X_b=\sum_t\sum_f \gamma^*_{b}(f,t)\frac{H(f)}{N_{f,t}}C_{f,t},
\end{equation}
are the components of the so-called `dirty' map, corresponding to the SGWB power on the sky as seen through the beam matrix of the two detectors (also referred to as the Fisher matrix), $\Gamma_{ab}$, given by
\begin{equation}
\label{matrix}
\Gamma_{ab}=\sum_t\sum_f \gamma^*_{a}(f,t)\frac{H^2(f)}{N_{f,t}}\gamma_{b}(f,t),
\end{equation}
which can be approximated for weak signals by
\begin{equation}
\label{approxmatrix}
\Gamma_{ab}\approx\braket{X_a X^*_b}-\braket{X_a}\braket{X^*_b}.
\end{equation}
One should note that the likelihood estimators $\hat{\mathcal{P}}_{a}$ of the
SGWB power across the sky are generally biased, since the assumed spectral shape of the SGWB, $H(f)$, does not-necessarily correspond to the real spectrum of the SGWB.
One therefore generally estimates  $\hat{\mathcal{P}}_a$ for various assumptions of the spectral index $\alpha$. 

    The biggest advantage of a directional search, however, is the potential signal-to-noise ratio (SNR) improvement with respect to an isotropic search~\cite{Ballmer_2006}, when the dominant source component is anisotropic. If all correlated signals were to come from a single point in the sky, denoted by $\hat{\bf{n}}$, the ratio between the isotropic and directional SNR values equals

\begin{equation}
\label{snrratio}
\frac{(SNR)_{iso}}{(SNR)_{dir}}=\frac{[\gamma_{iso},\gamma_{\hat{\bf{n}}}]}{\sqrt{[\gamma_{iso},\gamma_{iso}][\gamma_{\hat{\bf{n}}},\gamma_{\hat{\bf{n}}}]}},
\end{equation}
where the notation $[A,B]$ is shorthand for $[A,B]=\int df A^*(f)H^2(f)B(f)/P_IP_J$, for any quantity $A$ and $B$ (in this case the ORF for the istropic and directional cases), and $P_{I,J}$ are the respective PSDs for the two detectors used in the cross-correlation analysis. This ratio is bounded between -1 and 1 which implies not only that the directional search outperforms the isotropic search for a single point source, but also for a point source at an unfortunate position, the isotropic search can yield negative or zero correlated power.\\

In practice, the Fisher matrix, described in Eq.~\eqref{matrix}, is somewhat ill-conditioned due to the fact that the detector pair is limited in its spatial resolution. When choosing a basis with higher spatial resolution, the signal-to-noise ratio gets reduced for the deconvoluted sky maps. Secondly, there are certain power distributions, $\hat{\mathcal{P}}_{a}$ to which each detector pair is blind, resulting in corresponding dirty map values of $X_b\approx 0$. Both problems have to be treated with care and depend on the chosen decomposition basis. Some aspects will be discussed in the following sections.

These considerations imply the use of several strategies to optimally detect an anisotropic SGWB, that is either that of a broadband radiometer search that averages over frequency and that produces sky maps in a pixel or spherical harmonics basis in order to detect either point-like sources or large scale asymmetries, or a narrow-band radiometer search targeted at specific candidate sky locations.
The data used for these types of analyses are usually folded~\cite{PhysRevD.92.022003, PhysRevD.98.024001} into one sidereal day by taking advantage of the temporal symmetry of the earth's daily motion, which reduces the computation time by a factor equal to the total number of observation days.
The most general approach is that of an all-sky all-frequency (ASAF) search that produces sky maps in narrow frequency bands. These methods and their results will be briefly described below.

\subsubsection{Spherical harmonics decomposition}
\label{sec:AnalysisTechniques_Anisotropic_SHD}
If the SGWB contains a dipole intensity anisotropy whose origin is the motion of our local earth-bound coordinate system, or if the stochastic signal is (partly) produced by sources that follow a non point-like matter distribution, such as the distribution of luminous matter in our galaxy, one would expect large scale asymmetries in the angular distribution of the GW power $\mathcal{P}(\hat{\bf{n}})$. Following the techniques formerly applied to CMB analyses, Allen and Ottewill~\cite{PhysRevD.56.545} proposed to decompose a background of SGWB, dominated by a dipole or quadrupole distribution, into a basis of spherical harmonic functions, $Y_{lm}$:
\begin{equation}
\label{eigenfunc}
\mathcal{P}(\hat{\bf{n}})=\sum_{lm} \mathcal{P}_{lm}Y_{lm}(\hat{\bf{n}}) ,
\end{equation}
where
\begin{equation}
\label{eigenval}
\mathcal{P}_{lm}=\int_{S^2} d\hat{\bf{n}} \mathcal{P}(\hat{\bf{n}}) Y^*_{lm}(\hat{\bf{n}}).
\end{equation}

The analysis therefore reduces to determining the multi-pole coefficients $\mathcal{P}_{lm}$ from the frequency- and time-dependent cross-correlation test statistic
\begin{equation}
\label{croscorr_SH}
\braket{C(f,t)}=H(f)\int_{S^2}d\hat{\bf{n}}\gamma(\hat{\bf{n}}, f, t)\mathcal{P}(\hat{\bf{n}}),
\end{equation}
where $\gamma(\hat{\bf{n}}, f, t)$ is the general form of the ORF, taking into account the relative separation and orientation of the detector pair.
In the spherical harmonics basis, this function can be decomposed as well:
\begin{equation}
\label{ORF_SH}
\gamma(\hat{\bf{n}}, f, t)=\sum_{lm}\gamma_{lm}(f,t)Y^*_{lm}(\hat{\bf{n}}).
\end{equation}
The time (or frequency) scales involved in this type of analysis determine the time scale over which the observational data is averaged. Given the total observation periods of $\mathcal{O}$(1yr), one can naturally assume that the asymmetry induced by the relative motion of the detectors with respect to a fixed background reference will be that of the Earth's own rotation, corresponding to a time scale 
$T_e=8.6\times 10^4 \rm~s$. Comparing this to the travel time of a GW signal between two detector locations which is of order $\Delta T=10 \rm~ms$, it is clear that a choice of analysis window or averaging time scale, $\tau$, needs to be of the order $\tau = 10 \rm~s$, such that $\Delta T \ll \tau \ll T_e$. This also defines the best possible frequency resolution to be of order $(1/\tau)\approx 0.1 - 0.01 Hz$, which is further coarse grained to take into account the time-stability of the intrinsic detector noise. With these constraints, it is then possible to examine correlations between two detectors, as a function of time, averaged over periods of length $\tau$. In this case, the time dependence of $\gamma_{lm}(f,t)$ simplifies to

\begin{equation}
\label{ORF_time}
\gamma_{lm}(f,t)=\gamma_{lm}(f,0)\exp\left( im2\pi \frac{t_{sid}}{1 \rm~sid.~day}\right)
\end{equation}
The sidereal time, $t_{sid}$, provides the celestial coordinates of the analyzed sky location and is taken as the central time of each analysis segment of duration $\tau$. The $\gamma_{lm}(f,t)$ factors are the components of the ORF in Eq.~\eqref{ORF_SH}.
The actual expressions for $\gamma_{lm}(f,0)$ are derived in the original paper by Allen and Ottewill~\cite{PhysRevD.56.545}. However, it is important to note that, in contrast to the integrated ORF used in the isotropic analysis, due to the change in effective distance and orientation of the interferometer network with respect to a point source during one siderial day, there is no frequency where all $\gamma_{lm}(f,0)$ factors become zero for any given pair of detectors. In the isotropic case, the ORF can be expressed as $\gamma(f)=(5/\sqrt{4\pi})\gamma_{00}(f,t)$.\\

The Fisher matrix that can be calculated according to Eq.~\eqref{approxmatrix} specifies the correlation between the estimates of various multi-pole moments~\cite{PhysRevD.80.122002}. This matrix has several symmetries, which can be exploited.
 Almost half of the matrix elements are zero due to parity symmetry of the spherical harmonic eigenfunctions, implying that detector noise from antipodes cancels out so that $\Gamma_{lm,l'm'}=0$ for all odd $(l-l')$ combinations. In absence of daily variations in detector noise, $z$-axis rotational symmetry implies that $\Gamma_{lm,l'm'}\approx 0$ for $m \neq m'$. The regularization of the Fisher matrix is further aided by imposing a resolution cut-off that only allows values of $l<l_{max}$. This analysis is therefore well suited for detecting extended sources across the sky, but not for point-like sources, which require to take into account all $l$-modes with $l \to \infty$.
Recent analyses~\cite{PhysRevD.104.022005} chose the maximum $l$ value $l_{max}$ based on the diffraction limited angular resolution $\theta$ on the sky. This is determined by the distance $D$ between two detectors and the most sensitive frequency, $f$, in the analysis band:
\begin{equation}
\label{diffract}
\theta = \frac{c}{2Df},~~l_{max}=\pi/\theta.
\end{equation}
However, in recent considerations\cite{https://doi.org/10.48550/arxiv.2203.17141}, the achievable angular resolution can go beyond the diffraction limit, depending on whether one optimizes the search for localized source detection or optimal sky localization. 

All of these considerations make the regularization of the Fisher matrix in the spherical harmonic basis more robust and computationally less intensive compared to a pixel basis, where appropriate regularisation methods have been applied as well~\cite{10.1093/mnras/sty2546, PhysRevD.104.123018}. For practical purposes, the Fisher matrix used in the spherical harmonic analyses is diagonalized, and the eigenvalues which are smaller than a chosen threshold value are set to infinity (or their inverted values to zero). The threshold value is chosen by balancing the quality of deconvolution and the corresponding angular resolution with respect to the addition of noise due to poorly measured modes.

\begin{figure}[ht]
	\centering
    \includegraphics[width=0.95\textwidth]{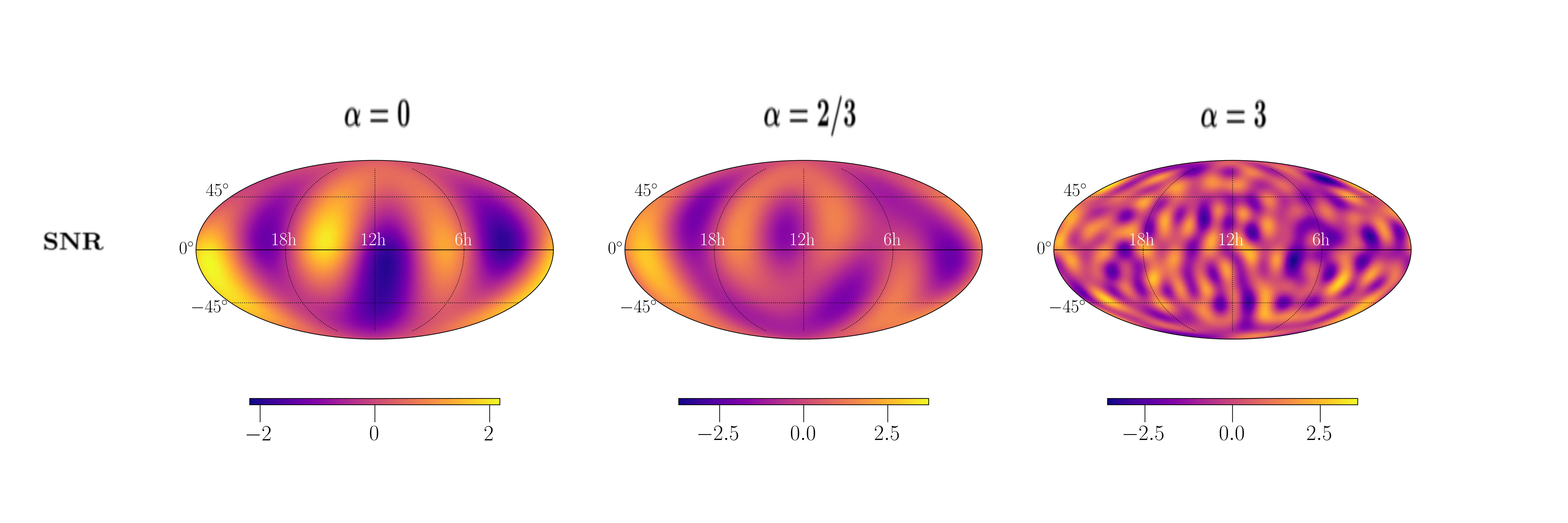}
    \includegraphics[width=0.95\textwidth]{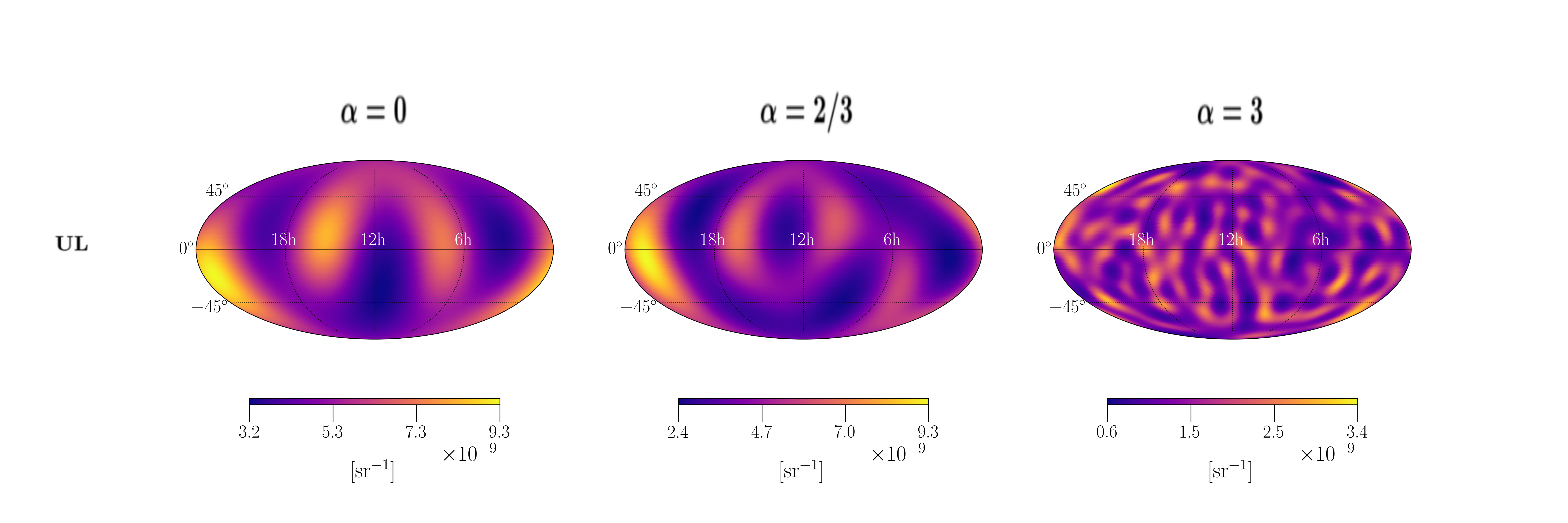}
	\caption{\textbf{Top}: SNR maps from the SHD search for extended sources. \textbf{Bottom}: sky maps representing 95\% upper limit on the
normalized gravitational-wave energy density $\Omega_{\alpha}[sr^{-1}]$. Both sets of maps, presented in equatorial coordinate system, are derived by
combining all three observing runs of LIGO-Virgo data (Virgo was incorporated only for O3). $\alpha = 0,2/3$, and 3 are represented from
left to right. The original version of this plot was presented in \cite{PhysRevD.104.022005}; the version shown here was obtained using open data published in \cite{O3DirectionalDataset}.}
	\label{Fig:SPH-maps}
\end{figure}

\begin{figure}[ht]
	\centering
	\includegraphics[width=0.7\columnwidth]{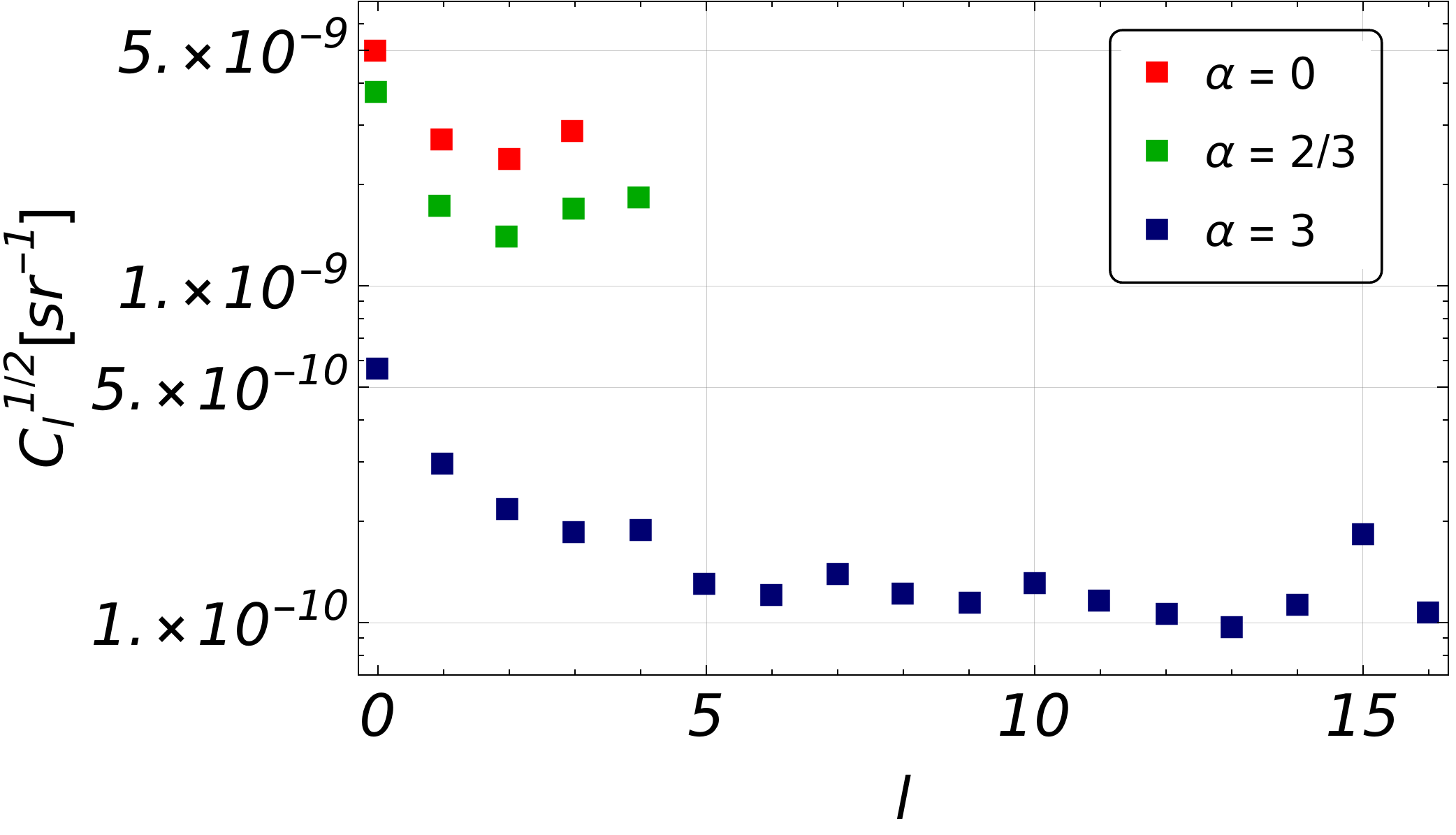}
	\caption{95\% upper limits on Cl for different $\alpha$ using combined
O1 + O2 +  O3 data. The original version of this plot was presented in~\cite{PhysRevD.104.022005}; the version shown here was obtained using open data published in \cite{O3DirectionalDataset}.}
	\label{Fig:SPH-coeff}
\end{figure}

The final result of a spherical harmonics decomposition search is either a sky map of $\Omega_{\hat{\bf{n}}}$ which is the frequency integrated version of $\Omega$ expressed in Eq.~\eqref{omega-spectral} and evaluated at a reference frequency $f_{\rm ref}$, by reconstructing it from the inferred $\mathcal{P}_{lm}$ coefficients, and Eq.~\eqref{eigenfunc} via
\begin{equation}
\label{Omega-SPH}
\hat{\Omega}_{\hat{\bf{n}}}=\frac{2\pi^2}{3H^2_0}f^3_{\textrm{ref}}\mathcal{P}_{\hat{\bf{n}}},
\end{equation}
or a set of estimators of the squared angular power in each mode $l$ given by
\begin{equation}
\label{Power-mode}
\hat{C_l}=\left(\frac{2\pi^2f^3_{\textrm{ref}}}{3H^2_0}\right)^2\frac{1}{1+2l}\sum_{m=-l}^l\left[|\hat{\mathcal{P}}_{lm}|^2-(\Gamma^{-1}_R)_{lm,lm}\right],
\end{equation}
where $\Gamma_R$ is the regularized Fisher matrix.
The results for the combined dataset of the first three observation runs O1-O3 by the LVK collaboration are shown in Fig~\ref{Fig:SPH-maps} and~\ref{Fig:SPH-coeff}.


\subsubsection{Radiometer - narrowband and broadband}
\label{sec:AnalysisTechniques_Anisotropic_Radiometer}
Radiometer searches have been introduced early on in the search for a SGWB~\cite{Ballmer_2006, PhysRevD.76.082003, PhysRevD.76.082001}.
Broadband radiometer searches (BBR) are optimized for resolvable point-like sources across the sky in a wide frequency band, while narrowband searches (NBR) can point at specific candidate sky locations and analyze the correlated strain spectrum in narrow bands of frequency. Broadband searches using current Earth-bound interferometric gravitational-wave detectors typically integrate over the frequency band between 20 and 500 Hz, a range that accounts for more than 90\% of the sensitivity for the typical power law spectral models. The narrowband analyses search over the frequency range between 20 and 1726 Hz using frequency bins of various sizes depending on the central frequency and the sky direction.

For radiometer searches, the spectral density from Eq.~\eqref{spectral_fact} is often redefined to obtain an energy flux
\begin{equation}
\label{energy_flux}
\mathcal{F}(f,\hat{\bf{n}})=\frac{c^3\pi}{4G}f^2\mathcal{P}(\hat{\bf{n}})H(f),
\end{equation}
which has units of erg cm$^{-2}$ s$^{-1}$ Hz$^{-1}$. 
The corresponding strain power spectrum, $H(f)$ can also be used to characterize the BBR results. In this case one again assumes a power-law spectrum, where the amplitude at the pivot point of $f=100~\textrm{Hz}$ is described by $H_\alpha$:
\begin{equation}
\label{strain_power}
H(f)=H_\alpha (f/100~\textrm{Hz})^\alpha
\end{equation}
The choice of pixel basis is generally based on the  HEALPix~\cite{Gorski_2005} (Hierarchical Equal Area isoLatitude Pixelation) PYTHON package~\cite{PhysRevD.103.083024}. The sky is typically divided in 12288 pixels, each with an area of 3 deg$^2$.
In radiometer searches one assumes that the GW power is contained within
a singe pixel of the sky map, and there is no signal covariance between neighboring pixels.
This simplifies the inversion of the Fisher matrix to the inversion of its diagonal elements only.
Due to the smearing induced by the detector response functions, this approximation is only valid for a small number of well-separated point sources.
As was noted earlier in this section, a radiometer search can drastically outperform an isotropic one, if the SGWB power is localized in one or a very limited number of point sources that are typically nearby (up to redshifts of $z\approx 3$).
The lowest value of strain power that can be detected with a significance of one sigma, by using a radiometer search at any given frequency band centered at frequency, $f$, is given by~\cite{Ballmer_2006}

\begin{equation}
\label{radiometer_sens}
H_{\textrm{sens}, \hat{\bf{n}}}=\frac{H(f)}{2\sqrt{T}\sqrt{\langle \int df \frac{|\gamma_{\hat{\bf{n}}}|^2H^2}{P_1P_2}\rangle}},
\end{equation}
which is mostly dependent on the declination, and independent on the right ascension, assuming an even coverage of the sidereal day in terms of up-time and stable sensitivity of the detector pair used. In the case of a flat source power spectrum, this translates for the LIGO Hanford-Livingston detector pair at design sensitivity to a GW energy flux density sensitivity of order
\begin{equation}
\label{radiometer_sens2}
\mathcal{F}(f,\hat{\bf{n}})df\approx 5 \times 10^{-11} \frac{\rm Watt}{\rm m^2Hz}\left(\frac{f}{100 \rm Hz}\right)^2\left(\frac{1 \rm yr}{T}\right)^{1/2} df,
\end{equation}
with a typical variation of 35\% on the declination of the source. Note that this value is of the same order of magnitude as the current best upper limits obtained with radiometer searches by the LVK collaboration~\cite{PhysRevD.104.022005}, as quoted below.

As demonstrated in~\cite{PhysRevD.89.084076}, a directed search for a localized source can outperform an isotropic sky search for SGWBs dominated by nearby sources, up to redshifts of 3.
Given the diffraction limited angular resolution from Eq.~\eqref{diffract} clusters of galaxies can appear as point sources in directional searches. It was shown in~\cite{PhysRevD.89.084076} that a large population of millisecond pulsars in the nearby Virgo cluster can produce a significantly stronger signal than the nearby isotropic background of unresolved sources of the same type. A typical Milky Way type galaxy contains at least 40,000 millisecond pulsars (MSP), each of which is expected to emit a narrowband GW signal, thus forming a `forest' of emission lines in a broad frequency domain. In galaxy clusters like the Virgo cluster, with an estimated amount of order $10^8$ MSPs, this forest will manifest itself as a continuum. The gain in sensitivity depends on the population distribution of the sources at different frequencies and redshifts, and on the cosmological history of our Universe. When the Universe has MSPs at high redshifts, typically in a younger universe, and is statistically isotropic at large scales, an isotropic search performs well. If the background is dominated by nearby sources of the type discussed in~\cite{PhysRevD.89.084076}, as is the case for an older universe, a localized search will most likely outperform an isotropic search. A directional search for a completely isotropic background will always underperform with respect to an isotropic search, which justifies the need to follow all strategies in parallel. So far, no point-like sources that emit a stochastic GW background have been identified. The LVK collaboration has recently set upper limits on the GW energy flux across the sky, based on three observation runs taken between September 2015 and March 2020~\cite{PhysRevD.104.022005}. These upper limits lie in the range [0.013-7.6]$\times 10^{-8}\rm erg~cm^{-2}s^{-1}Hz^{-1}$, depending on the sky position and on the assumptions made for the spectral index of the frequency dependence of the signal.

The narrowband radiometer analysis searches for stochastic GW signals originating from specific candidate locations on the sky. These correspond to either nearby galaxy clusters that are bound to contain a large population of binary pulsars such as the Virgo cluster, or the center of our galaxy, or well-known and identified low-mass X-ray binaries with powerful X-ray signatures, such as Scorpius-X.  The X-ray luminosity, $L_X$ of such systems can be related to the GW  luminosity, $L_{GW}$ via
\begin{equation}
\label{xraylumi} 
L_{\textrm{GW}}\approx \frac{f_{\textrm{spin}}}{f_{\textrm{Kepler}}}L_{\textrm{X}},
\end{equation}       
where $f_{\textrm{Kepler}}$ is the final orbital frequency of the accreting matter, which is roughly 2kHz for a neutron star, and $f_{\textrm{spin}}$ is the spin frequency. In the same reference (\cite{PhysRevD.76.082003}) the X-ray luminosity  of all low-mass X-ray binaries (LMXBs) in the nearby Virgo cluster is estimated to be roughly $10^{40} \rm erg/sec/galaxy$ which corresponds to a strength of the strain power spectrum, introduced in Eq. \ref{energy_flux}, of roughly $H(f)\approx 10^{-55} \text{Hz}^{-1} (100 \text{Hz}/f_{\text{center}})(100 \text{Hz}/\Delta f)$, where $f_{\rm center}$ is the typical frequency of the $\Delta f$ wide search band of interest.  This value can be compared with the current best limits set by the LVK collaboration, which are of order $10^{-50}$ for a flat power spectrum, and of order $10^{-47}$ for a power spectral index of $\alpha=3$.

It is therefore fair to assume that accreting neutron stars, which are the brightest X-ray emitters, are also the brightest emitters in GWs, which motivates the choice of Scorpius-X as a target location. Searches targeting the location of these binaries can provide competitive limits on the characteristic parameters of these systems, as was first demonstrated in~\cite{PhysRevD.76.082003}.

In addition to X-ray binaries, neutron stars in young supernova remnants are also prime candidates for a targeted directional search. This is motivated by several reasons~\cite{PhysRevD.94.082004}. Indirect wave strain upper limits are proportional to the spin-down rate of these remnants, or inversely proportional to their age. Less time has passed in young objects for their crusts and interiors to settle down and erase nonaxisymmetries that were present at birth. Young objects also spin down rapidly, which can excite nonaxisymmetric flows in their interior. For this reason, the location of a possible neutron star in one of the youngest and closest known supernova remnants, SNR 1987A in the Large Magellanic Cloud, is chosen in recent directed searches. In the past, other supernova remnants have also been investigated, such as the Crab pulsar, Cassiopeia A, and other young pulsars with radio or X-ray ephemerides. As a byproduct, upper limits can be placed on parameters of astrophysical interest, such as the maximum ellipticity, internal magnetic field strength, and the amplitude of r-mode oscillations~\cite{PhysRevD.94.082004}.

\begin{figure}[ht]
	\centering
	\begin{subfigure}[b]{0.28\textwidth}
         \centering
         \includegraphics[width=\textwidth]{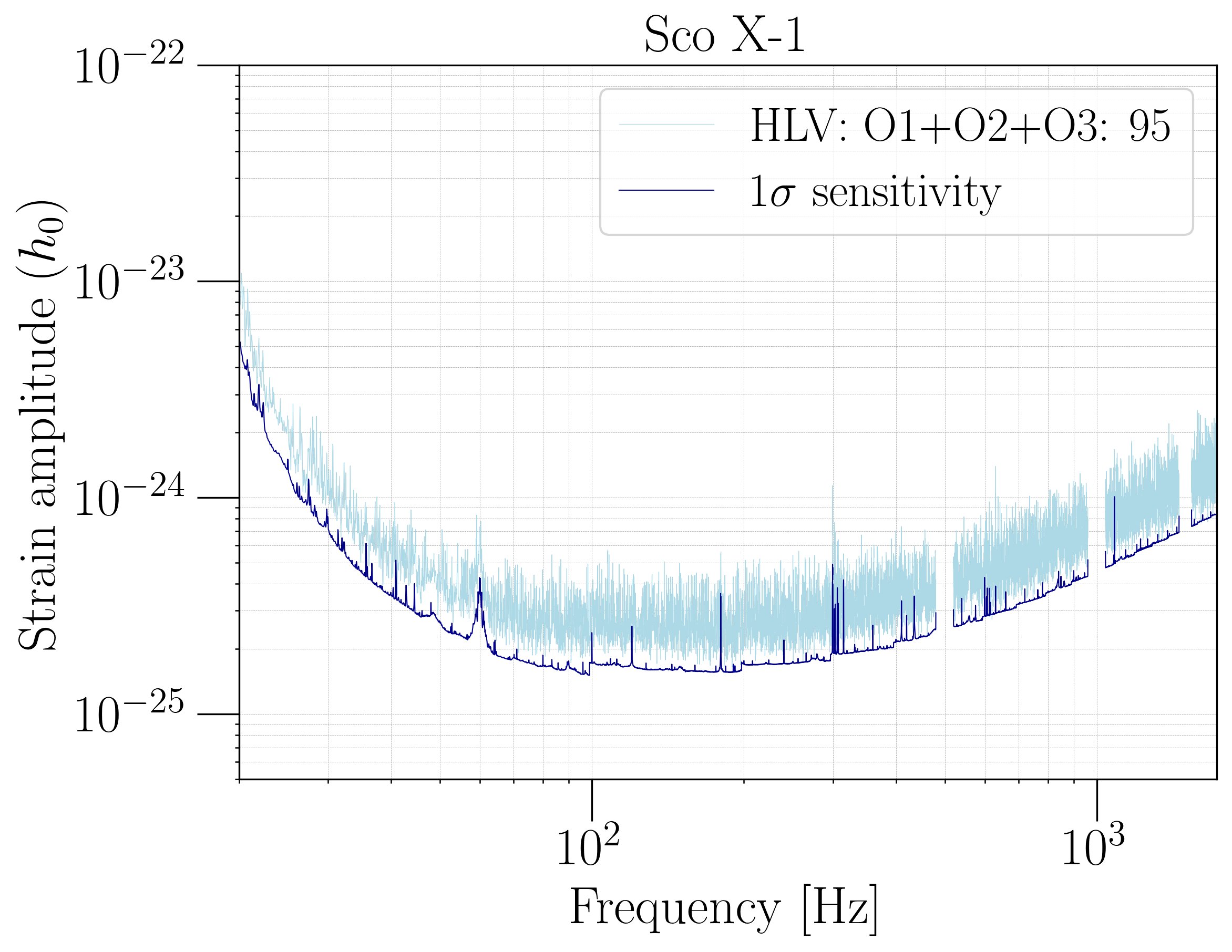}
     \end{subfigure}
     \hfill
     \begin{subfigure}[b]{0.28\textwidth}
         \centering
         \includegraphics[width=\textwidth]{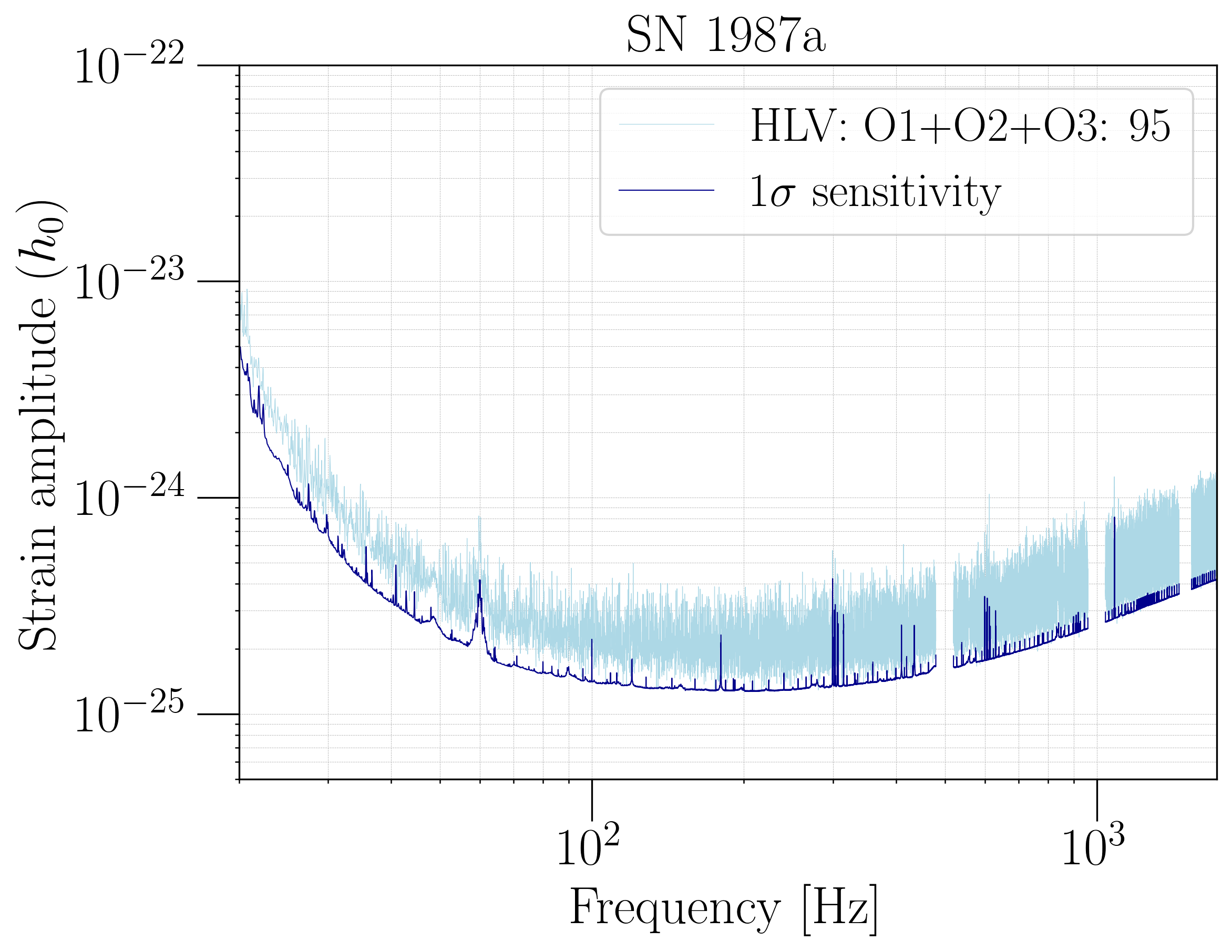}
     \end{subfigure}%
     \hfill
     \begin{subfigure}[b]{0.28\textwidth}
         \centering
         \includegraphics[width=\textwidth]{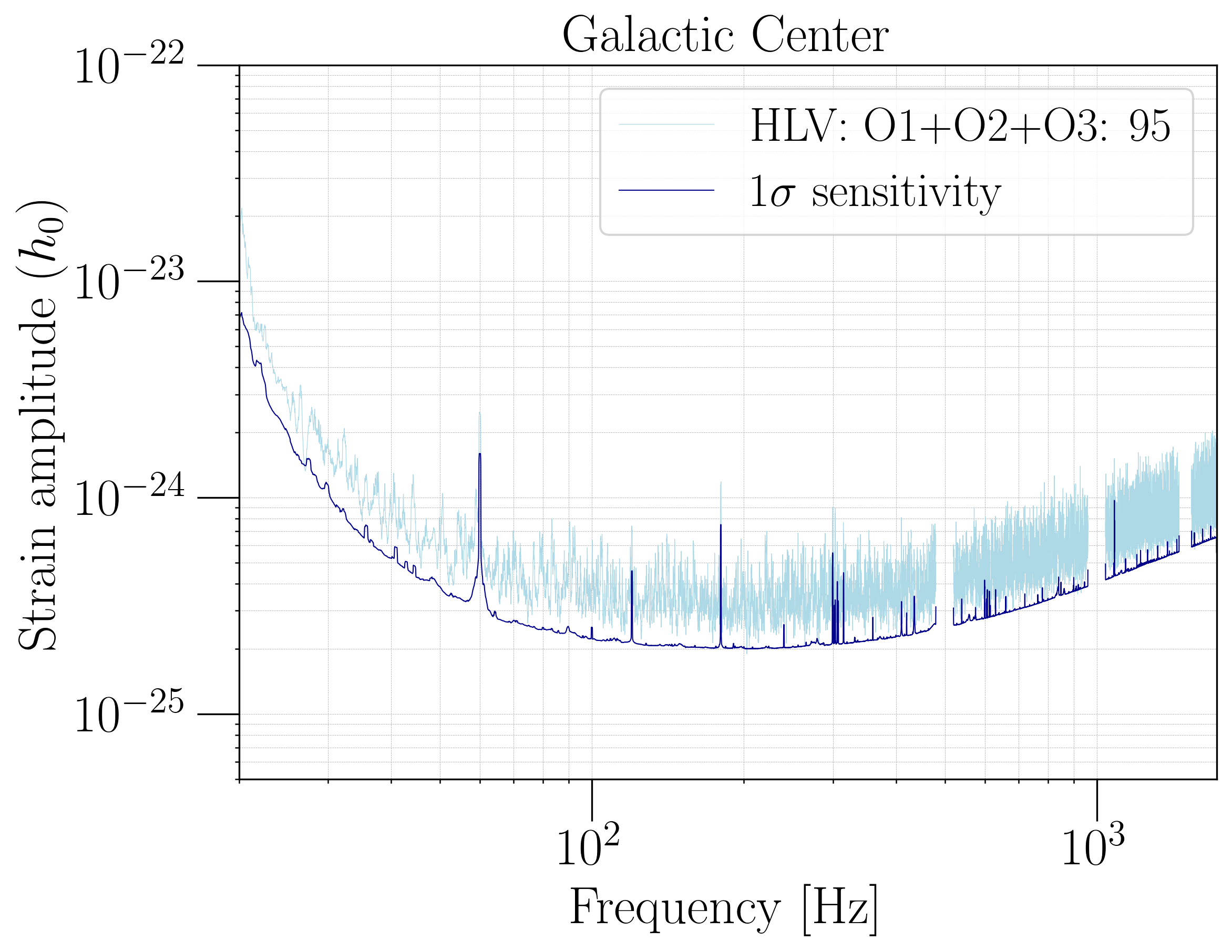}
     \end{subfigure}
	\caption{Upper limits on the dimensionless strain amplitude $h_0$, using the data from three observing runs of the LIGO-Virgo detectors, at the 95\% confidence level for the narrow-band radiometer search are indicated by the gray bands for Scorpius X-1 (left), SN 1987A (middle) and the Galactic Center (right). The dark line shows the 1$\sigma$ sensitivity of the search for each direction. The original version of this plot was presented in~\cite{PhysRevD.104.022005}; the version shown here was obtained using open data published in \cite{O3DirectionalDataset}.}
	\label{Fig:NBR-spectra}
\end{figure}

In the absence of detection, NBR searches determine upper limits on the peak strain amplitude $h_0 = \sqrt{\hat{\mathcal{H}}(f)}$, extracted from the measured GW strain power
$\hat{\mathcal{H}}(f)=X(f)$, based on the time and frequency dependent DOFs for each baseline pair, their noise PSDs and the cross-correlation statistic, as defined in Eq.~\eqref{xbeta}, where the subscript $b$ is omitted because we point at a specific sky location, and where no summation over the frequencies is performed. The shape of the frequency spectrum is fixed to  a flat frequency power spectrum, corresponding to a fixed power spectral index $\alpha=0$ in Eq.~\eqref{plaw} for $H(f)$. Source-dependent effects such as frequency broadening due to the binary motion of the source and the orbital motion of Earth during the observation time are taken into account by grouping multiple frequency bins in optimally-sized combined bins at each frequency, as is described in detail in~\cite{PhysRevLett.118.121102}. The latest upper limits from the LVK collaboration~\cite{PhysRevD.104.022005} using this method are shown in Fig~\ref{Fig:NBR-spectra} for three search directions: Scorpius-X, SN 1987A, and the galactic center.

\subsubsection{All sky all frequency}
\label{sec:AnalysisTechniques_Anisotropic_ASAF}
\begin{figure}[ht]
	\centering
	\includegraphics[width=0.28\columnwidth]{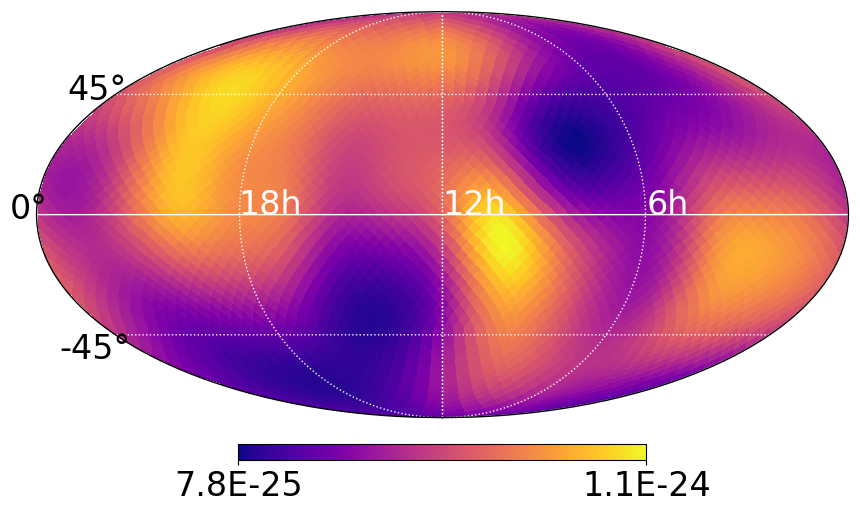}
	\includegraphics[width=0.28\columnwidth]{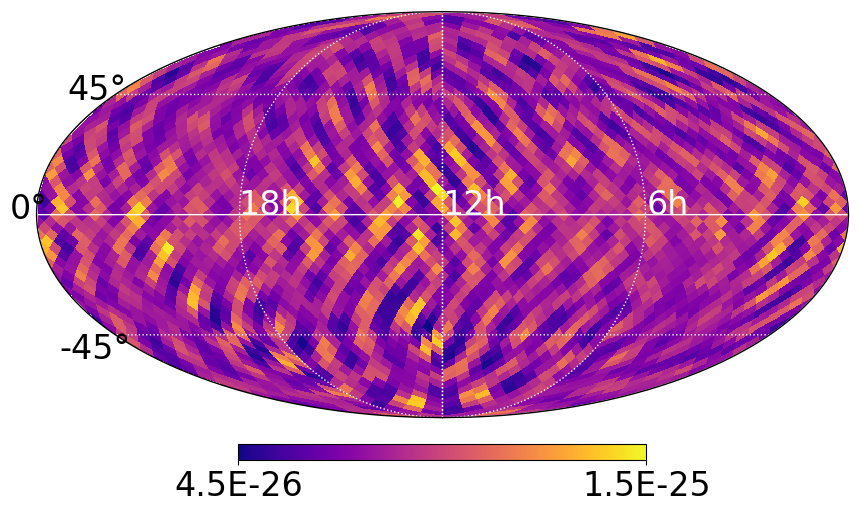}
	\includegraphics[width=0.28\columnwidth]{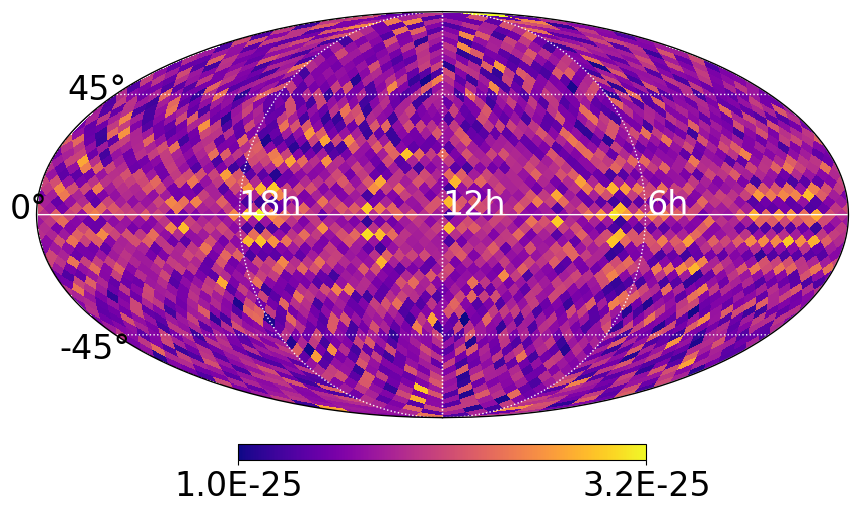}
	\caption{The Bayesian upper limit maps on the strain amplitude
with 95\% confidence for three frequencies 23.0625 Hz, 423.0635 Hz, 1223.0625 Hz, based on the O1-O3 joint observation runs with the three Adv LIGO and Virgo detector baselines. The original version of this plot was presented in~\cite{https://doi.org/10.48550/arxiv.2110.09834}; the version shown here was obtained using open data published in \cite{ASAFDataset}.}
	\label{Fig:ASAF-map}
\end{figure}

The broadband and narrowband radiometer searches described above are not optimized to detect an unknown narrowband anisotropic SGWB. The narrowband search is clearly limited by its choice of targeted sky locations, while the broadband search, due to its integration over a large frequency band, is limited by the noise from many frequency bins that do not contain any signal, thus diluting the signal-to-noise ratio. The all-sky all-frequency search \cite{https://doi.org/10.48550/arxiv.2110.09834}, as it is currently performed by the LVK collaboration overcomes these limitations by performing a directional search in the pixel eigenbasis at all narrowband frequency bins separately~\cite{PhysRevD.91.124012, PhysRevD.98.064018}. Up to recently, this type of analysis faced computational limitations and utilizes only the diagonal elements of the Fisher information matrices that transform the dirty sky maps, as described above. The larger available datasets, combined with better computational tools, including the folding of data over one sidereal day, will overcome most of these problems in the near future. Since no matched filtering is applied, the analysis is model independent and has less parameters to infer, again assisting in computational efficiency.
The formalism used by the ASAF is the most general one and is described in the beginning of Sec.~\ref{sec:AnalysisTechniques}. Equations~\eqref{omega-spectral} and~\eqref{PSD-spectral} describe respectively  the dimensionless GW energy density (now in units of sr$^{-1}$), and the spectral density obtained from the cross-correlation of Fourier transformed strain outputs of two detectors. The spectral density is again decomposed in an eigenbasis of pixels, as shown in Eq.~\eqref{PSD-spectral}. The spectral density in each pixel is estimated by the maximum likelihood estimator based on the deconvolution of the dirty pixel map in each frequency bin
\begin{equation}
\label{pixmap}
\hat{\mathcal{P}}(f)=\Gamma(f)^{-1}X(f),
\end{equation}
which can be identified with Eq.~\eqref{estimators}, where the only difference is an explicit frequency dependence. The dirty map is obtained in the same way as in Eq.~\eqref{xbeta}, but is now frequency dependent:
\begin{equation}
\label{eq:dirtymap}
X_p(f)=\tau \Delta f \Re{\rm e}\left(\sum_{I,t}\frac{\gamma^{I*}(f,t,p)C^I(f,t)}{P_{I_1}(f,t)P_{I_2}(f,t)}\right),
\end{equation}
where $\tau$ is the duration of each analyzed time segment, and $\Delta f$ the coarse-grained frequency bin size. The index $p$ refers to a specific pixel in the sky map, and the index $I_{1,2}$ refers to a specific detector pair. $P_{I_{1,2}}(f,t)$ are the noise power spectral densities, and $C^I(f,t)$ is the cross-correlation test statistic for any baseline, $I$. The sum in eq.~\ref{eq:dirtymap} is made both over the time segments and over all available baselines, $I$. The ORF for each of the baselines consisting of two detectors (1 and 2) now admits its most general, unintegrated form
\begin{equation}
\label{orfgeneral}
\gamma^I(f,t,p) = \sum_A F^A_{I_1}(\hat{\textbf{n}}_p,t)F^A_{I_2}(\hat{\textbf{n}}_p,t) e^{2\pi if\Delta\textbf{x}_I(t)\cdot\hat{\textbf{n}}_p/c},
\end{equation}
with $\Delta\textbf{x}_I(t)$ the separation vector between the detectors. The ORF characterizes the joint response to a signal coming from direction $\hat{\textbf{n}}_p(t)$ in the pixel sky map when cross-correlating data streams from the detector pair $I_{1,2}$ with time varying phase delay, along with the sky modulation induced by the polarization dependent antenna pattern functions of the detectors, $F^A_{1,2}$.
The corresponding Fisher information matrix expression for the covariance between two pixels $p$ and $p'$ now generalizes to
\begin{equation}
\label{eq:fishergeneral}
\Gamma_{p,p'}=\frac{\tau \Delta f}{2}\Re{\rm e}\left( \sum_{I,t}\frac{\gamma^{I*}(f,t,p)\gamma^{I}(f,t,p')}{P_{I_1}(f,t)P_{I_2}(f,t)}\right).
\end{equation}
The time segment duration is chosen to be $\tau = 192~\textrm{s}$ and the frequency bin size $\Delta f$ is taken as to ensure that $\tau \Delta f > 1$. The sky maps, constructed for each frequency bin, are again based on the HEALpix scheme with a total of 3072 pixels. In the absence of a detection, Bayesian upper limits are set on an equivalent strain amplitude of a circularly polarized signal, $h(f,\hat{\textbf{n}})=\sqrt{\mathcal{P}(f,\hat{\textbf{n}})df}$ without correcting for Doppler modulation induced by Earth's motion. These upper limits are in the range $(0.030 - 9.6) \times 10^{-24}$ for the three joint observational runs with the Advanced LIGO and Advanced Virgo detectors when combining all three possible baselines~\cite{https://doi.org/10.48550/arxiv.2110.09834}. These limits are the strongest in the frequency range between 25 Hz and 1kHz, where the intrinsic noise PSDs of the individual detectors is the smallest.
Figure~\ref{Fig:ASAF-map} shows these upper limits in the shape of three sky maps for three frequencies: 23.0625 Hz, 423.0636 Hz, and 1223.0625 Hz. Note that the features in these upper limit maps become broader at low frequencies due to the slower modulation of the detector ORFs.

%-------------------------%
%-- Analysis techniques --%
%---------- TBS ----------%
%-------------------------%

\subsection{The Bayesian Search (TBS)}
\label{sec:AnalysisTechniques_Isotropic_TBS}
The cross-correlation analysis technique for an isotropic SGWB introduced in Section \ref{sec:AnalysisTechniques_Isotropic} is best suited for GW signals that are continuous in time. Indeed, because this technique relies on averaging over long time periods, the contribution from non-continuous signals would get diluted by the noise due to this averaging. However, the binary black hole merger contribution to the SGWB is expected to be non-continuous as these events happen every few minutes and only last for a few seconds \cite{PopCornNatureBBH2018}. Thus, if one wants to define an optimal analysis technique for the detection of such a background, the intermittent nature of the signal needs to be taken into account.

Drasco and Flanagan \cite{PhysRevD.67.082003} first explored alternatives to the standard cross-correlation search to account for the non-continuous nature of the signal contributing to the SGWB. Their approach was then revisited by Smith and Thrane \cite{PhysRevX.8.021019}, who named the method \textit{The Bayesian Search} (TBS), as it relies on a Bayesian inference framework. The non-continuity of the signal is captured by introducing the notion of the duty cycle $\xi$, which parameterizes the probability of a data segment containing a binary black hole merger signal on top of noise. Furthermore, the TBS method attempts to match concrete waveforms to the strain data, instead of looking for cross-correlated power as done in the standard cross-correlation search described in Sec. \ref{sec:AnalysisTechniques_Isotropic}.

The method consists of first dividing the data into smaller segments, labeled $i$, such that the probability of a segment containing two black hole mergers is negligible. By taking 4s segments of data, the probability of a segment containing two mergers is around $10^{-4}$ \cite{PhysRevX.8.021019}. One proceeds by defining the signal likelihood for the $i$-th segment, which takes the form
\begin{equation}
\label{Eq: TBS likelihood signal}
	\log \mathcal{L}\left(s_i|\theta_i\right)\propto-\frac{1}{2}\braket{s_i-h(\theta_i), s_i-h(\theta_i)},
\end{equation}
where $h(\theta_i)$ is the signal model which depends on the parameters $\theta_i$ that describe the compact binary merger, e.g. the component masses, spin etc. The angle brackets represents the following inner product
\begin{equation}
	\braket{a,b}=\Re{\rm e}\left(4\Delta f \sum_{k}\frac{a^*(f_k)b(f_k)}{P(f_k)}\right),
\end{equation}
where $P(f_k)$ denotes the noise power spectral density. Note that for more than one detector, the signal likelihood for the \textit{i}-th segment corresponds to
\begin{equation}
	\mathcal{L}\left(\vec{s}_i|\theta_i\right)=\prod_{j=1}^{M}\mathcal{L}(s_i^{(j)}|\theta_i),
\end{equation}
where the product runs over the number of detectors $M$ and $\vec{s}_i$ is a vector of length M containing the data from all detectors in segment $i$. To properly model the intermittent, non-Gaussian nature of an astrophysical background, the previous likelihood can be generalized to include the duty cycle $\xi$ which, as introduced previously, represents the probability of a segment containing a binary black hole merger signal. Concretely, the likelihood now admits the form
\begin{equation}
	\mathcal{L}(\vec{s}_i|\theta_i,\xi)=\xi\mathcal{L}(\vec{s}_i|\theta_i)+(1-\xi)\mathcal{L}(\vec{s}_i|0),
\end{equation}
where the likelihood $\mathcal{L}(\vec{s}_i|0)$ is the likelihood given that no signal is present, i.e. $h(\theta_i)=0$ in Eq.~\eqref{Eq: TBS likelihood signal}. The next step consists of marginalizing over each of the segment-dependent astrophysical parameters $\theta_i$ which describe the merger. To this end, the following signal and noise evidences are introduced:
\begin{equation}
	\mathcal{Z}_S^i=\int d\theta\mathcal{L}(\vec{s}_i|\theta_i)\pi(\theta_i)~~~~~~~~\text{and}~~~~~~~~\mathcal{Z}_N^i=\mathcal{L}(\vec{s}_i|0),
\end{equation}
after which the marginalized likelihood can be written as
\begin{equation}
	\mathcal{L}\left(\vec{s}_i|\xi\right)=\xi\mathcal{Z}_S^i+\left(1-\xi\right)\mathcal{Z}_N^i.
\end{equation}
Finally, the data from a number of segments $n$ is combined by multiplying the individual segment likelihoods:
\begin{equation}
	\mathcal{L}^{\rm tot}(\{\vec{s}\}|\xi)=\prod_i^n\left(\xi\mathcal{Z}_S^i+(1-\xi)\mathcal{Z}_N^i\right).
\end{equation}
The signal hypothesis is described by the signal evidence marginalized over the duty cycle $\xi$
\begin{equation}
	\mathcal{Z}_{\rm SGWB}=\int d\xi \mathcal{L}\left(\{\vec{s}\}|\xi\right)\pi\left(\xi\right),
\end{equation}
whereas the null-hypothesis is characterized by the null evidence, i.e. where the duty cycle $\xi=0$:
\begin{equation}
\mathcal{Z}_0=\mathcal{L}\left(\{\vec{s}\}|\xi=0\right).
\end{equation}
As a last step, one then defines an optimal detection statistic for the astrophysical background as the Bayes factor comparing the two hypotheses: $\mathcal{B}=\mathcal{Z}_{\rm SGWB}/\mathcal{Z}_0$,where a log Bayes factor of $\approx$ 8 is used as a threshold for a statistically significant preference of one hypothesis over the other \cite{Jeffreys61}.

The efficiency of the above formalism compared to the standard cross-correlation search is demonstrated using simulated data. In \cite{PhysRevX.8.021019}, it is found that an astrophysical background with a realistic duty cycle $\xi\approx4\times10^{-4}$ corresponding to what is expected for a binary black hole background can be detected with 20 hours of data (created assuming a two-detector LIGO network operating at design sensitivity), which is an extremely small time compared to the detection time of over a year needed using the standard cross-correlation method described in the previous section \cite{PopCornNatureBBH2018}. This improvement in sensitivity is due to the accurate modeling of the deterministic nature of the waveforms, as well as the inclusion of the intermittent character of the astrophysical background. However, one notes that this increase in sensitivity comes at a very high computational cost. Indeed, the marginalization over the many astrophysical parameters $\theta_i$ increases the computational needs tremendously. Although the detection of an astrophysical background might be achieved within a day using this method, there could be strong motivation to analyze all the data, as this would allow to infer properties of the binary black hole population. Even though a one day analysis seems feasible with current computational resources, a full analysis of the data would come at a huge computational cost. Therefore, one of the biggest tasks remains identifying ways to reduce the cost \cite{PhysRevX.8.021019}. Another challenge that presents itself is running the search on real data. Indeed, contrarily to simulated data, real detector data is expected to contain glitches and non-stationarities that will have an impact on the search. These challenges need to be addressed before being able to apply the method to real data. \\~\\
To conclude, note that the above arguments were made in the context of attempting to formulate an optimal detection method for a SGWB arising from popcorn-like binary-black hole mergers, which are not continuous in time. However, \cite{PhysRevX.8.021019} mentions that the method described above could be generalized in order to be used for different backgrounds, such as the one arising from binary neutron stars. Indeed, binary neutron star mergers last much longer than BBH mergers (roughly 100s instead of a few seconds), such that their signals overlap from segment to segment. However, all is not lost and one could remedy this by generalizing the TBS method to have the number of binary neutron star mergers within a segment as an extra parameter. For more details on various generalizations of the method, we refer the reader to \cite{PhysRevX.8.021019}. 
\paragraph{Disentangling a cosmological from an astrophysical background}~\\
The TBS search can be generalized to disentangle a cosmological background from an astrophysical one \cite{PhysRevLett.125.241101}. Consider a signal which is composed of noise, an intermittent binary black hole background and a continuous cosmological background, whose GW energy density is given by a simple power-law, as given by Eq.~\eqref{eq:OmegaGW_PL}. The difference with the regular TBS search is that, although still modeling the intermittent nature of the astrophysical background from BBH by using the duty cycle $\xi$, the "noise likelihood" now contains noise and the continuous cosmological background which is always 'on'. Concretely, the likelihood is given by
\begin{equation}
	\mathcal{L}(s_i|\Omega_\alpha,\alpha,\xi)=\xi\mathcal{L}_S(s_i|\Omega_\alpha,\alpha)+(1-\xi)\mathcal{L}_N(s_i|\Omega_\alpha,\alpha),
\end{equation}
where
\begin{align}
	\mathcal{L}_S(s_i|\Omega_\alpha,\alpha)&=\int d\theta\mathcal{L}(s_i|\theta,\Omega_\alpha,\alpha)\pi\left(\theta\right),\\
	\mathcal{L}_N(s_i|\Omega_\alpha,\alpha)&=\mathcal{L}(s_i|\theta=0,\Omega_\alpha,\alpha).
\end{align}
Remember that $\theta$ was the set of parameters describing the GW signal coming from binary black hole mergers, as introduced below Eq.~\eqref{Eq: TBS likelihood signal}. Note that as for the TBS search, these parameters are marginalized over. Thus, the signal likelihood $\mathcal{L}_S$ assumes the presence of noise, a cosmological background and an intermittent astrophysical background, whereas the noise likelihood $\mathcal{L}_N$ only assumes noise and a cosmological background.\\
It is shown that this method forms an effective way of disentangling the cosmological background from an intermittent astrophysical foreground by applying the method to mock data. The data set consisted of 101 segments of 4 seconds each, with Gaussian noise and a cosmological background with an amplitude $\log\Omega_\alpha=-6$ and spectral index $\alpha=0$, corresponding to the spectral index expected for a SGWB from slow-roll inflation \cite{PhysRevLett.125.241101}. Binary black hole mergers were added to 11 of the 101 segments, corresponding to a duty cycle $\xi=.11$. By performing the search on this simulated data set, it is possible to recover the duty cycle $\xi$ of the injected astrophysical background, as well as the parameters $\Omega_\alpha$ and $\alpha$ describing the continuous cosmological background. Although performed on unrealistic data, i.e. large cosmological SGWB amplitude and Gaussian noise, this already illustrates the effectiveness of the method to disentangle a cosmological from an astrophysical SGWB.

Many other methods have been investigated to achieve the spectral separation of a cosmological and astrophysical background since this will be an important feature for future generation Earth-based as well as spaced based laser interferometric gravitational-wave detectors. Although we will not discuss these studies in this review paper, we refer the interested reader to the following references on spectral separation in the context of LIGO and Virgo \cite{Ungarelli_2004,Parida_2016}, third generation Earth-based interferometric gravitational-wave detectors \cite{PhysRevLett.118.151105,PhysRevD.102.063009,PhysRevD.103.043023} and the context of LISA \cite{2021PhRvD.103j3529B, 10.1093/mnras/stab2575,PhysRevD.105.023510,Pieroni_2020}. We also note that \cite{2021PhRvD.103j3529B} summarizes the different efforts pursued in the context of spectral separation. 

%-------------------------%
%-- Analysis techniques --%
%- Validation techniques -%
%-------------------------%
\newpage
\section{Validation techniques}
\label{sec:AnalysisTechniques_ValidationTechniques}

Observing a SGWB both from astrophysical and/or cosmological sources would provide us information about phenomena in the distant universe. Observing any of the possible sources could fundamentally and drastically change our knowledge of the physics at play. Therefore, when claiming a detection of a SGWB, it is of the utmost importance to exclude the possibility that the observed signal is not a SGWB, but is due to e.g. some noise source.

We will start by giving some examples of noise sources that can affect the search for a SGWB (Sec. \ref{sec:AnalysisTechniques_ValidationTechniques_GlobalCorrelatedNoise}) and how we can monitor them (Sec. \ref{sec:AnalysisTechniques_ValidationTechniques_ENVmonitoring}). Afterwards, we focus on various methods (Sec. \ref{sec:AnalysisTechniques_ValidationTechniques_Geodesy}-\ref{sec:AnalysisTechniques_ValidationTechniques_NullChannel}), that could help us make an informed decision when claiming the detection of a SGWB. Some of these methods are already developed and studied in various circumstances whereas others could still benefit from more investigations in the future. We will focus on Earth-based interferometric gravitational-wave detectors, but some of these methods can also be used for detections claimed by other instruments.

\subsection{Typical noise sources at Earth-based gravitational wave detectors}
\label{sec:AnalysisTechniques_ValidationTechniques_GlobalCorrelatedNoise}

One could categorize the noise sources which can affect searches for a SGWB into: transient, (long duration) narrowband and (long duration) broadband noise sources. Each of these noise sources will be discussed in more detail below.

Using multiple, widely separated detectors has already the advantage that many of the noise sources are uncorrelated since these are often of local nature. This implies they don't affect correlation studies searching for a SGWB.
However, many past studies have indicated that there are several sources that are correlated at a level that impacts GW searches either at the current sensitivities or might do so in the future.
Therefore it is of utmost important to be monitor those noise sources carefully. In Sec. \ref{sec:AnalysisTechniques_ValidationTechniques_ENVmonitoring}, we will discuss in more detail how these disturbances are monitored and how the coupling mechanisms to the detectors can be determined. 

\subsubsection{Transient noise sources}

As mentioned earlier, transient noise sources or glitches in the detector break down the stationarity of the data. Such noise sources can be particularly problematic for searches for transient GW sources or searches for the astrophysical SGWB which rely on the intermittent behaviour of the GW sources such as the TBS search introduced in Sec. \ref{sec:AnalysisTechniques_Isotropic_TBS}. However they will introduce a bias in any search for a SGWB. Even though these glitches are uncorrelated most of the time, they can bias searches for an SGWB by introducing a bias in the PSD estimation of segments contaminated by the glitches. Howver, a recent study shows that lightning strikes could lead to correlated glitches in future observing runs of Earth-based interferometric gravitational-wave detectors \cite{https://doi.org/10.48550/arxiv.2209.00284}.

A first method to deal with glitches an more generally, all forms of non-stationarity in the data is to apply a cut removing all times where the standard deviation, i.e. the square root of the variance introduced in Eq. \ref{eq:optimal-filter}, between adjacent segments varies too much. The LVK collaborations have chosen to limit this variation between segments to be at most 20\% in the analysis of O3 data \cite{PhysRevD.104.022004}.

However, due to the high rate of glitches during O3 in the LIGO detectors, more than 50\% of the data was lost after applying the non-stationarity cut. Therefore, a method called gating \cite{gatingDocument,y_stochasticGatingDocument} was implemented. The procedure can be understood as zeroing out the times of the glitch by using an inverse Tukey window \cite{gatingDocument}. Gating applied to O3 data effectively removed glitches by zeroing out a small amount of data $\lesssim$ 1\% \cite{PhysRevD.104.022004}. The effect of gating on the search for an isotropic SGWB was investigated using mock data \cite{y_stochasticGatingDocument}. No significant bias from gating for stochastic searches was observed. Furthermore, known GW events were not removed and introduced spectral artefacts were found to be minimal \cite{y_stochasticGatingDocument}.

\subsubsection{Narrowband noise sources}
When analyzing long stretches of data, as is done for the search for a SGWB, spectral lines or artifacts can have a significant impact on the search. These are (very) narrow spectral features that often need to be studied with long duration time segments ($\geq$ 100-1000 s) and over a long time period ($\geq$ 1 day) to be able to observe them. 
Although both the search for an isotropic as well as anisotropic SGWB are affected by these spectral lines, the anisotropic narrowband radiometer search discussed in Section \ref{sec:AnalysisTechniques_Anisotropic} can be affected more since one looks in detail at the frequency spectrum of a specific sky location. Therefore, one needs to understand which features are from experimental/environmental origin and which could be from a GW signal.

In addition to individual narrow spectral lines, spectral artifacts can also appear in a set of lines with equal spacing which is called a comb of lines.

Individual spectral lines can have various origins of which we present the most important categories below.

\begin{itemize}
\item \textbf{Calibration lines:} A set of lines that are purposely injected into the detector to control and calibrate the detector \cite{Aasi_2012}. 
\item \textbf{Mechanical resonances:} Detector components have intrinsic resonances and their spectral artifacts are therefore an intrinsic part of detector design and cannot be completely removed \cite{Aasi_2012}.
\item \textbf{Instrumental lines:} Another part of the lines originates from operating instruments (e.g. air conditioning, vacuum equipment, ...). Often they operate at a certain frequency, inducing spectral lines matching that frequency. The coupling mechanism (e.g. vibrational, magnetic, ...) depends on the type of noise and equipment \cite{Aasi_2012}. Combs most often fall under this last category.
\end{itemize}

Whereas this narrowband noise sources are often not correlated among detectors, this isn't always the case. An example is a 1Hz comb caused by blinking LEDs which are synchronised to the same time, i.e. Global positioning system (GPS) \cite{PhysRevD.97.082002}.

The frequencies of calibration lines and mechanical resonances are typically always excluded from the analysis. Instrumental lines are often only removed from the analysis if they show up significantly in the searches \cite{PhysRevD.104.022004,PhysRevD.104.022005}. These noise lines are eliminated from the analysis by notching them \cite{PhysRevD.104.022004,PhysRevD.104.022005}. 
 
More concrete examples of lines and combs are discussed for the LIGO detectors in \cite{Davis_2021} and for the Virgo detector in \cite{https://doi.org/10.48550/arxiv.2204.03566,Aasi_2012}. The full list of lines for O3 for both LIGO and Virgo can be found in \cite{LVKNoiselines}. The list of spectral lines that was identified for the search for a SGWB using data from LIGO's and Virgo's thrid observing run can be found in \cite{PhysRevD.104.022004,G2001287} for an isotropic SGWB and in \cite{PhysRevD.104.022005,G2002165} for an anisotropic SGWB.
As explained in \cite{PhysRevD.104.022005} the search for an anisotropic GSWB includes some additional notches with respect to the isotropic search. More specifically, the anisotropic analysis has wider notches around (already notched) loud lines e.g. calibration lines. These wider notches aim to remove small spectral artifacts which are caused by gating \cite{PhysRevD.104.022005}.

\subsubsection{Broadband noise sources} 
Apart from narrow spectral features, there are also noise sources which can affect a broader frequency range. The example best studied in the context of the search for a SGWB are the Schumann resonances \cite{Schumann1,Schumann2}, which can couple magnetically to the detectors and induce a correlated signal from terrestrial origin. Schumann resonances are electromagnetic excitations in the cavity formed by the Earth's surface and the ionosphere, sourced by lightning strikes across the globe \cite{Schumann1,Schumann2}. Given their global character, the Schumann resonances are correlated over distances of at least several thousands of kilometers.\\
The Schumann resonances have magnetic strengths of the order of 0.5 - 1 pT at the fundamental mode which has a frequency of 7.8Hz \cite{Schumann1,Schumann2}. However, studies at the site of the underground KAGRA detector have shown that local geographics can magnify the resonances \cite{Atsuta_2016}. Furthermore, the strength varies over the course of the day as well as depending on the time of the year \cite{ZHOU201386}. In addition, the geographical location has an effect as well \cite{Wiener2018}.

\begin{figure}[ht]
\centering
\includegraphics[width=0.48\textwidth]{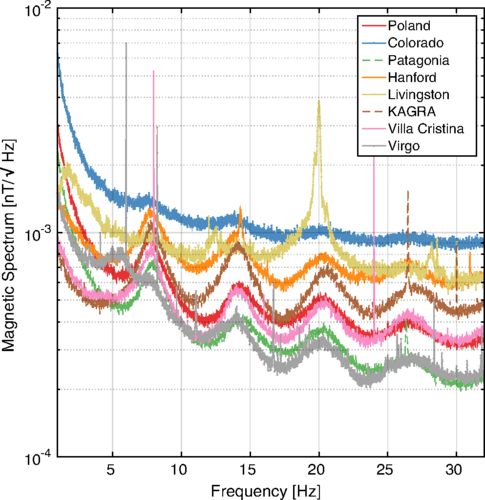}
\caption{Median power spectral density of the North-South direction Poland, Colorado, Villa Cristina and Patagonia magnetometers, as well as the KAGRA, LIGO Hanford and LIGO Livingston X-arm direction magnetometers, as described in \cite{Wiener2018}. These are computed using 128s segments. In addition to the sharp instrumental line features in the Villa Cristina magnetometer at 8 Hz and 24 Hz, the Schumann resonances are visible in all of the magnetometers. The 20 Hz line at LIGO Livingston is likely due to power lines which cross the site on the Y-arm. Figure first presented in \cite{Wiener2018}. \copyright APS. Reproduced with permission.}
\label{fig:SchumannResonances}
\end{figure}

As an example, Fig \ref{fig:SchumannResonances} shows the Schumann resonances as measured at multiple measurement stations around the world. The first four Schumann resonances are clearly visible around 7.8 Hz, 14 Hz, 21 Hz and 27 Hz.

\subsection{Instrumental and environmental monitoring}
\label{sec:AnalysisTechniques_ValidationTechniques_ENVmonitoring}
The LIGO, Virgo and KAGRA detectors have many witness-sensors installed to continuously monitor the instrumental and environmental conditions. The first check to complete when dealing with a possible detection is to look for indications the signal could be caused by instrumental or environmental transients or ambient noise.

Such studies are already performed by the LIGO, Virgo and KAGRA collaborations when searching for a SGWB \cite{PhysRevD.104.022004,PhysRevD.104.022005}. When performing these analyses, spectral features are thoroughly investigated and if witness channels give proof that these spectral features are of environmental or instrumental origin, they are added to a list of known instrumental/environmental lines \cite{LVKNoiselines,G2001287,G2002165}. Based on this list, such lines can be excluded from the analysis.

Another example of an additional study investigating environmental noise, is the construction of the magnetic budget to investigate the impact of global magnetic fields, and more specifically the Schumann resonances that were introduced in the previous section. Given the broadband character of the Schumann resonances - as can be seen in Fig \ref{fig:SchumannResonances} - excluding these frequencies from the analysis would lead to a significant decrease in sensitivity. Therefore, this is not desirable and the magnetic budget informs us whether the amplitude is negligible compared the sensitivity of the search for a SGWB or not. If one finds that the Schumann resonances are affecting our sensitivity, one needs to resort to other techniques such as subtraction (Sec. \ref{sec:AnalysisTechniques_ValidationTechniques_noiseSubtraction}) or Bayesian parameter estimation (Sec. \ref{sec:AnalysisTechniques_ValidationTechniques_BayesianSchumann}).

To construct this magnetic budget one needs to know how the magnetic fields couple to the detector. As in \cite{https://doi.org/10.48550/arxiv.2209.00284} one can conceptually separate this into the product of two different contributions: the inside-to-DARM\footnote{DARM is the GW sensitive channel of the interferometer.} magnetic coupling and the outside-to-inside coupling.
The former is measured by injecting strong magnetic fields in the main buildings of the detector and observe the impact of these injections on the GW-sensitive channel of the detector \cite{Nguyen_2021,galaxies8040082,Cirone_2018,Cirone_2019}. The latter part, as was measured in \cite{https://doi.org/10.48550/arxiv.2209.00284}, describes the reduction/amplification caused by the building.

\begin{figure}[ht]
\centering
\includegraphics[width=0.65\textwidth]{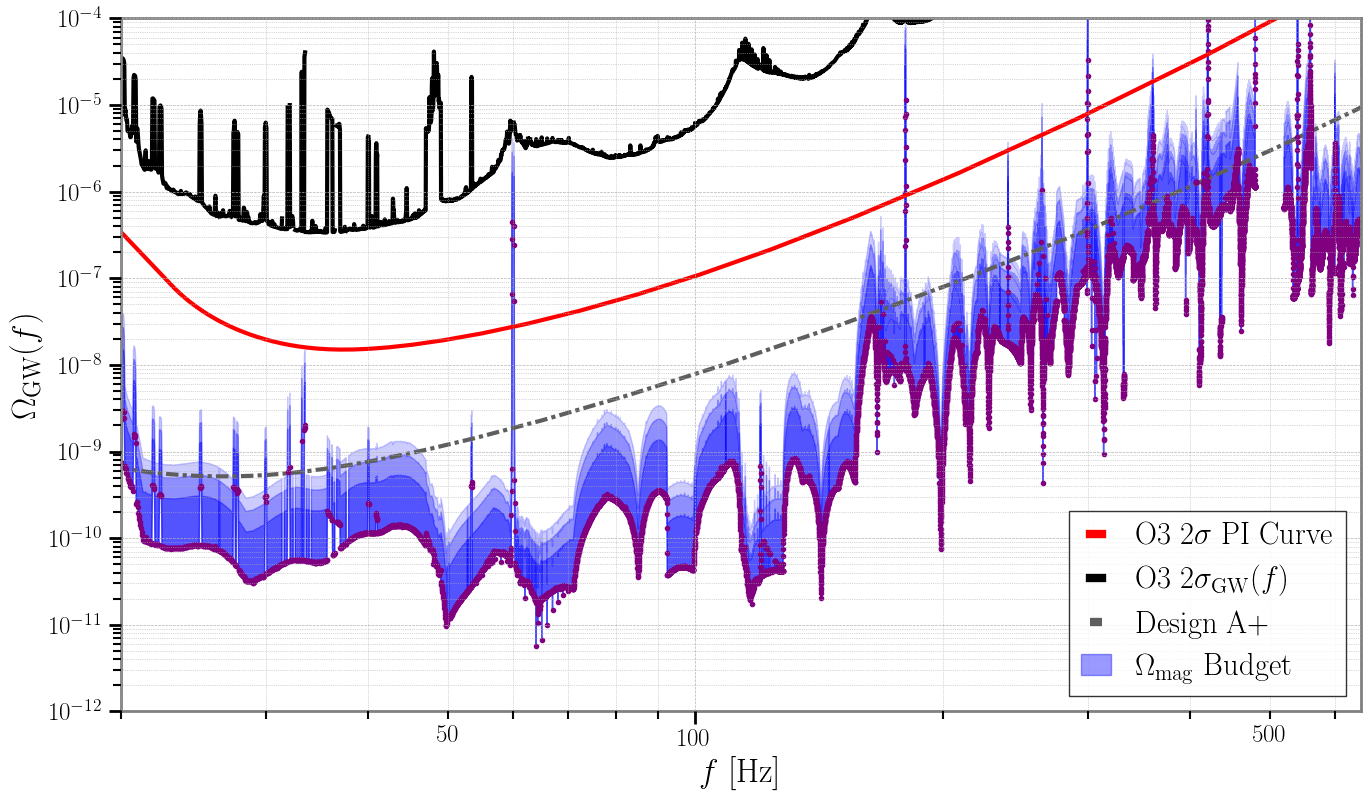}
\caption{
Magnetic noise budget, as presented in Sec. V.B of \cite{https://doi.org/10.48550/arxiv.2209.00284}, represented by the blue band. The purple dots represent the budget without any errors included. The dark and lighter blue bands represent the upper 1$\sigma$-3$\sigma$ uncertainty as described in Appendix B of \cite{https://doi.org/10.48550/arxiv.2209.00284}. No error is included for the weekly variation, however as explained in the text of \cite{https://doi.org/10.48550/arxiv.2209.00284} this effect was found to be minimal.
The lower errors are not shown in this figure since the error propagation as described in Appendix B of \cite{https://doi.org/10.48550/arxiv.2209.00284} leads to negative lower limits.
Also represented are the O3 sensitivity for narrowband features for an isotropic GWB, given by its standard deviation $\sigma(f)$; the O3 broadband sensitivity, given by its power-law integrated (PI) curve (red); and the broadband sensitivity expected to be reached with the LIGO A+ and Advanced Virgo Plus network, Design A+ (grey dot-dashed).
Figure first presented in \cite{https://doi.org/10.48550/arxiv.2209.00284}, which is submitted to Physical Review D.
}
\label{fig:O3MagBudget}
\end{figure}

There are several known mechanisms through which these magnetic fields can couple to an interferometric gravitational-wave detector. A first coupling mechanism is by directly acting on the electromagnetic actuators on the test mass mirrors or their suspensions, causing a physical movement of the test masses \cite{galaxies8040082,Cirone_2019}. Another possible coupling mechanism is by interacting with signal cables, such as the ones connected to the mirror magnets \cite{Cirone_2019}. A third mechanism is by acting on other detector components such as suspended benches. The external magnetic field can couple to magnets or other magnetically sensitive components mounted on the bench, leading to an exerted force on the bench. This force results in a movement of the bench which in turn can enhance other noise sources such as scattered light coupling to the detector and affecting the sensitivity \cite{galaxies8040082}. 
Fig \ref{fig:O3MagBudget} shows the magnetic budget for the search for an isotropic GWB using data from LIGO and Virgo, as computed in \cite{https://doi.org/10.48550/arxiv.2209.00284}. A similar budget was constructed as part of the analysis searching for an isotropic SGWB in \cite{PhysRevD.104.022004}. The more recent budget computed in \cite{https://doi.org/10.48550/arxiv.2209.00284} yields a more up to date budget where the outside-to-inside magnetic coupling is taken into account. Furthermore error propagation is used to construct a final error based on the multiple errors taken into account.

It is concluded there is no effect from magnetic contamination, neither broadband, nor narrowband for an O3 search using LIGO and Virgo data. Broadband features of the magnetic budget have to be below the PI-curve, which, as explained in Sec. \ref{sec:AnalysisTechniques_Isotropic}, is a measure of the sensitivity of an isotopic SGWB search to broadband signals. The narrowband magnetic features on the other hand are well below the sensitivity to GWs in each frequency bin, $\sigma_{\rm GW}$, as defined in Sec. \ref{sec:AnalysisTechniques_Isotropic}, Eq. \eqref{eq:total_crossCorr}.

Whereas the broadband magnetic budget is about one to two orders of magnitude below the latest sensitivity of the third observing run by LIGO and Virgo, correlated magnetic noise might affect stochastic searches in the future. As indicated by Fig. \ref{fig:O3MagBudget} there is a significant risk of contamination from correlated magnetic noise when LIGO and Virgo reach respectively LIGO A+ and Advanced Virgo Plus sensitivities \cite{10.1007/s41114-020-00026-9}. This assumes the coupling function of the detectors remain the same as during O3.
Here we want to point out that the inside-to-DARM magnetic coupling at frequencies above $\sim$160 Hz are often not measured values at the LIGO detectors but rather upper limits.
Therefore it remains crucial to accurately measure this magnetic coupling during future observing runs and assess the importance of the magnetic budget.
Furthermore, as mentioned in \cite{https://doi.org/10.48550/arxiv.2209.00284}, more research might be needed on methods to reduce the magnetic coupling among others magnetic shielding\cite{Cirone_2019}.

Recently a study investigated the impact of correlated magnetic noise in the case of ET, indicating ET has to significantly lower the magnetic coupling below 30 Hz by factors of $10^2$ to $10^4$ compared to the magnetic coupling measured at LIGO and Virgo. In case of insufficient reduction, correlated magnetic noise could be a limiting noise source for the ET performing stochastic searches below 30Hz \cite{PhysRevD.104.122006}.

\subsection{Gravitational wave geodesy - a validation tool for the SGWB}
\label{sec:AnalysisTechniques_ValidationTechniques_Geodesy}
As mentioned before, many external factors can affect the signal, and thus, the claim of a detection as well. Therefore, GW geodesy was proposed as an additional tool to validate the detection of an isotropic SGWB \cite{GWGeodesy}. This method is complementary to using instrumental and environmental monitoring as explained in Sec \ref{sec:AnalysisTechniques_ValidationTechniques_ENVmonitoring}, as well as the methods that will be discussed in Sec. \ref{sec:AnalysisTechniques_ValidationTechniques_noiseSubtraction} and Sec. \ref{sec:AnalysisTechniques_ValidationTechniques_BayesianSchumann}.\\
GWs couple in a specific way to each interferometric gravitational-wave detector, as described by their detector response function, introduced in section Sec. \ref{sec:AnalysisTechniques_Isotropic}. When looking for an isotropic SGWB, these response functions enter via the ORF as defined in Eq. \eqref{eq:ORF}. This ORF uniquely depends on the distance between the two detectors, as well as their rotation angle around their respective vertex point.\\
When a SGWB is observed, its signal has to be consistent with the ORF associated to the observing baseline, i.e. pair of detectors. A priori, there is no reason at all for any non-GW signal to be consistent with this specific ORF or geometrical setup of the detectors. The GW-geodesy tool uses this assumption to differentiate between SGWB signals and other sources of correlation. To this end, two hypotheses are constructed:\\
\begin{itemize}
\item Hypothesis \( \mathcal{H}_{\gamma} \): The observed cross-correlation is consistent with the ORF / geometry of the observing baseline.
\item Hypothesis \( \mathcal{H}_{\rm Free} \): The observed cross-correlation is consistent with an ORF linked to a baseline model with unconstrained geometry on the Earth's surface. The orientation of the individual detectors and the distance between them are inferred for this model.
\end{itemize}

In order to compare these two hypotheses one constructs a Bayes factor,
\begin{equation}
 \label{eq:logBayes}
     \mathcal{B} = \frac{p(\hat{C}|\mathcal{H}_{\mathrm{\gamma}})}{p(\hat{C}|\mathcal{H}_{\mathrm{\mathrm{Free}}})},
\end{equation} 
where $p(\hat{C}|\mathcal{H}_{\mathrm{\gamma}})$ and $p(\hat{C}|\mathcal{H}_{\mathrm{\mathrm{Free}}})$ are the probabilities of finding the observed cross-correlation -- as defined in Eq. \eqref{eq:methods:bin_by_bin_estimator} -- given hypothesis $\mathcal{H}_{\mathrm{\gamma}}$ and $\mathcal{H}_{\mathrm{\mathrm{\rm Free}}}$, respectively. In a Bayesian analysis, the \( \mathcal{H}_{\rm Free} \) hypothesis will be penalized by the ''Occam's factor'' given its more complex model, leading to a preference of the \( \mathcal{H}_{\gamma} \) hypothesis for a SGWB signal. Globally correlated signals from non-GW origin are expected to be consistent with \( \mathcal{H}_{Free} \), due to the additional degrees of freedom in this model.

\begin{figure}[ht]
\centering
     \begin{subfigure}[b]{0.45\textwidth}
         \centering
         \includegraphics[width=\textwidth]{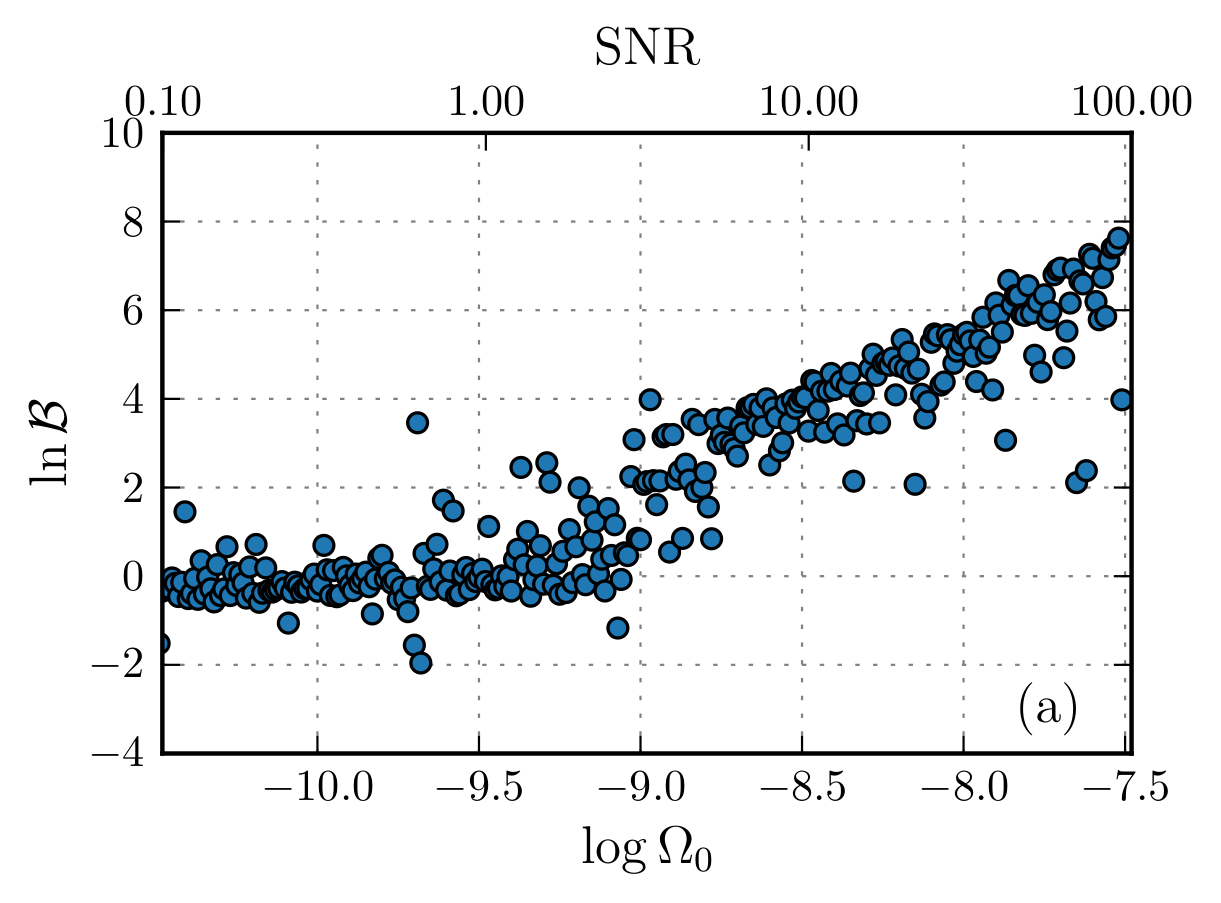}
         \label{fig:GeodesyBayes_PL}
     \end{subfigure}
     \hfill
     \begin{subfigure}[b]{0.45\textwidth}
         \centering
         \includegraphics[width=\textwidth]{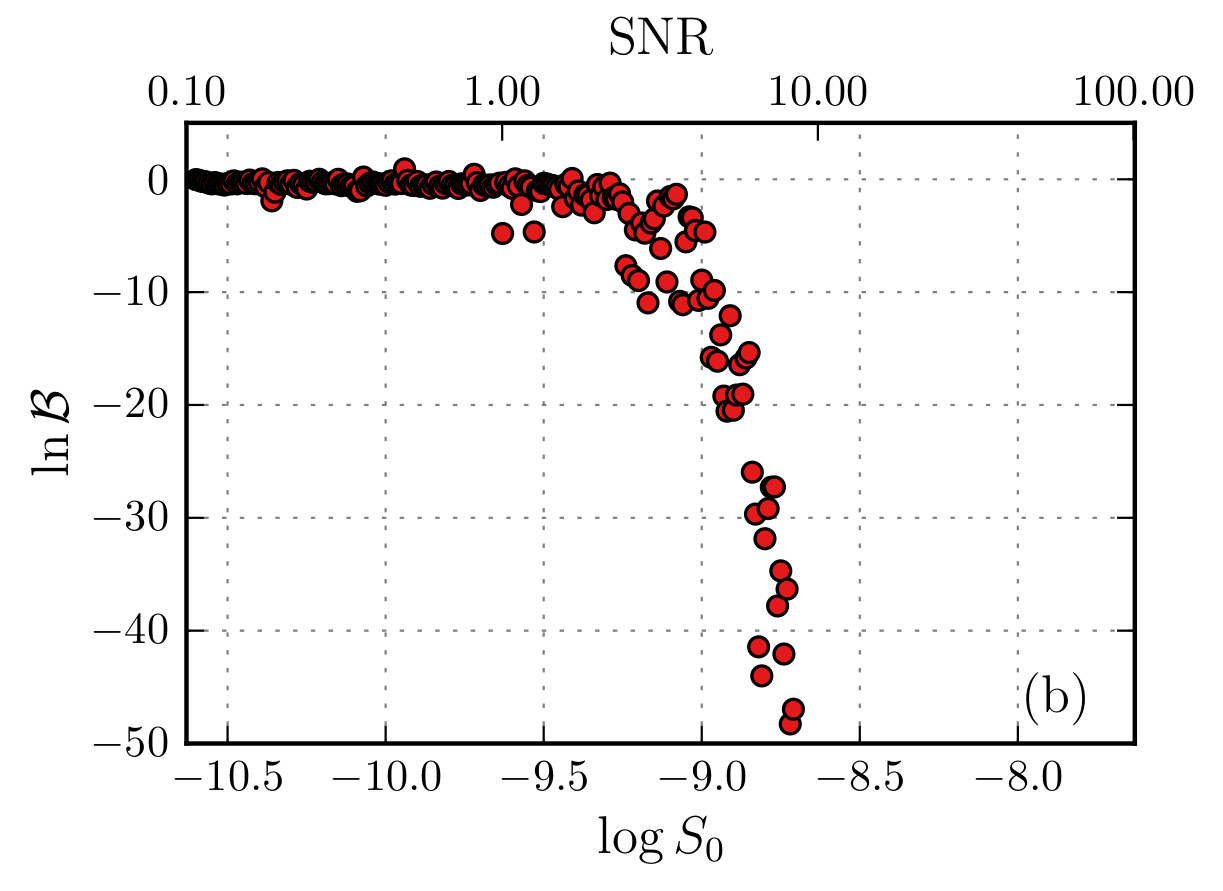}
         \label{fig:GeodesyBayes_Schumann}
     \end{subfigure}%
     \hfill
     \begin{subfigure}[b]{0.45\textwidth}
         \centering
         \includegraphics[width=\textwidth]{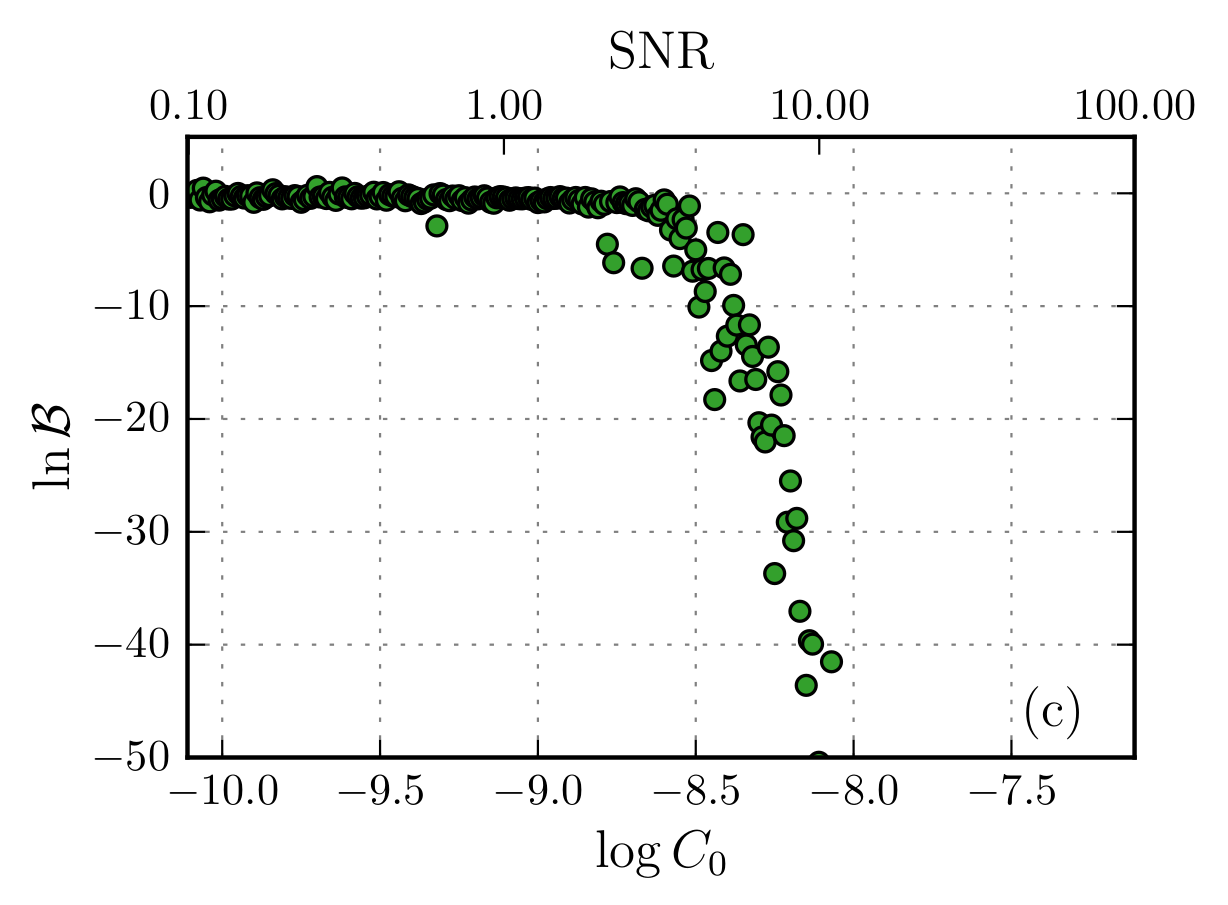}
         \label{fig:GeodesyBayes_comb}
     \end{subfigure}
		\caption{Log-Bayes factors between the physical and un-
physical hypotheses \( \mathcal{H}_{\gamma} \) and \( \mathcal{H}_{Free} \) as a function of injection
strength for isotropic astrophysical backgrounds, Schumann
resonances, and correlated combs [Eqs. (11), (12), and (13) of \cite{GWGeodesy}].
To enable a direct comparison between injection types, the
upper horizontal axes show the signal-to-noise ratios of these
injections. ln\( \mathcal{B} \) increases linearly with the strength of an astrophysical SGWB injection, indicating consistency with the correct
(known) detector geometry. Meanwhile, ln\( \mathcal{B} \) decreases exponentially for the terrestrial sources of correlation, disfavoring
the correct geometry. In the considered cases, ln\( \mathcal{B} \) therefore
successfully discriminates between astrophysical and terrestrial sources of measured cross-correlation. Figure first presented in \cite{GWGeodesy}. \copyright AAS. Reproduced with permission.}
		\label{fig:GeodesyBayes}
\end{figure}

The geodesy tool was demonstrated to be effective in differentiating a correlated signal coming from a comb of lines as well as a signal from Schumann resonances with respect to a SGWB as expected from unresolved CBC events, as shown in Fig \ref{fig:GeodesyBayes} \cite{GWGeodesy}.

\begin{figure}
\centering
\includegraphics[width=\textwidth]{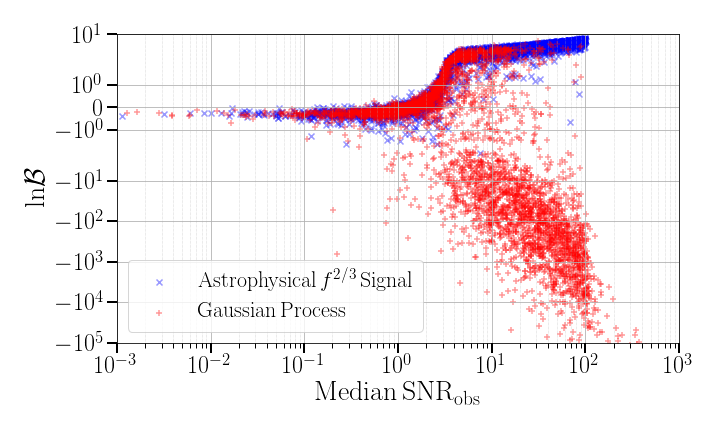}
\caption{The Log-Bayes factor comparing a 2/3-power-law signal with a Gaussian process is plotted versus the median of the SNR-posterior, assuming the signal has been observed by the HL-baseline. There are 40 events with ln$\mathcal{B} < 10^{-5}$, which are not shown in this figure. The smallest Log-Bayes factor is ln$\mathcal{B}= -5.3 \times 10^7$. Each point in this scatter plot represents an injected signal. For each signal type 5000 injections were performed with a logarithmically spaced injection strength between SNR=0.1 and SNR=100. Figure first presented in \cite{PhysRevD.105.082001}. \copyright APS. Reproduced with permission.}
\label{fig:SNRVar_LogBayes}
\end{figure}

In general, this tool can be used for any correlated noise source that might affect the isotropic SGWB search. The only requirement is to know the cross-correlation spectrum. Furthermore, in a recent paper, false alarm probabilities (FAP) and detection probabilities were constructed in the framework of GW-geodesy to make quantitative statements on the origin of a signal being correlated noise or a GW signal \cite{PhysRevD.105.082001}. To achieve this, Gaussian processes were used to generate smooth functions similar to the expected cross correlation statistic from a GW signal. The set of these signals forms a conservative estimation of the background of all correlated noise sources, both known and unknown. Although the conservative nature of the tool makes that the discriminating power of the tool is not excellent, it is a first approach to construct a FAP related to correlated noise. Furthermore the method shows that for increasingly high SNR, the tool becomes better at separating a GW signal from the Gaussian process signal -- the proxy for correlated noise -- as shown in Fig \ref{fig:SNRVar_LogBayes} \cite{PhysRevD.105.082001}. The power of the tool also increases for steeper power-law GW signals, in which case it becomes easier to differentiate GW signal from correlated noise \cite{PhysRevD.105.082001}.  When comparing to known sources such as Schumann resonances, the discriminating power of the tool drastically increases \cite{PhysRevD.105.082001}.

Although the tool is currently only able to validate an isotropic SGWB, investigations are ongoing to explore if the same ideas can be used for validating an anisotropic SGWB \cite{PhysRevD.105.082001}.

\subsection{Subtraction of noise sources}
\label{sec:AnalysisTechniques_ValidationTechniques_noiseSubtraction}

If one is able to measure the noise sources that couple to the detector using dedicated sensors, it could be possible to subtract these noise sources. One needs two key ingredients: an accurate measurement of the noise source, as well as of the transfer function describing how this noise source couples to the channel that the subtraction needs to be applied to.\\
A first method to get the transfer function is to directly measure it. In Sec. \ref{sec:AnalysisTechniques_ValidationTechniques_ENVmonitoring}, we explained how the (inside-to-DARM) magnetic coupling function or transfer function is measured on a regular basis by injecting known magnetic fields and studying the impact on the GW-sensitive channel. However, if one wants to implement successful subtraction, this transfer function should be known accurately. Often this is not the case since the coupling mechanisms are complex and depend on many variables -- such as directionality, homogeneity or phase of the noise source and coupling mechanism -- which might be hard to control during the injections.

A second, more general technique is to use a Wiener-filter to construct the transfer function. First we will explain some general concepts of the Wiener filtering formalism. Afterwards, we will use the Schumann resonances as an example and discuss the investigations that were performed to apply Wiener filtering to reduce/mitigate the impact of Schumann resonances on the search for a SGWB if they were to couple significantly.

\subsubsection{Wiener filter}
If the channel of interest does not only contain the target signal, but also a noise source, one could use witness sensors observing this same noise source to construct a transfer function using a Wiener filter \cite{Wiener2018,Wiener2014,Wiener2016,Wiener2013}. 

Let us assume we have a strain channel $s$, which contains a GW component $h$ which we want to measure. However, assume $s$ also contains the noise source we want to subtract $m$, whose coupling is described by the transfer function $t$. We allow for other noise source(s) $n$ to be present in $s$ as well. Furthermore, we have a witness channel $w$, which monitors the noise source $m$ directly. Our witness channel also contains a noise source $\eta$ which is independent of the $s$ channel. In such a scenario the following equations hold:

\begin{equation}
	\label{eq:channels}
	\begin{aligned}
	\tilde{s}(f) &= \tilde{h}(f) + t(f) \tilde{m}(f) + \tilde{n}(f)  \\
	\tilde{w}(f) &= \tilde{\eta}(f) + \tilde{m}(f),
	\end{aligned}
\end{equation}
where the tildes denote Fourier transforms. The transfer function between the witness channel $w$ and target channel $s$, can be estimated as \cite{Wiener2014}

\begin{equation}
	\label{eq:WienerFilter}
	\hat{t}(f) = \frac{\overline{\tilde{s}(f)\tilde{w}^*(f)}}{\overline{|\tilde{w}(f)|^2}},
\end{equation}
where $\hat{t}$ is the statistical estimator of the transfer function $t$, and the overline in Eq. \eqref{eq:WienerFilter} implies time-averaging.\\

\subsubsection{Noise subtraction}
Once one has an estimate for the transfer function $\hat{t}(f)$, either because one measured it, or because one constructed it using the Wiener filtering formalism as in Eq. \eqref{eq:WienerFilter}, one can construct the subtracted data \cite{Wiener2014}

\begin{equation}
	\label{eq:subtractedData}
	\tilde{s}^\prime(f) = \tilde{s}(f) - \hat{t}(f) \tilde{w}(f).
\end{equation}
When the estimator $\hat{t}(f)$ equals the true value $t(f)$, the correlated noise between the target channel and witness channel $m$ will be successfully subtracted. This limit is reached in the regime where $m$ is the dominant component of the witness channel $w$, and $\eta$ is negligible. If $\eta$ dominates the witness channel $w$, Wiener filtering will fail.

If one defines $M(f) = \langle \tilde{m}^*(f) \tilde{m}(f)\rangle$ and $\mathcal{N}(f) = \langle \tilde{\eta}^*(f) \tilde{\eta}(f)\rangle$, one can define a ''witness signal-to-noise ratio'' $\rho_w$, 
\begin{equation}
        \label{eq:SNRw}
        \rho_w = \sqrt{\frac{M(f)}{\mathcal{N}(f)}},
\end{equation}
which plays an important role in determining whether Wiener filtering is successful or not \cite{Wiener2014}. Although partial subtraction might already take place when $\rho_w<1$, the witness signal-to-noise ratio should be large enough ($\rho_w > 4$) to achieve more complete noise subtraction.

\subsubsection{Case study of noise subtraction -- Schumann resonances}
Noise subtraction has been tested using simulated data in the context of Schumann resonances, both with measuring the transfer function as well as constructing it using a Wiener-filter. Furthermore a proof of concept analysis has shown Wiener filtering is effective in (partial) subtraction when applied to magnetometers \cite{Wiener2018,Wiener2014,Wiener2016,Wiener2013}.

\begin{figure}[ht]
\centering
\includegraphics[width=0.48\textwidth]{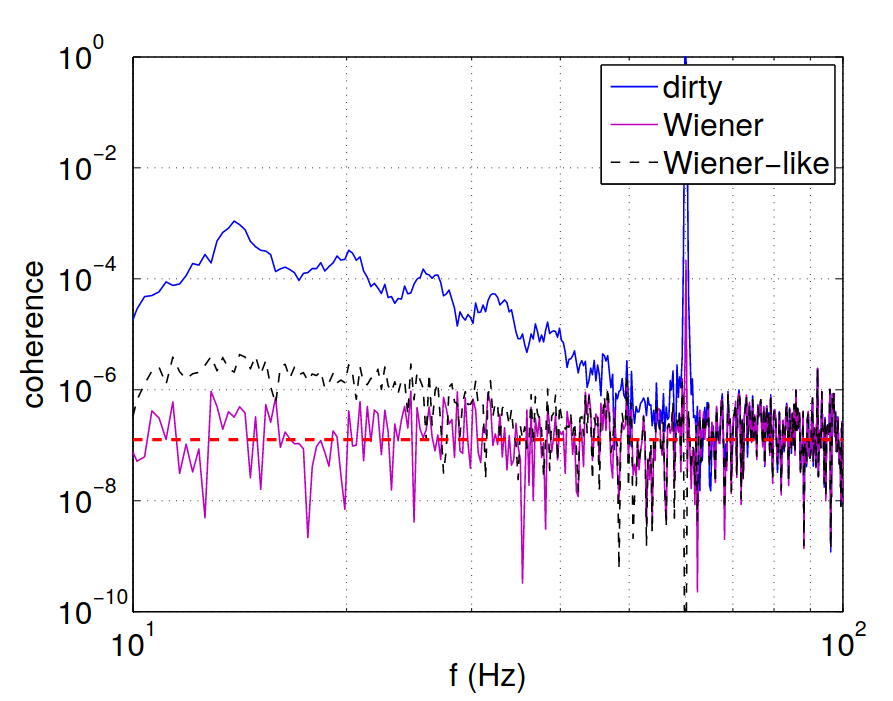}
\caption{Coherence spectra showing the contaminated strain channels (blue), the somewhat cleaner spectrum obtained by Wiener-like filtering (black), and the spectrum obtained with true Wiener filtering (magenta). The Wiener-filtered spectrum is significantly cleaned, but measurable residual contamination remains. 1 yr of integration time was assumed. Figure first presented in \cite{Wiener2014}. \copyright APS. Reproduced with permission.}
\label{fig:SchumannWienerFilter}
\end{figure}

Fig \ref{fig:SchumannWienerFilter} shows the effect of Wiener and Wiener-like filtering -- that is the formalism where one estimates $\hat{t}(f)$ relying on other methods than the Wiener filter, e.g. direct measurement-- on simulated strain data containing Schumann resonances. Both formalisms are able to subtract a part of the correlated Schumann noise, where the Wiener filter performs the best. However residual contamination remains \cite{Wiener2014}.

It has been shown Wiener filtering could be used to significantly subtract the effect of Schumann resonances \cite{Wiener2018,Wiener2014,Wiener2016,Wiener2013}. Since the Schumann resonances are currently not coupling significantly to detectors, the method could not be applied to real GW data. However, the simulations combined with proof of concept applied to magnetometers illustrate Wiener filtering would be promising once Schumann resonances couple significantly to the interferometers \cite{Wiener2018,Wiener2014,Wiener2016,Wiener2013}.\\
Another noise source for which Wiener filtering is investigated is Newtonian noise \cite{PhysRevD.86.102001,PhysRevD.92.022001,Coughlin_2014_NN,Badaracco_2019}. Although Newtonian noise will mainly become an important issue for third generation interferometric gravitational-wave detectors such as the Einstein Telescope \cite{NN_Sardinia2020,10.1785/0220200186,Bader_2022,Koley_2022}, Newtonian noise subtraction has also been investigated for second generation detectors Advanced LIGO \cite{Coughlin_2016,PhysRevLett.121.221104} and Advanced Virgo \cite{Tringali_2019,Badaracco_2020}.

\subsection{Joint Bayesian modeling of noise sources and a SGWB}
\label{sec:AnalysisTechniques_ValidationTechniques_BayesianSchumann}

Recently, a new method was proposed to take Schumann resonances into account using a Bayesian model selection \cite{BayesianGWMag}. However, one could imagine using this technique for any known source of correlated noise.\\
\indent

In the context of taking Schumann resonances into account, one can construct a likelihood as follows\footnote{Note our Eq. \eqref{eq:PE_likelihood_withSchumann} uses a slightly different notation compared to Eq. (25) of \cite{BayesianGWMag}.}

\begin{equation} \label{eq:PE_likelihood_withSchumann}
p(\hat{C}_k^{IJ}|\mathbf{\Theta}_{\rm GW},\mathbf{\Theta}_{\rm Mag}) \propto \exp \left[ -\frac{1}{2} \sum_{IJ} \sum_k \left( \frac{\hat{C}_k^{IJ} - \Omega_{\rm GW}(f_k|\mathbf{\Theta}_{\rm GW}) - \Omega_{\rm Mag}(f_k|\mathbf{\Theta}_{\rm Mag})}{\sigma^2_{IJ}(f_k)}\right) \right] .
\end{equation}
Compared to Eq. \eqref{eq:PE_likelihood}, Eq. \eqref{eq:PE_likelihood_withSchumann} takes the Schumann resonances $\Omega_{\rm Mag}(f_k|\mathbf{\Theta})$ into account in addition to the SGWB model $\Omega_{\rm GW}(f_k|\mathbf{\Theta})$.\\
This framework has been applied to a typical magnetic Schumann spectrum \cite{BayesianGWMag}, where the inside-to-DARM magnetic coupling functions were approximated as a power-law and the coupling strength and the power-law slope were treated as nuisance parameters.\\

It was shown that the method is able to differentiate between an isotropic SGWB and correlations arising from Schumann resonances \cite{BayesianGWMag}. However, strong magnetic coupling or noisy magnetic measurements can decrease the significance of the detection of a SGWB. Furthermore, the simulations show that going from a two to a three detector network significantly increases the effectiveness of the method, regardless of whether the third detector contributes much to constraining the SGWB power or not. This strongly supports using a larger detector network even though their contribution to observing a SGWB might be small/negligible due to a lower sensitivity or effects of the ORF, as is the case for adding Virgo and KAGRA to the LIGO Hanford and Livingston network. Since correlated noise sources, e.g. Schumann resonances, couple differently to the network compared to a SGWB, a larger network helps differentiating a SGWB from correlated noise.

While this method can be used as an alternative for the subtraction of the Schumann resonances with Wiener filtering as explained in Sec. \ref{sec:AnalysisTechniques_ValidationTechniques_noiseSubtraction}, both methods could also be used simultaneously. One could imagine first applying (partial) Wiener filtering and afterwards use this Bayesian parameter estimation model to allow for a remaining contribution coming from the Schumann resonances. Depending on the situation, one could use these two methods in parallel and cross check the two independent results with each other, or adopt this combined approach. This leads to a large flexibility in applying these methods and ensuring to efficiently take globally cross-correlated signals such as the Schumann resonances into account.

\subsection{The null channel: a gravitational wave insensitive channel for a triangular configuration of interferometers}
\label{sec:AnalysisTechniques_ValidationTechniques_NullChannel}
%detectorresponse_a

For future instruments it might be challenging to estimate the PSD due to a large amount of signals present in the data. A first method to deal with this is to perform a joined estimation of noise and signal parameters~\cite{PhysRevD.92.064011,Christensen:2022bxb}. However, in the case of a triangular configuration of interferometers one can also use the so-called (sky independent) null channel (also know as null stream or symmetrized Sagnac) \cite{PhysRevD.86.122001,Tinto2005,PhysRevD.100.104055}.  
For a network consisting of $N$ detectors one can construct $N - 2$ sky location dependent null streams \cite{PhysRevD.40.3884,Wen_2005,Wen_2008,PhysRevD.81.082001,Sutton:2009gi}. However, in the case of a triangular configuration of three interferometers, there is one unique null channel which is insensitive to GWs from every direction. 

As explained in Sec. \ref{sec:DetectionMethods_LaserITF}, the proposal of the third generation ET interferometric gravitational-wave detector -- as well as the space-based LISA -- uses an equilateral triangular composition. 
Therefore, this method has been studied for the ET in the context of the SGWB \cite{PhysRevD.86.122001} as well as to understand the effect of glitches \cite{PhysRevD.105.122007}.
It also has been studied extensively in the context of the space based LISA detector \cite{PhysRevD.100.104055,PhysRevD.82.022002,2014PhDT286A,2021PhRvD.103j3529B,10.1093/mnras/stab2575,PhysRevD.105.023510}, and to a smaller extent in context of the space based TianQin detector \cite{PhysRevD.105.022001}. 

In the context of the SGWB, which targets unresolved sources with an isotropic or extended sky location, one can only use the unique null channel of a triangular configuration of interferometers (e.g. ET, LISA, ...) which is insensitive to GW sources from all directions.
If one assumes an equilateral triangular formation of three interferometers, their detector response functions, first introduced in Sec. \ref{sec:DetectionMethods_LaserITF}, are given by
\begin{equation}
\label{eq:detectorresponse_b}
\begin{aligned}
\bf{d^1} &= \frac{1}{2}(\bf{e^1}\otimes\bf{e^1} - \bf{e^2}\otimes\bf{e^2})\\
\bf{d^2} &= \frac{1}{2}(\bf{e^2}\otimes\bf{e^2} - \bf{e^3}\otimes\bf{e^3})\\
\bf{d^3} &= \frac{1}{2}(\bf{e^3}\otimes\bf{e^3} - \bf{e^1}\otimes\bf{e^1})\\
\end{aligned}
\end{equation}

The sum of the strain output of the three interferometers becomes insensitive to any GW signal \cite{PhysRevD.86.122001,2022PhRvD.105h4002W}
\begin{equation}
\label{eq:nullchannel}
\begin{aligned}
\sum_{n=1}^3 s_n(t) &= \sum_{n=1}^3 \sum_m d_{n,ij} e_m^{ij} h_m(t) \\
&= \sum_m \left( \sum_{n=1}^3 d_{n,ij}\right)e_m^{ij} h_m(t)
&= 0,
\end{aligned}
\end{equation}
since from Eq.~\eqref{eq:detectorresponse_b} follows $\sum_{n=1}^3 \bf{d}_{n}=\bf{0}$. Regardless of the waveforms and polarization of the GWs, the sum of three interferometers which form an equilateral triangle, as is the case for ET and LISA, becomes insensitive to the GWs. Since this sum, the null channel, is insensitive to any GW signal, one can use it to estimate the noise sources present in the detectors. 

The formalism has been demonstrated for the ET \cite{PhysRevD.86.122001}, as well as for LISA \cite{PhysRevD.82.022002,2014PhDT286A,2021PhRvD.103j3529B,10.1093/mnras/stab2575,PhysRevD.105.023510}. However, especially in the context of the ET, more work is needed to go from the relatively simple example use in \cite{PhysRevD.86.122001} to more complex and realistic situations.

While the null channel might be a necessary tool for PSD estimation, it might also be valuable for data quality and the identification and characterisation of correlated noise sources.
The basic search methods for the SGWB rely on cross-correlations between detectors. Nevertheless, as discussed in Sec. \ref{sec:AnalysisTechniques_Isotropic_TBS} there are also other analysis techniques being investigated. However, correlated noise can limit the sensitivity of these searches in both cases. A large distance between the detectors of the baseline can greatly reduce the amount of correlated noise. Only some global sources as explained in Sec. \ref{sec:AnalysisTechniques_ValidationTechniques_GlobalCorrelatedNoise} remain correlated on the scale of thousands of kilometers.\\
When using (nearly) co-located detectors, cross-correlating data from the detectors no longer suppresses the same amount of noise, since both detectors are to a large extent affected by the same local noise sources. Such searches were conducted in the past using the two co-located LIGO Hanford detectors: H1 and H2 \cite{Fotopoulos_2008,PhysRevD.91.022003}. During the analysis of the H1-H2 data from the fifth science run of initial LIGO \cite{PhysRevD.91.022003}, the frequency band 40Hz to 460Hz was not used in the analysis due to insufficient mitigation of correlated noise. Several methods, such as time shifting the data and computing coherence with environmental noise monitoring channels were used to ensure the other frequency regions were not significantly affected \cite{Fotopoulos_2008,PhysRevD.91.022003}.

The equilateral triangular detector configuration implies the different detectors making up this triangular configuration are (almost) co-located. Due to their co-located nature correlated noise could be expected. There have been several studies investigating correlated magnetic noise at the ET \cite{PhysRevD.104.122006,https://doi.org/10.48550/arxiv.2209.00284} as well as correlated seismic and Newtonian noise \cite{PhysRevD.106.042008}. Based on LISA pathfinder data also some preliminary noise correlation studies have been performed \cite{2022arXiv220403867B}.

This implies the three interferometers of the ET configuration might have many (local) correlated noise sources, which will lead to a non-negligible impact of these noise sources on the cross-correlation spectrum used in searches for a SGWB.

One could use a similar technique as used for the H1-H2 analysis \cite{Fotopoulos_2008,PhysRevD.91.022003} and rely on environmental and instrumental monitors. However, as was the case for the H1-H2 analysis \cite{Fotopoulos_2008,PhysRevD.91.022003}, this method might be insufficient to ensure a clean spectrum in the entire sensitive frequency range of the ET detector. Furthermore, noise that couples non-linearly to the detector (e.g. scattered light) would not be taken into account by these methods \cite{Fotopoulos_2008,PhysRevD.91.022003}. While the environmental and instrumental monitoring will be very important, one should aim to do better by using additional techniques.

The null channel can also be used as a tool to understand the levels of correlated noise between the different detectors as was investigated by a recent study \cite{https://doi.org/10.48550/arxiv.2205.00416}.

    \newpage
\section{Implications}
\label{sec:Implications}

As various detection methods were introduced throughout this paper, upper limits provided by these detection methods were given as well. Based on the upper limits on the SGWB itself, it is possible to derive various astrophysical and cosmological implications. An overview of some of these implications are respectively given in Sec. \ref{sec:Implications_Astrophysical} and Sec. \ref{sec:Implications_Cosmological}.

%------------------%
%-- Implications --%
%-- Astrophysical -%
%------------------%

\subsection{Astrophysical Implications}
\label{sec:Implications_Astrophysical}
\begin{figure}[ht]
	\centering
	\includegraphics[scale=.65]{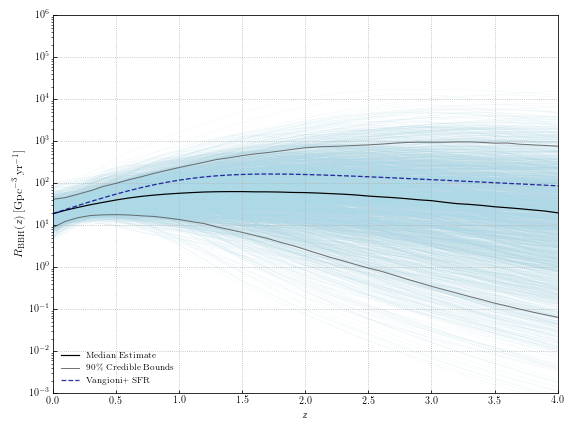}
	\caption{Constraints on the BBH merger rate coming from the isotropic stochastic O3 search \cite{PhysRevD.104.022004}, combined with BBH direct detections from the GWTC-2 catalog \cite{PhysRevX.11.021053}. The gray lines denote the 90\% credible bounds, while the solid black line denotes the median estimate for the merger rate $R_{\rm BBH}(z)$. The original version of this plot was presented in~\cite{PhysRevD.104.022004}; the version shown here was obtained using open data published in \cite{O3IsotropicDataset}.}
	\label{Fig:MergerRateO3}
\end{figure}

\begin{figure}[ht]
	\centering
	\includegraphics[scale=.6]{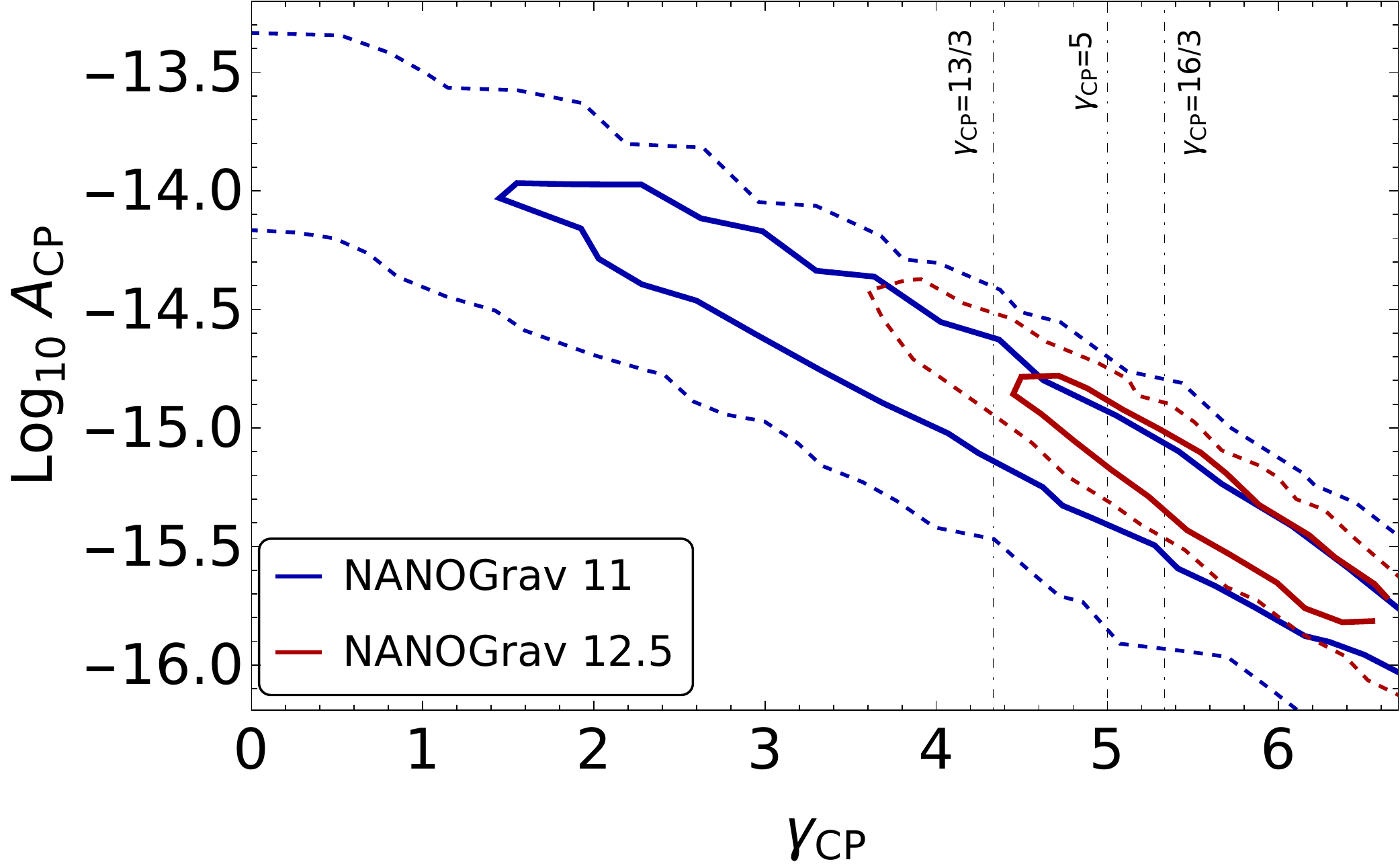}
	\caption{Posterior for a common-spectrum process with spectral index $\gamma_{\rm CP}$ and amplitude $A_{\rm CP}$ across all pulsars \cite{2020}. The blue and red lines represent the results from the 11 and the 12.5 year data set, respectively. Both colors show the 1$\sigma$ and 2$\sigma$ contours. The vertical lines show the value of $\gamma_{\rm CP}$ expected for a SGWB from SMBHBs ($\gamma_{\rm CP}=13/3$), from primordial backgrounds ($\gamma_{\rm CP}=5$), and from cosmic strings ($\gamma_{\rm CP}=16/3$) \cite{phinney2001practical}. The original version of this plot was presented in~\cite{2020}.}
	\label{Fig:SMBHNanoGrav}
\end{figure}

\paragraph{Binary black hole merger rate}~\\
From the direct observation of binary black hole mergers, it is possible to construct the most plausible BBH merger rate $R_{\rm BBH}(z)$ as a function of redshift $z$. Using this estimate for the merger rate density of binary black holes, one can predict the GW energy density for the astrophysical background from BBHs, as was shown in Fig.~\ref{Fig:pop-omega} \cite{PhysRevD.104.022004}. Concretely, one uses the estimated merger rate to integrate the energy density radiated by each individual source over the redshift $z$, thus obtaining an overall energy density for the SGWB from BBHs in the Universe.\\ However, one can reverse the process and attempt to directly measure the binary black hole merger rate $R_{\rm BBH}(z)$ by combining direct observations of binary black hole mergers from the GWTC-2 catalog \cite{PhysRevX.11.021053}, as well as the upper limits on the SGWB obtained in the LVK collaboration's third observing run (O3) \cite{PhysRevD.104.022004}. This method was first proposed and applied to O2 data in \cite{Callister_2020}. A broken power-law is assumed for the BBH mass distribution and the following phenomenological parameterization is assumed for the rate:
\begin{equation}
\label{Eq:MergerRate}
	R_{\rm BBH}(z)=\mathcal{C}\left(\lambda_1,\lambda_2,z_{\rm peak}\right)\frac{R_0(1+z)^{\lambda_1}}{1+\left(\frac{1+z}{1+z_{\rm peak}}\right)^{\lambda_1+\lambda_2}},
\end{equation}
where $\mathcal{C}$ is a normalization constant and $R_0$ is the local merger rate. This parameterization describes a merger rate that evolves as $R_{\rm BBH}(z)\sim(1+z)^{\lambda_1}$ for $z<z_{\rm peak}$ and $R_{\rm BBH}(z)\sim(1+z)^{-\lambda_2}$ for $z>z_{\rm peak}$. By combining direct observations and results from the isotropic stochastic search, it is possible to perform Bayesian inference on the parameters (e.g. $z_{\rm peak}$, $\lambda_1$, and $\lambda_2$) describing the merger rate given by Eq.~\eqref{Eq:MergerRate}, yielding posterior distributions for each of them. One constructs the following likelihood:
\begin{equation}
\label{eq:LikelihoodsMergerRate}
    p\left(\hat{C}
    (f), \{d_i\}| R_{BBH}\right)=p_{\rm BBH}\left(\{d_i\}|R_{BBH}\right)p_{\rm SGWB}\left(\hat{C}(f)|R_{BBH}\right),
\end{equation}
where the $p_{\rm BBH}$ represents the probability of detecting a direct BBH observation $\{d_i\}$ given a merger rate $R_{\rm BBH}$ and $p_{\rm SGWB}$ that of detecting a SGWB given merger rate $R_{\rm BBH}$. For each parameter draw (e.g. $\lambda_1$, $z_{\rm peak}$, etc.), one obtains a different merger rate $R_{\rm BBH}$ in Eq. \eqref{eq:LikelihoodsMergerRate}, eventually resulting in posterior constraints on the rate $R_{\rm BBH}(z)$, depicted by the blue lines in Fig~\ref{Fig:MergerRateO3}. The gray lines represent the 90\% credible bounds on the rate $R_{\rm BBH}(z)$, whereas the black line represents the median value of the rate. For redshifts larger than $z\simeq 2$, the merger rate can be bounded to be below $\sim10^3$ Gpc$^{-3}$yr$^{-1}$ at 90\% credibility, constituting an improvement of an order of magnitude compared to the first two observing runs of the LVK collaboration \cite{Callister_2020}. It is interesting to note that the GWTC-2 direct BBH detection catalog is able to estimate the $\lambda_1$ parameter to be $\lambda_1=1.3\pm 2.1$, whereas current SGWB searches are unable to further constrain its value  \cite{PhysRevX.11.021053}. Thus, Fig. \ref{Fig:MergerRateO3} is largely dominated by direct BBH detections. However, it is expected that next observing runs will allow SGWB constraints to offer more information and constrain the value of $\lambda_1$ and $z_{\rm peak}$ further \cite{PhysRevD.104.022004}. Furthermore, one observes that no constraints can be placed on the parameter $\lambda_2$ as of now, since it is related to high redshift behavior. For more details on the Bayesian inference and results, we refer the reader to \cite{PhysRevD.104.022004}.
\paragraph{Supermassive black holes binaries}~\\
As mentioned in Sec. \ref{sec:DetectionMethods_PTA}, the 12.5 year data set of NANOGrav exhibits evidence for a common-spectrum process across all pulsars \cite{2020}. However, as the expected Hellings and Downs correlation is not found within this data, they are unable to claim the detection of a SGWB. Nevertheless, the posterior distribution plot of the spectral index $\gamma_{CP}$ and amplitude $A_{CP}$ hinting at such a common-spectrum process is displayed in Fig~\ref{Fig:SMBHNanoGrav} \cite{2020}. The possibility that this common-spectrum process comes from a SGWB is discussed in \cite{2020}. Various candidate sources are considered, among which a SGWB of supermassive black hole binaries (SMBHBs) would be the dominant contribution \cite{Sesana_2004}. Although black hole mergers have been studied extensively, the question remains whether SMBHB mergers can occur. The amplitude of such a background from SMBHBs would be determined by the number of binaries as well as their mass distribution. The detection of a SGWB from SMBHBs would be the first evidence that SMBHBs can form and eventually coalesce after the emission of GWs \cite{2020}. Furthermore, if one attributes this common-spectrum process to a GWB from SMBHBs, the recovered amplitude could imply that the BH mass function was underestimated \cite{10.1093/mnras/sty2849}. For more details, we refer the reader to \cite{2020}.
~\\
However, one notes that the posterior on the spectral index $\gamma_{\rm CP}$ depicted in Fig~\ref{Fig:SMBHNanoGrav} is fairly broad. Thus, the common process could be attributed to another background different than the one coming from SMBHBs ($\gamma_{\rm CP}=13/3$), such as cosmic strings ($\gamma_{\rm CP}=16/3$) or a primordial SGWB ($\gamma_{\rm CP}=5$). Longer pulsar measurements will constrain the parameter space even further and will provide more evidence for or against the detection of a SGWB.

%------------------%
%-- Implications --%
%-- Cosmological --%
%------------------%

\subsection{Cosmological Implications}
\label{sec:Implications_Cosmological}
\begin{figure}[ht]
	\centering
	\hspace{-0.5cm}
	\includegraphics[scale=.4]{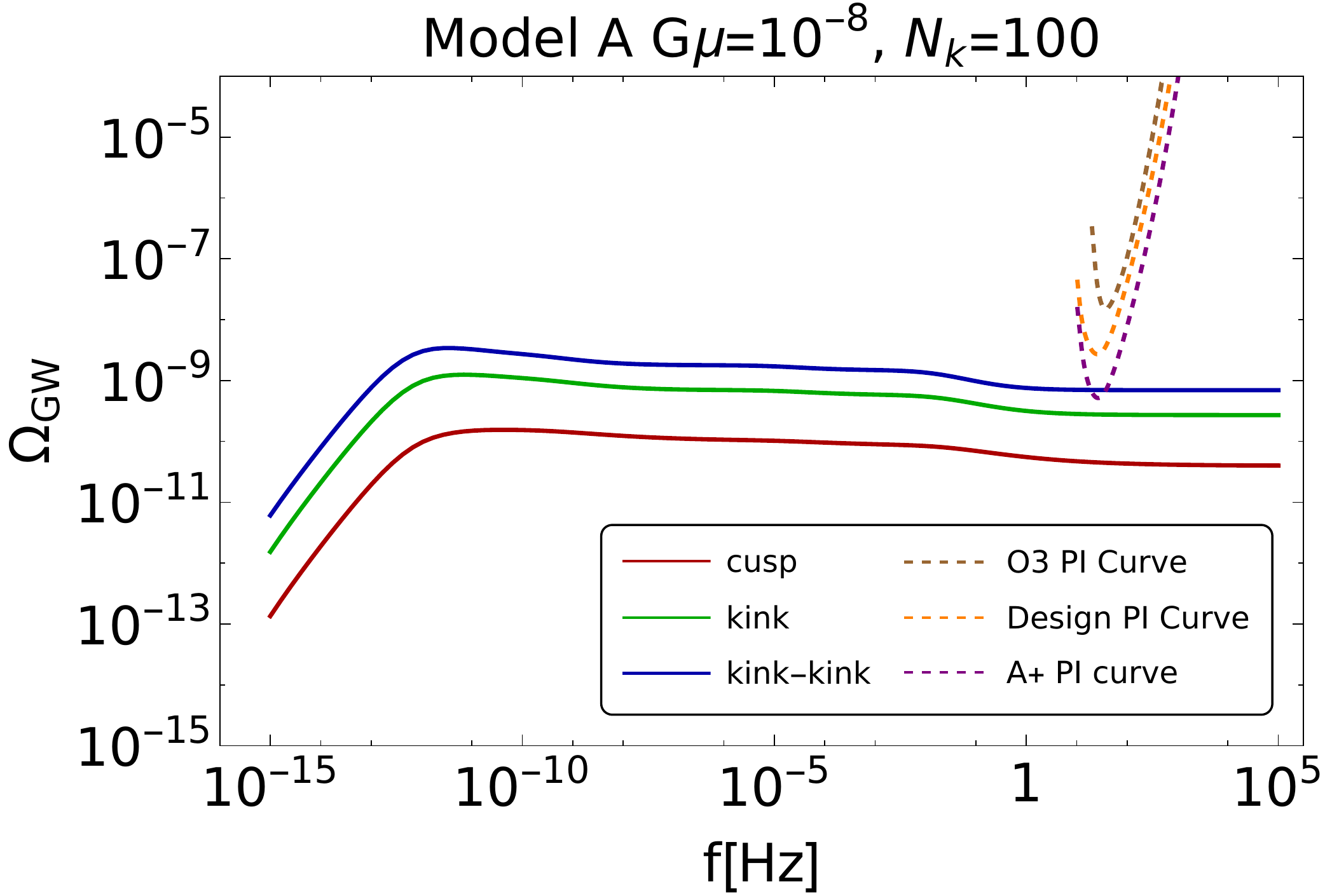}
	\includegraphics[scale=.4]{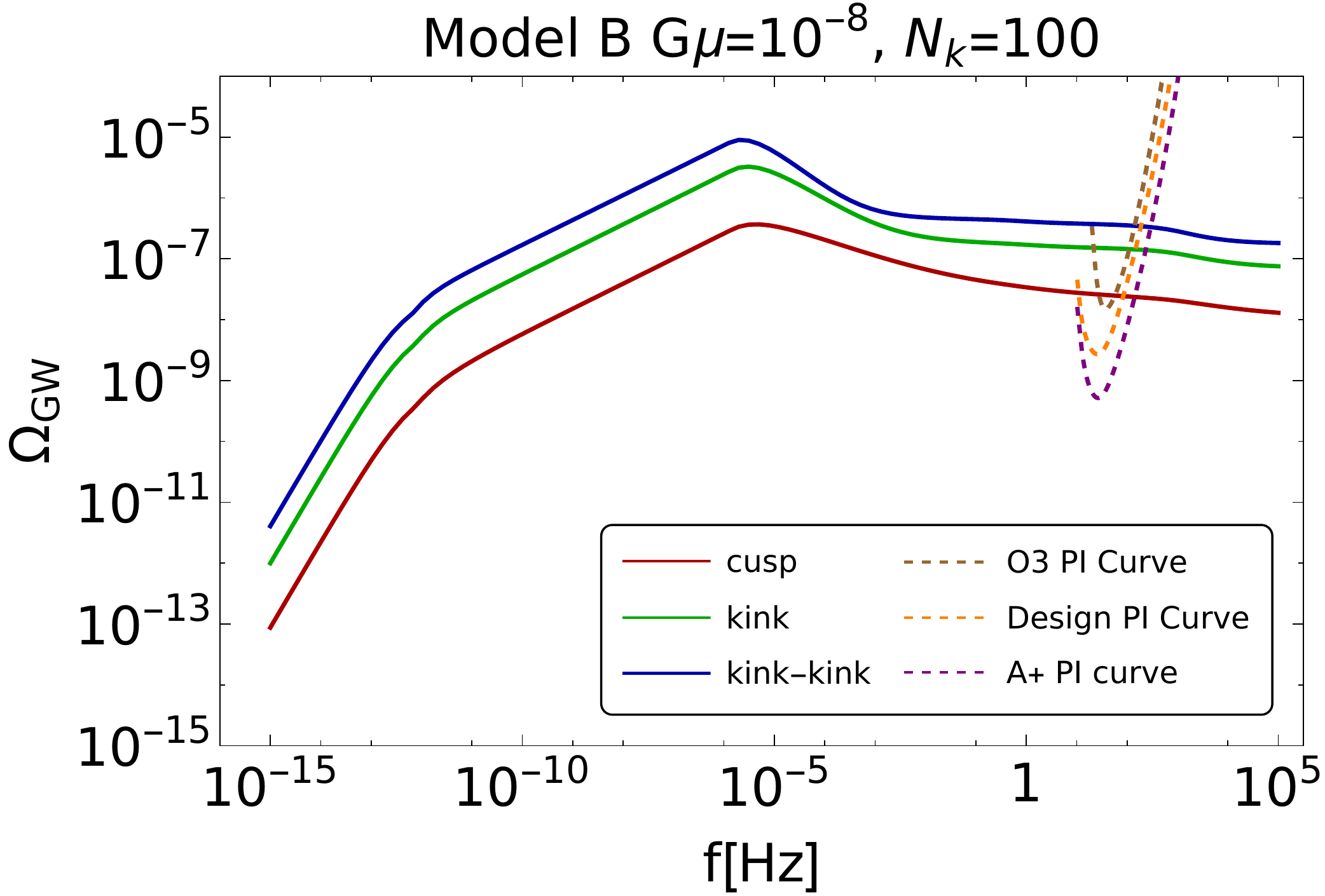}
	\includegraphics[scale=.4]{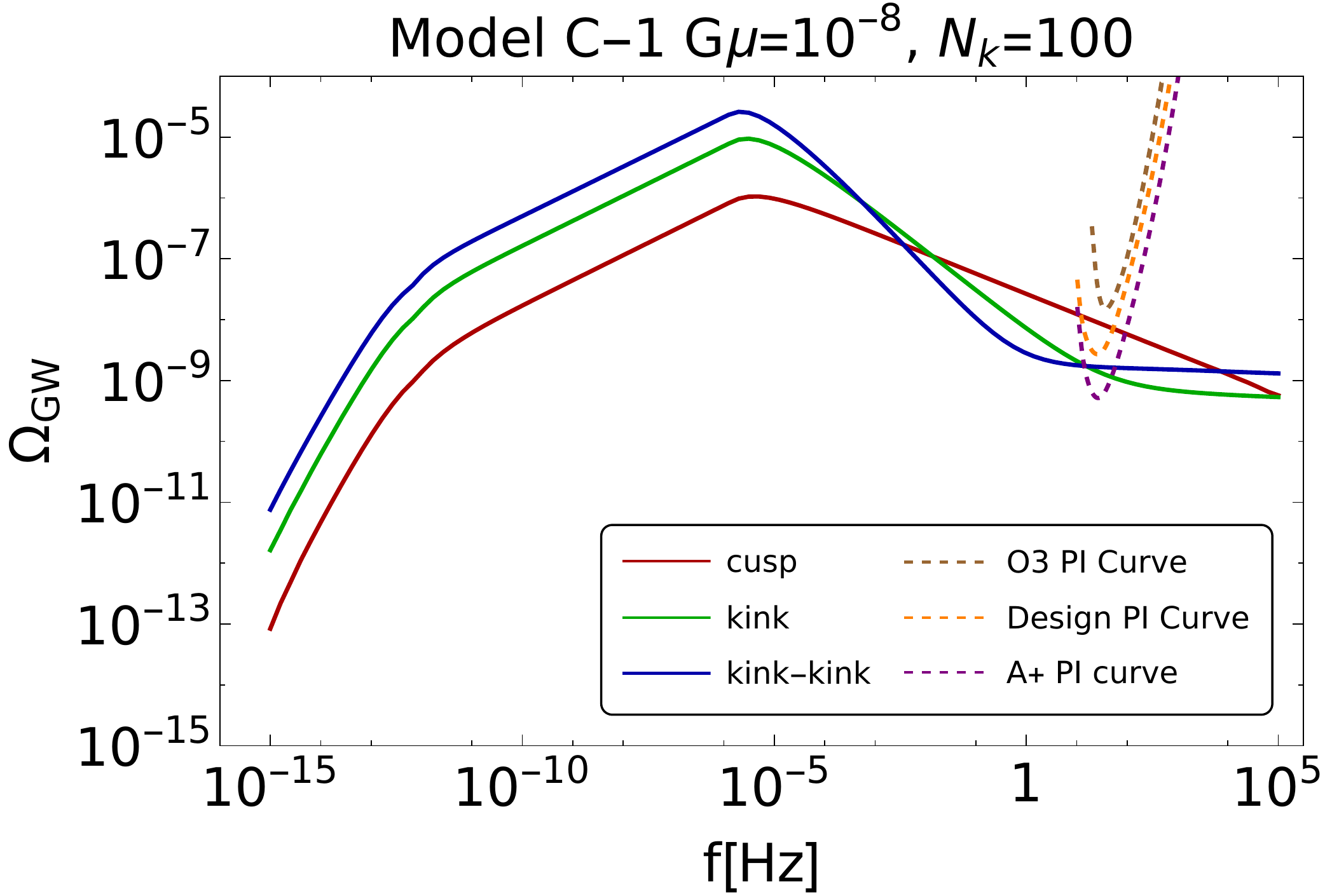}
	\includegraphics[scale=.4]{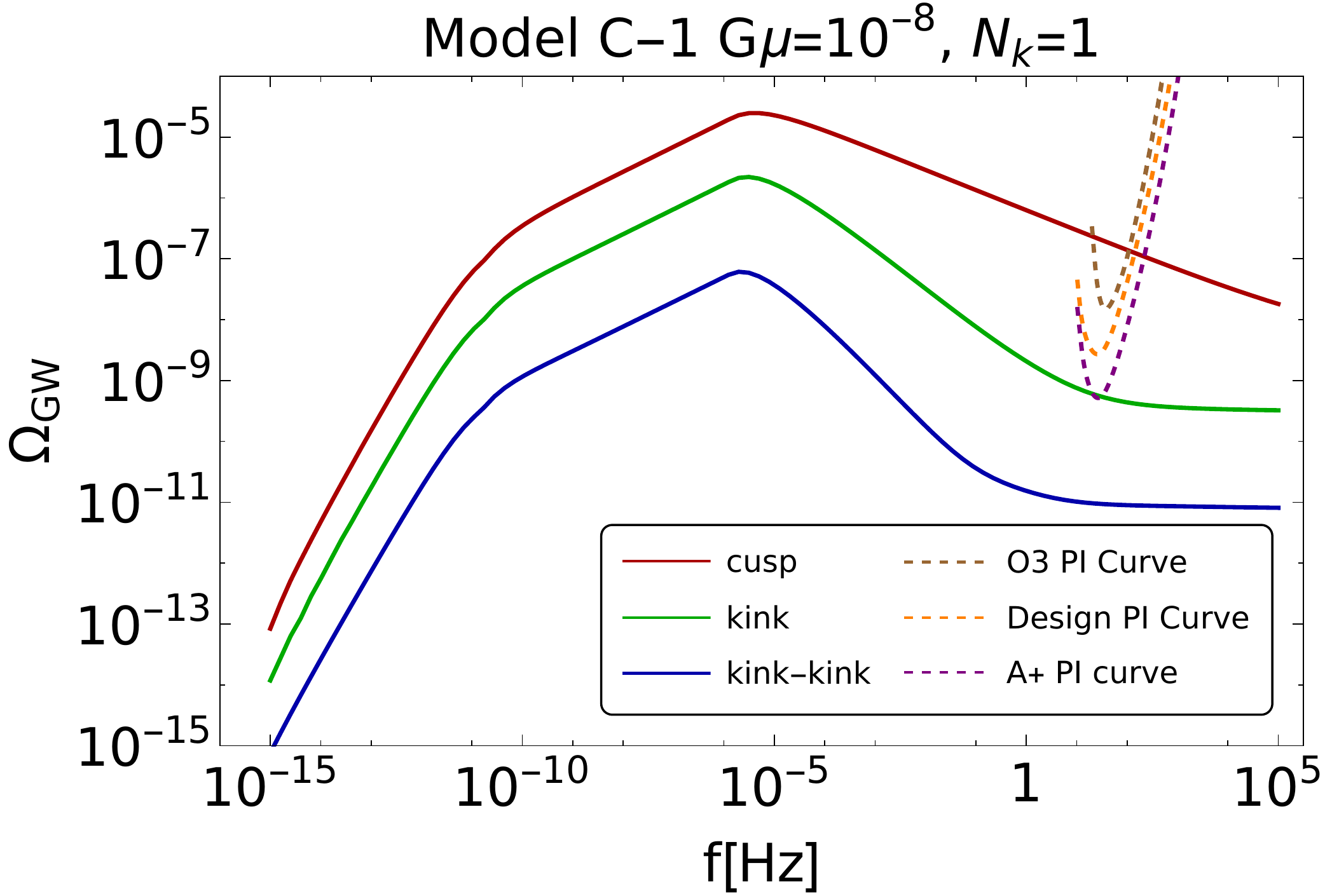}
	\caption{Predicted gravitational wave energy density spectra for the different models outlined in the main text, with the various contributions coming from cusps, kinks, and kink-kink interactions. The dashed curves represent the power-law integrated (PI) sensitivity curves for current and future sensitivities of the LVK detectors, as introduced in Section~\ref{sec:AnalysisTechniques_Isotropic}. The original version of this plot was presented in~\cite{cosmic2021}; the version shown here was obtained using open data published in \cite{cosmicStringsDCC}.}
	\label{Fig:CosmicStringsSpectra}
\end{figure}

\begin{figure}[ht]
	\hspace{-0.5cm}
	\centering
	\includegraphics[scale=.4]{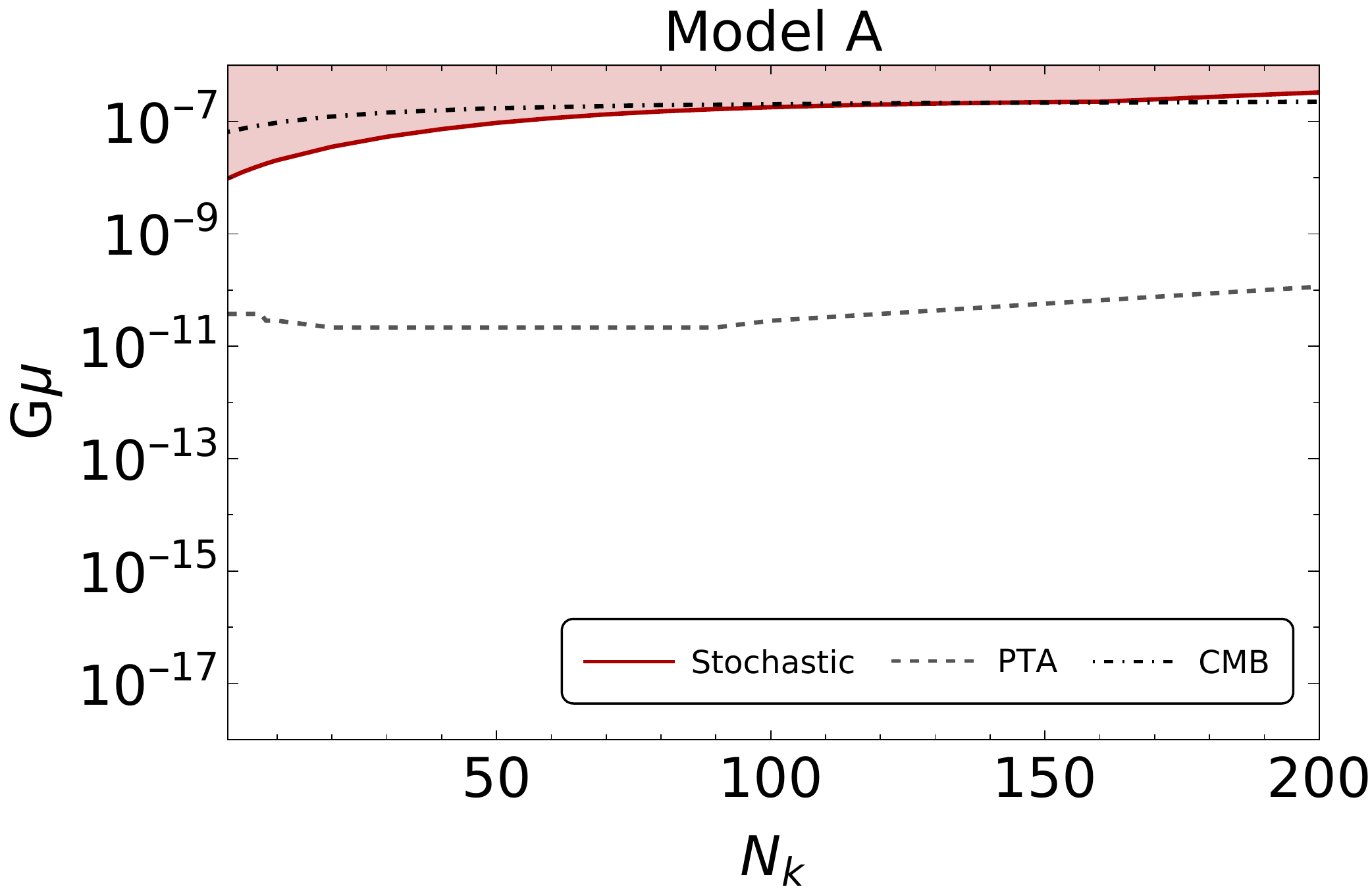}
	\includegraphics[scale=.4]{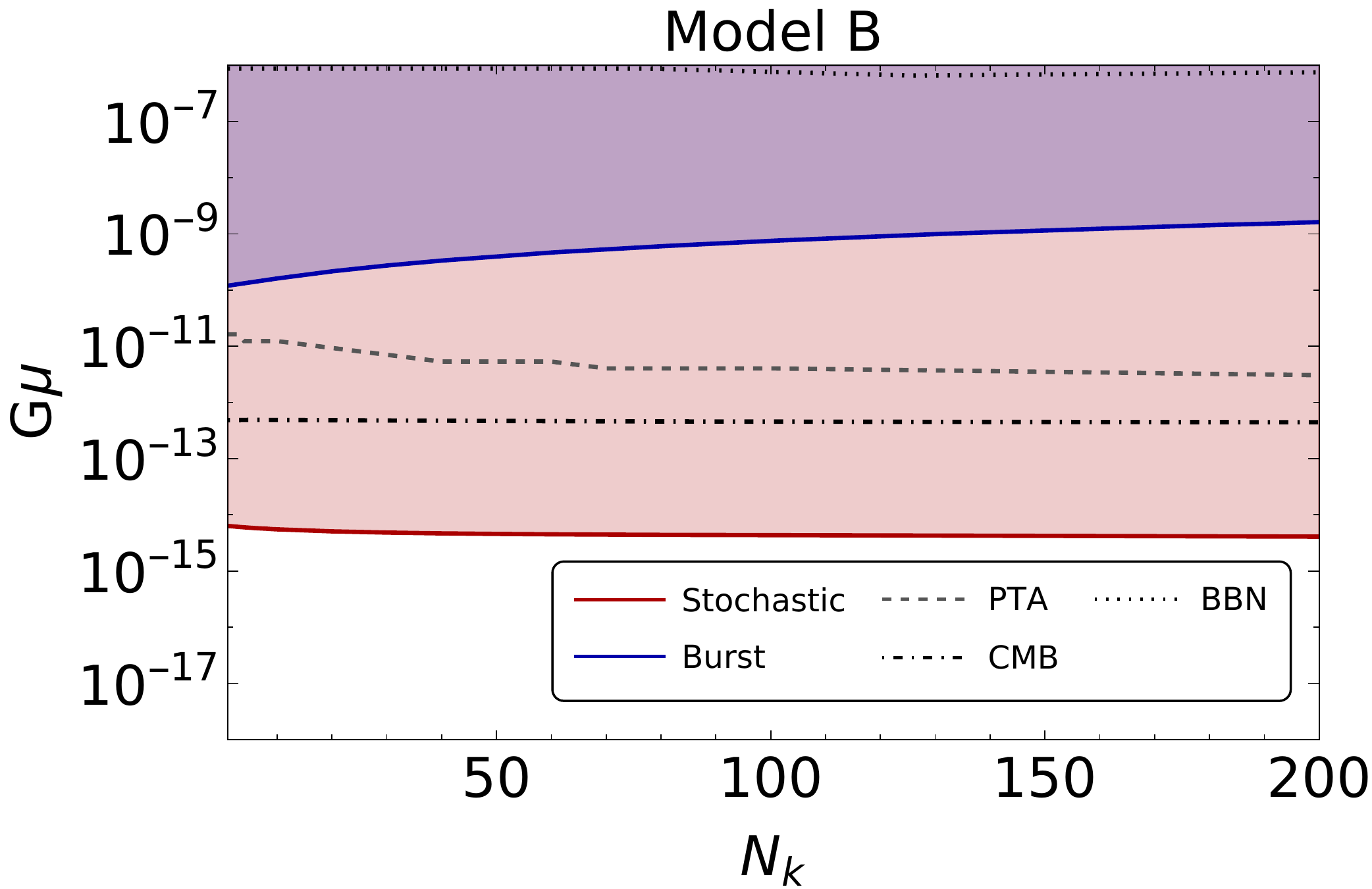}
	\includegraphics[scale=.4]{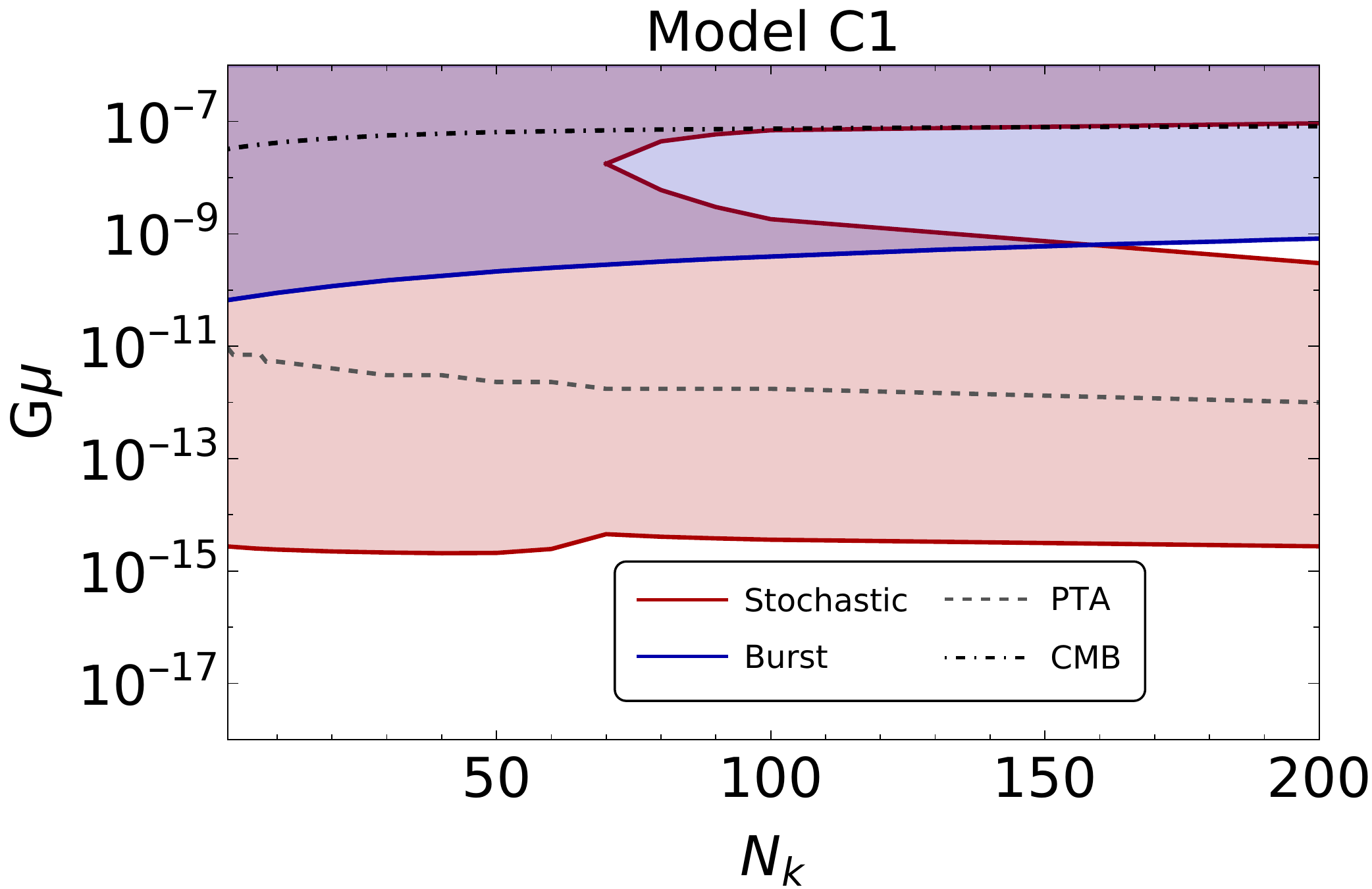}
	\includegraphics[scale=.4]{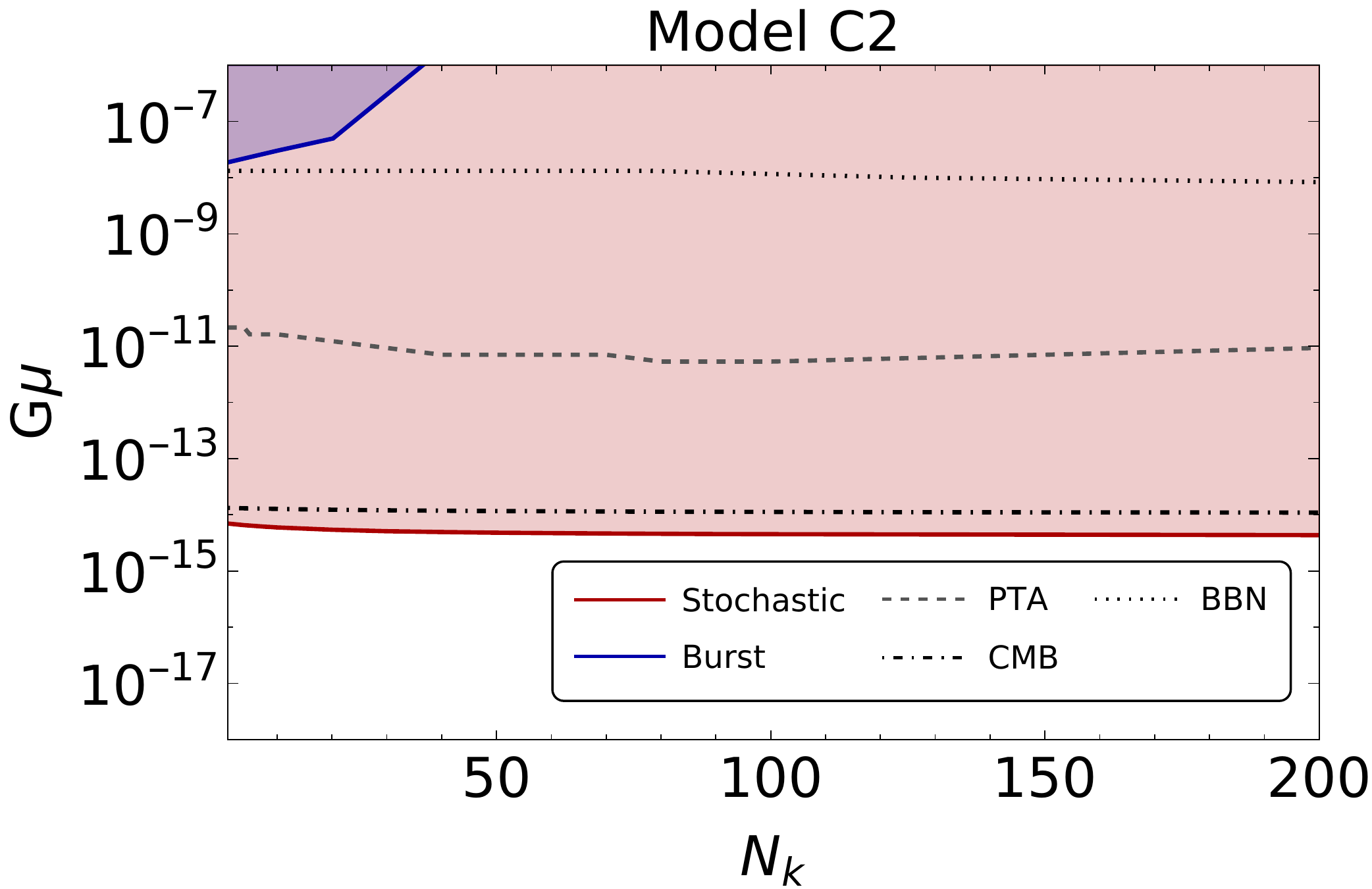}
	\caption{Constraints from the first three observing runs of the LVK collaboration on the number of kinks $N_k$ and the string tension $G\mu$ for the various cosmic strings models models described in the main text of \cite{cosmic2021}. Both the constraints from a stochastic search and a burst are given in red and blue, respectively. The original version of this plot was presented in~\cite{cosmic2021}; the version shown here was obtained using open data published in \cite{cosmicStringsDCC}.}
	\label{Fig:CosmicStrings}
\end{figure}

\begin{figure}[ht]
	\centering
	\includegraphics[scale=.7]{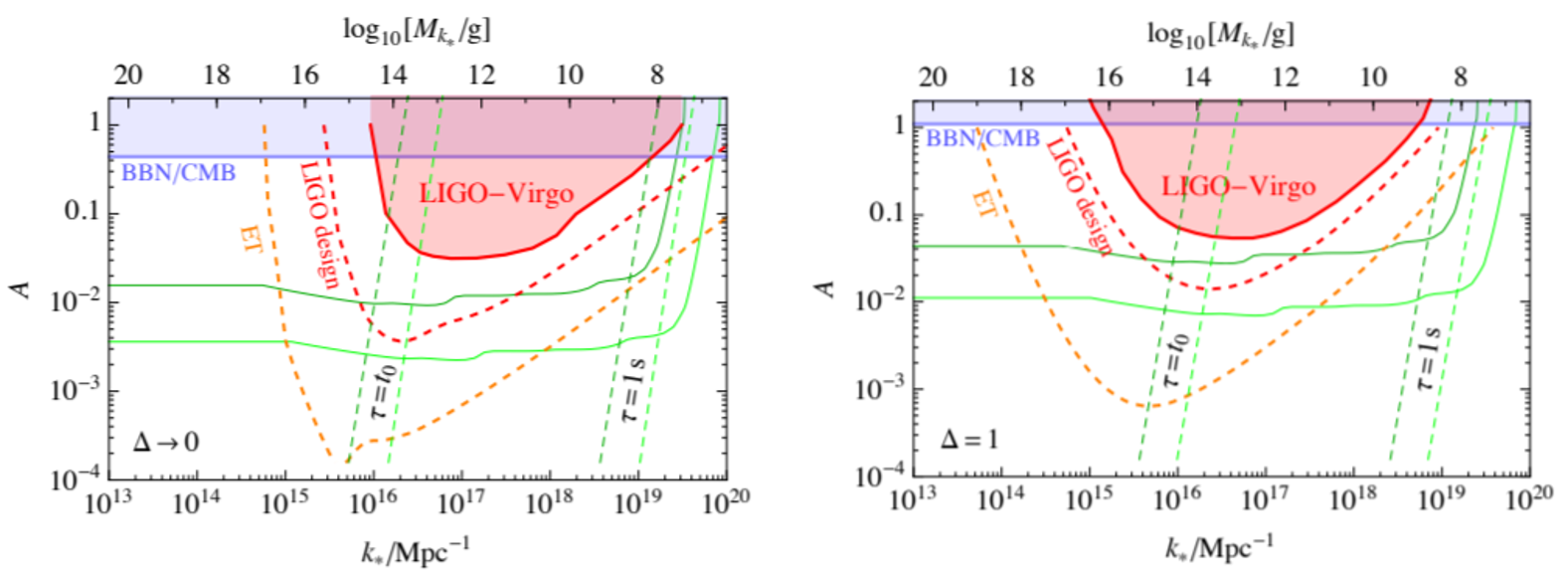}
	\caption{Current 95\% upper limits on the primordial black hole integrated power of the peak $A$ from \cite{PBHRomero2022} for $\Delta=0$ (left) and $\Delta=1$ (right) as a function of the peak position $k_*$ are shown in red using data from the O3 LVK run. The green lines show the constraints from PBH formation based on \cite{pbh2019} (the region in between them reflects the theoretical uncertainty in the PBH formation), whereas the blue band shows the exclusion region from BBN and CMB \cite{PBH2018, PAGANO2016823}. The projected sensitivities for design LVK as well as ET are displayed in dashed red and orange, respectively.}
	\label{Fig:PBH constraints}
\end{figure}

Various processes during the cosmological history of the Universe are expected to give rise to a SGWB. Among these are cosmic string models, first order phase transitions, and primordial black hole formation. For a complete review on these sources and their GW generation, we refer the reader to \cite{Caprini_2018}.\\
Although a SGWB from any of these cosmological models has not yet been detected, it is possible to place constraints on parts of the parameter space of each model using the most recent GW data. Future observing runs of the LVK detectors, or upcoming detectors such as the Einstein Telescope and Cosmic Explorer, will allow to constrain this parameter space even further if not claim a detection of a cosmological SGWB.
\paragraph{First order phase transitions}~\\
During its evolution, the Universe might have gone through a series of phase transitions. If this process is of first order, the Universe will transition from a meta-stable vacuum to a stable vacuum through the nucleation of bubbles. During a first order phase transition (FOPT), these bubbles of stable vacuum will expand and will eventually collide with other bubbles. The collision of such bubbles and their subsequent evolution in the plasma of the Universe is expected to give rise to a SGWB. Although turbulent motion in the plasma is expected to give rise to a SGWB as well, its contribution is expected to be subdominant \cite{Caprini_2016}. Thus, the two main mechanisms of generation that are considered are bubble collision and sound waves sourced in the plasma. Both result in $\Omega_{\rm GW}(f)$ spectra, which are expected to obey a broken power-law, where the slopes of the spectra are different for the bubble collision contribution and the sound wave contribution. The peak frequency of these spectra is largely determined by the overall energy scale of the phase transition. For temperatures at which the phase transition happens $\mathcal{O}(10^7-10^{10}~\text{GeV})$, the peak frequency will be within the frequency range of the LVK detectors. On the other hand, for phase transitions happening at the electroweak scale ($\mathcal{O}\sim 100$ GeV), the resulting FOPT signal will be with in the frequency range of LISA \cite{Caprini_2016}.\\
In \cite{PhysRevLett.126.151301}, two methods are adopted to search for a broken power-law signal in the LVK collaboration's most recent data from O3. The first method consists of looking for a generic broken power-law signal in the data, where one does not assume a specific slope in the spectrum, which takes the form
\begin{equation}
	\Omega_{\rm BPL}(f)=\Omega_*\left(\frac{f}{f_*}\right)^{n_1}\left(1+\left(\frac{f}{f_*}\right)^{\Delta}\right)^{\frac{n_2-n_1}{\Delta}},
\end{equation}
where the spectral indices $n_1$, $n_2$ and $\Delta$ depend on the contribution (sound waves or bubble collisions). By marginalizing over these spectral indices, as well as the peak frequency (which is related to $f_*$), it is possible to derive an upper limit on the amplitude $\Omega_{\rm BPL}$ of a broken power-law signal, as no detection is made. This is done while simultaneously fitting for an astrophysical SGWB from unresolved CBC events, as this background is expected to be present as a foreground of any detected spectrum. The contribution from compact binary coalescences is modeled by
\begin{equation}
	\Omega_{\rm GW}(f)=\Omega_{\rm CBC}\left(\frac{f}{f_{ \rm ref}}\right)^{2/3},
\end{equation}
where $\Omega_{\rm CBC}$ is the GW energy density at a reference frequency $f_{\rm ref}=25$ Hz. For this case, the amplitude of the broken power-law background is constrained to be $\Omega_{\rm BPL}< 4.4\times 10^{-9}$ at a frequency $f=25$ Hz, while the upper limit on an astrophysical background reads $\Omega_{\rm CBC}<6.1\times10^{-9}$ at the same frequency $f=25$ Hz. Note that the upper limit on the astrophysical background from compact binary coalescences is consistent with the upper limits discussed previously (see Section~\ref{sec:AnalysisTechniques_Isotropic}). \\
A second approach consists of considering a more precise phenomenological model for both the bubble collisions and sound waves contributions. For the exact parameterization of the spectra used, we refer to \cite{PhysRevLett.126.151301}. In this case, one is able to derive upper limits on the underlying parameters of the phase transition. These parameters include the temperature at which the phase transition happened $T_{\rm pt}$, the inverse duration of the phase transition in units of Hubble time $\beta/H_*$ and the latent heat $\alpha$. Each of these parameters enters in the spectrum, such that an upper limit on these can be translated into an upper limit on the amplitude of the broken power-law spectrum for both contributions. The 95\% confidence upper limits for the bubble collision and sound wave contributions read $\Omega_{\rm bc}<5.0\times10^{-9}$ and $\Omega_{\rm sw}<5.8\times10^{-9}$, respectively, at a frequency $f=25$ Hz, for phase transitions occurring at temperatures above $T_{\rm pt}=10^8$ GeV.
\paragraph{Cosmic strings}~\\
Another way to generate a stochastic background of GWs is through cosmic strings. These strings are topological defects that are formed during phase transitions at energy scales as high as $10^{16}$ GeV \cite{Kibble_1976}. GWs are expected to be emitted by strings through various mechanisms \cite{PhysRevD.31.3052}. Indeed, periodically oscillating cosmic string loops will emit GWs. Furthermore, bursts of GWs will be emitted by features of the cosmic strings called cusps and kinks \cite{PhysRevD.64.064008}. These cusps are points on the string that instantly travel at the speed of light, whereas kinks are discontinuities in the tangent vector of the string which also travel at the speed of light. The LVK collaboration considered the generation of GWs produced by cusps, kinks and kink-kink interactions and was able to put constraints on various cosmic string models using the data from their first three observing runs \cite{cosmic2021}. As no SGWB was detected by the LVK collaboration (see upper limits placed in Section~\ref{sec:AnalysisTechniques_Isotropic}), the upper limits found by the LVK collaboration were used to constrain the string tension $G\mu$ as a function of the number of kinks $N_k$. Note that the GW spectrum due to cosmic strings is expected to be approximately flat within the frequency range of the LVK detectors, i.e. with spectral index $\alpha=0$, such that the upper bound reads $\Omega_{\rm GW}<5.8\times10^{-9}$ at the 95\% credible level \cite{cosmic2021}. However, to derive upper limits on the string tension $G\mu$ as a function of the number of kinks $N_k$, the present search performs a Bayesian analysis that takes into account the exact shape of the spectrum rather than assuming a power-law. 
Concretely, three models were considered to describe the distribution of cosmic string loops, which were labeled by A, B and C in \cite{cosmic2021}, all yielding a different expected GW spectrum. Models A and B were obtained from numerical Nambu-Goto string simulations, where the main qualitative difference between the two models is larger amount of tiny loops produced in the latter \cite{PhysRevD.89.023512, Lorenz_2010}. On the other hand, model C extends and encompasses both models by interpolating between the two \cite{Auclair_2019}. A few examples of the GW spectra obtained in each of the models are given in Fig~\ref{Fig:CosmicStringsSpectra}.
~\\
As each model yields a different GW energy density spectrum, different upper limits are found for the string tension $G\mu$, displaying little to no dependence on the number of kinks $N_k$. The LVK collaboration places the following constraints on the string tension, where the various ranges encompass a range of number of kinks values $N_k$: $G\mu<9.6\times10^{-9}-10^{-6}$ for model A, $G\mu<(4.0-6.3)\times10^{-15}$ for model B, $G\mu<(2.1-4.5)\times10^{-15}$ for model C-1 and $G\mu<(4.2-7.0)\times10^{-15}$ for model C-2, where model C-1 and C-2 refer to model C in the regime it corresponds to model A and B, respectively. These upper limits are illustrated in Fig~\ref{Fig:CosmicStrings}. Note that constraints from CMB and PTA measurements are given as well, although weaker than the ones placed by the LVK collaboration (expcept for model A). Furthermore, the LVK collaboration performed a search to look for bursts of GW coming from cosmic strings as well, for which the constraints are reported in the same figure, Fig. \ref{Fig:CosmicStrings} \cite{cosmic2021}. 
\paragraph{Alternative polarizations}~\\
Within the search for an isotropic SGWB by the LVK collaboration \cite{PhysRevD.104.022004}, deviations from general relativity are also sought for. In general relativity, only tensor polarizations are allowed for GWs. However, other polarizations such as vector and scalar polarizations may appear in alternative gravity theories \cite{SGWBRevRomano}.\\
Concretely, in a search for a SGWB, such different polarizations would have an effect on the ORF that enters in the analysis. Indeed, different polarizations lead to different ORFs, as is apparent from the summation over polarizations in the definition of the ORF in Eq.\eqref{eq:ORF}. Thus, to search for alternative polarizations, one can define a model that accounts for these various polarizations: $\Omega_{\rm GW}=\sum_p\beta^{(p)}_{IJ}(f)\Omega_{\rm ref}^{(p)}\left(\frac{f}{f_{\rm ref}}\right)^{\alpha_p}$, where $\beta^{(p)}_{IJ}(f)=\gamma_{IJ}^{(p)}(f)/\gamma_{IJ}(f)$ is the ratio of the ORF for baseline $IJ$ with polarization $p$ to the standard, i.e. tensor, ORF for that baseline. The sum runs over the possible polarizations (tensor(T), vector(V), and scalar(S)) with corresponding amplitudes $\Omega_{\rm ref}^{(p)}$ and spectral indices $\alpha_p$ for each polarization. The goal is then to measure, or place upper limits on in the absence of a detection, the three amplitudes $\Omega_{\rm ref}^{(p)}$ and spectral indices $\alpha_p$ for $p\in\{T,V,S\}$, the various polarizations. Using data from their first three observing runs, the LVK collaboration was unable to find evidence for alternative polarizations \cite{PhysRevD.104.022004}, indicating consistency with general relativity. After marginalizing over the spectral index, upper limits on the amplitudes are found at a reference frequency $f_{\rm ref}=25$ Hz: $\Omega_{\rm GW}^{(S)}<2.1\times10^{-8}$, $\Omega_{\rm GW}^{(V)}<7.9\times10^{-9}$, $\Omega_{\rm GW}^{(T)}<6.4\times10^{-9}$ for scalar, vector, and tensor polarizations respectively. For comparison, we recall that the current upper limit on a tensor-polarized SGWB is of the same order of magnitude, e.g. $\Omega_{\rm ref}\le5.8\times 10^{-9}$ for a flat power-law spectrum with $\alpha=0$, as was discussed in Sec. \ref{sec:AnalysisTechniques_Isotropic}.
\paragraph{Primordial black holes}~\\
Primordial black holes (PBH) could be yet another source of a SGWB. These primordial black holes have gained a lot of interest over the last few years as they constitute a potential candidate for dark matter. They could have formed during inflation due to scalar fluctuations in the early Universe, which would have sourced a scalar induced SGWB. Such a SGWB is searched for in the data of the first three observing runs of the LVK collaboration in \cite{PBHRomero2022}. The GW spectrum resulting from scalar fluctuations is completely determined by the curvature power spectrum. The authors of \cite{PBHRomero2022} decide to parameterize the curvature power spectrum by
\begin{equation}
	\label{Eq: scalar power spectrum}
	P_\zeta(k)=\frac{A}{\sqrt{2\pi}\Delta}\exp\left(-\frac{\ln^2(k/k_*)}{2\Delta^2}\right),
\end{equation}
where $A$ is the integrated power of the peak, $\Delta$ determines the width of the peak and $k_*$ the position of the peak. The expected GW spectrum from such a curvature power spectrum can be found in \cite{GWSpectrumFromPBH2020} and is shown to have a peak at a frequency $f_*=1.6\times10^{-15}k_*$/Mpc$^{-1}$ Hz with an amplitude that is given by $\Omega_{\rm GW}=\mathcal{O}\left(10^{-5}\right)A^2$ (for $\Delta\ll1$). The authors \cite{PBHRomero2022} then proceed with a Bayesian analysis, searching for a GW signal in the O3 data of the LVK collaboration. This signal is assumed to contain both a SGWB from compact binary coalescences (regular, non-PBH, astrophysical SGWB), as well as a contribution from the GWs induced by the scalar fluctuations (PBH). As the Bayes factor is found to be $\ln\mathcal{B}_{\rm noise}^{\rm CBC+scalar}=-0.8$, no evidence is found for a stochastic background of GWs containing a signal from compact binary coalescences and/or scalar induced GWs. Nevertheless, in the absence of a detection, 95\% upper limits can be derived. However, these upper limits on the parameters describing the scalar power spectrum in Eq.~\eqref{Eq: scalar power spectrum} are less stringent than the exclusions coming from primordial black hole abundance, as illustrated in Fig~\ref{Fig:PBH constraints}. Indeed, using the prescription outlined in \cite{pbh2019}, it is possible to relate the power spectrum to the PBH abundance, which is constrained in \cite{PBHConstraints2010} using constraints from dark matter abundance and CMB/BBN measurements. These constraints on the integrated power $A$ are given by the green lines in Fig~\ref{Fig:PBH constraints}. As one can see, the current sensitivity of Earth-based interferometric gravitational-wave detectors does not allow to place better constraints on primordial black hole formation. Nevertheless, this is expected to change with the upcoming third generation detectors such as the Einstein Telescope, as is illustrated in Fig~\ref{Fig:PBH constraints}.
\paragraph{Parity violation}~\\
To conclude this section on cosmological models, we give an overview of the search for a SGWB coming from parity violating models conducted in \cite{Parity2021}, based on the formalism of \cite{PhysRevLett.99.121101,Seto_2008}. Several early Universe processes, e.g. turbulence due to first order phase transitions \cite{PhysRevD.49.2837} or axion inflation \cite{PhysRevLett.106.181301}, could generate asymmetric amounts of left- and right-handed circularly polarized GWs, as a result from parity violation. By observing the amount of polarization asymmetry in the GW signal, one can constrain parity violating models.\\~\\
Concretely, one considers the plane-wave expansion of the metric
\begin{equation}
	h_{ab}(t,\vec{x})=\sum_A\int_{-\infty}^{+\infty}df\int_{S^2}d\hat{\bf{n}}h_A(f,\hat{\bf{n}})e^{-2\pi i f (t-\vec{x}\cdot\hat{\bf{n}})}e_{ab}^A(\hat{\bf{n}}),
\end{equation}
where $f$ denotes the frequency and $e^A_{ab}$ the polarization tensor for a wave traveling in the $\hat{\bf{n}}$ direction. Using a circularly polarized basis, one can then write down the right- and left-handed correlators:
\begin{equation}
	\begin{pmatrix}
		\braket{h_R(f,\hat{\bf{n}})h^*_R(f^\prime,\hat{\bf{n}}^\prime)}\\
		\braket{h_L(f,\hat{\bf{n}})h^*_L(f^\prime, \hat{\bf{n}}^\prime)}
	\end{pmatrix}
	=\frac{\delta(f-f^\prime)\delta^2(\hat{\bf{n}}-\hat{\bf{n}}^\prime)}{4\pi}
	\begin{pmatrix}
		I(f,\hat{\bf{n}})+V(f,\hat{\bf{n}})\\
		I(f,\hat{\bf{n}})-V(f,\hat{\bf{n}})
	\end{pmatrix},
\end{equation}
where $I$ and $V$ are the Stokes parameters and $V$ characterizes the amount of asymmetry between left- and right-handed modes. Thus, if $V=0$, one would simply have an unpolarized background. The standard cross-correlation estimator is modified to account for parity violation:
\begin{equation}
	\braket{\hat{C}_{IJ}}=\frac{3H_0^2T}{10\pi^2}\int_0^\infty df\frac{\Omega^{\prime}_{\rm GW}(f)\gamma^{IJ}_I(f)\tilde{Q}(f)}{f^3}
\end{equation}
where $\tilde{Q}$ is a filter function, and
\begin{equation}
	\Omega_{\rm GW}^\prime=\Omega_{\rm GW}\left(1+\Pi(f)\frac{\gamma_V^{IJ}(f)}{\gamma_I^{IJ}(f)}\right),
\end{equation}
where $\gamma_I^{IJ}(f)$ and $\gamma_V^{IJ}(f)$ respectively denote the standard ORF and the ORF associated with the parity violation, which are defined in \cite{Parity2021}. The factor $\Pi(f)=V(f)/I(f)$ runs between -1 and 1 and is referred to as the polarization degree. One sees that for the case $V=0$, i.e. without asymmetry between left- and right-handed modes, the cross-correlation estimator reduces to the usual one used in the standard isotropic search as introduced in Eq.~\eqref{eq:methods:bin_by_bin_estimator}.\\
A search for such parity-violating signals was conducted on the first three observing runs of the LVK collaboration \cite{Parity2021}. Concretely, the GW spectrum is assumed to obey a power-law: $\Omega_{\rm GW}(f)=\Omega_{\rm ref}\left(\frac{f}{f_{\rm ref}}\right)^{\alpha}$. Furthermore, two different cases are considered for the polarization degree: $\Pi(f)=f^{\beta}$ and $\Pi=\rm const$. For both cases, no evidence is found for a SGWB, nor for parity violation. Nevertheless, upper limits are placed on the amplitude of the background for both cases:
$\Omega_{\rm ref}^{95\%}=4.9\times10^{-9}$ (for a frequency-dependent polarization degree, when marginalizing over $\beta$) and $\Omega_{\rm ref}^{95\%}=7\times10^{-9}$ (for a constant polarization degree).
\\
As no evidence was found for a SGWB from parity violating sources, data is simulated to perform further analyses on the detectability of parity violation within a SGWB \cite{Parity2021}. By performing studies on simulated data, it was found that the parameters describing the parity violation model are better estimated for stronger SGWBs, as well as when the Virgo and KAGRA detectors are included in the analysis. This emphasizes the importance of having a network consisting of multiple detectors in order to probe the signal from parity violating models.

    \newpage
%%%%%%%%%%%%%%%
%%% Outlook %%%
%%%%%%%%%%%%%%%

\section{Outlook}
\label{sec:Outlook}
We started this review with an introduction to gravitational waves and the stochastic gravitational wave background by giving an overview of various definitions and properties. We then covered various detection methods: resonant gravitational wave detectors, pulsar timing arrays and interferometric gravitational-wave detectors. For each of these we summarized current upper limits, as the stochastic background of gravitational waves remains unobserved to this day. Furthermore, we discussed analysis techniques that are used or are in development for data from Earth-based interferometric gravitational-wave detectors. Additionally we provided an overview of various validation techniques that can be used when claiming the detection of a background.

Although the stochastic background of gravitational waves has not been observed, current upper limits are already able to put constraints on astrophysical and cosmological models, as summarized in this review. The observation of gravitational waves started the field of multi-messenger astronomy and the observation of a stochastic gravitational wave background would allow for even more synergies between different fields. Indeed, as illustrated in the various examples of possible implications, particle physics models and early Universe cosmology can be constrained through the (absence of) observation of a stochastic gravitational wave background. The current sensitivity of Earth-based interferometric gravitational-wave detectors, although insufficient to make a detection of the stochastic gravitational wave background, already allows one to place constraints on the parameter space of various models. These are expected to be improved upon in the coming years with the current instruments and even more drastically with the advent of so-called third generation experiments, such as the Einstein Telescope or Cosmic Explorer, and space-based detectors, such as LISA.

	\newpage
	\section*{Acknowledgements}
The authors are grateful for their colleagues within the LIGO-Virgo-KAGRA collaborations. The authors are also grateful to Nelson Christensen for fruitful discussions as well as Tom Callister and Boris Goncharov for their suggestions concerning the paper. 
Kamiel Janssens and Kevin Turbang are supported by FWO-Vlaanderen through grant numbers 11C5720N and 1179522N, respectively.

	\bibliography{references}
	%Please use Bib\TeX\ to generate your bibliography and include DOIs whenever available. Example of bib file: 
	
	%%%%%%%%%%%%%%%%%%%%%%%%%%%%%%%%%%%%%%%%%%%%%%%%%%%%%%%%%%%%%%%%%%%
	% Encoding: ISO-8859-1

	%@Article{Eichmann:2016yit,
	%author        = {Eichmann, Gernot and Sanchis-Alepuz, Helios and Williams, Richard and Alkofer, Reinhard and Fischer, Christian S.},
	%title         = {{Baryons as relativistic three-quark bound states}},
	%journal       = {Prog. Part. Nucl. Phys.},
	%year          = {2016},
	%volume        = {91},
	%pages         = {1-100},
	%archiveprefix = {arXiv},
	%doi           = {10.1016/j.ppnp.2016.07.001},
	%eprint        = {1606.09602},
	%owner         = {chfi},
	%primaryclass  = {hep-ph},
	%slaccitation  = {%%CITATION = ARXIV:1606.09602;%%},
	%timestamp     = {2018.08.02},
	%}

	%@Comment{jabref-meta: databaseType:bibtex;}
	%%%%%%%%%%%%%%%%%%%%%%%%%%%%%%%%%%%%%%%%%%%%%%%%%%%%%%%%%%%%%%%%%%%
	
	%\newpage
	%\appendix
	%\renewcommand*{\thesection}{\Alph{section}}
	
	%\section{Appendices, if necessary}\label{appendix}

\end{document}